\documentclass{aa}  

\usepackage{graphicx}
\usepackage{txfonts}
\usepackage{tabularx}
\usepackage{color}
\usepackage{amsmath}
\usepackage{tikz,graphics,color,fullpage,float,epsf,caption,subcaption}
\usepackage[colorlinks=true,linkcolor=blue,citecolor=blue]{hyperref}
\usepackage{silence}
\begin{document}

   \title{Galaxy mergers in Subaru HSC-SSP: a deep representation learning approach for identification and the role of environment on merger incidence}

   \subtitle{}

   \author{Kiyoaki Christopher Omori
          \inst{1}
                    \and
          Connor Bottrell\inst{3,4,5}
          \and
          Mike Walmsley\inst{6}
          \and
          Hassen M.\ Yesuf\inst{3, 7}
          \and
          Andy D.\ Goulding\inst{8}
          \and
          Xuheng Ding\inst{3}
          \and
          Gerg\"{o} Popping\inst{13}
          \and
          John D.\ Silverman\inst{3, 9}
          \and
          Tsutomu T.\ Takeuchi\inst{1, 2}
          \and
          Yoshiki Toba\inst{10,11,12}
          }

    \institute{$^1$Division of Particle and Astrophysical Science, Nagoya University, Furo-cho, Chikusa-ku, Nagoya 464--8602, Japan\\
    $^2$The Research Centre for Statistical Machine Learning, the Institute of Statistical Mathematics, 10--3 Midori-cho, Tachikawa, Tokyo 190--8562, Japan\\
    $^3$Kavli Institute for the Physics and Mathematics of the Universe (WPI), UTIAS, University of Tokyo, Kashiwa, Chiba 277-8583, Japan\\    
    $^4$International Centre for Radio Astronomy Research, University of Western Australia, 35 Stirling Hwy, Crawley, WA6009, Australia\\
    $^5$Center for Data-Driven Discovery, Kavli IPMU (WPI), UTIAS, The University of Tokyo, Kashiwa, Chiba 277-8583, Japan\\
    $^6$Jodrell Bank Centre for Astrophysics, Department of Physics \& Astronomy, University of Manchester, Oxford Road, Manchester M13 9PL, UK\\
    $^7$Kavli Institute for Astronomy and Astrophysics, Peking University, Beijing 100871, People’s Republic of China\\
    $^8$Department of Astrophysical Sciences, Princeton University,Princeton, NJ 08544, USA\\
    $^9$Department of Astronomy, School of Science, The University of Tokyo, 7-3-1 Hongo, Bunkyo, Tokyo 113-0033, Japan\\
    $^{10}$National Astronomical Observatory of Japan, 2-21-1 Osawa, Mitaka, Tokyo 181-8588, Japan\\
    $^{11}$Academia Sinica Institute of Astronomy and Astrophysics, 11F of Astronomy-Mathematics Building, AS/NTU, No.1, Section 4, Roosevelt Road, Taipei 10617, Taiwan\\
    $^{12}$Research Center for Space and Cosmic Evolution, Ehime University, 2-5 Bunkyo-cho, Matsuyama, Ehime 790-8577, Japan\\
    $^{13}$European Southern Observatory, Karl-Schwarzschild-Str. 2, D-85748, Garching, Germany}

   \date{Received September , 2023; accepted September, 2023}

% \abstract{}{}{}{}{} 
% 5 {} token are mandatory
 
  \abstract
  % context heading (optional)
  % {} leave it empty if necessary  
   {Galaxy mergers and interactions are an important process within the context of galaxy evolution, however a method which identifies pure and complete merger samples is still not definitive. A method to create such a merger sample is required, so that studies can be conducted to deepen our understanding on the merger process and its impact on galaxy evolution.}%To create a merger galaxy catalogue for Subaru HSC-SSP by fine-tuning a deep learning model pre-trained on galaxy images, and to evaluate its performance. We also evaluate the relationship between galaxy mergers and environment.}
  % aims heading (mandatory)
   {In this work, we take a deep learning-based approach for galaxy merger identification in Subaru HSC-SSP, specifically through the use of deep representation learning and fine-tuning, with the aim of creating a pure and complete merger sample within the HSC-SSP survey. We can use this merger sample to conduct studies on how mergers affect galaxy evolution.}
  % methods heading (mandatory)
   {We use Zoobot, a deep learning representation learning model pre-trained on citizen science votes on Galaxy Zoo DeCALS images. We fine-tune Zoobot for the purpose of merger classification of images of SDSS and GAMA galaxies in HSC-SSP public data release 3. Fine-tuning is done using $\sim1200$ synthetic HSC-SSP images of galaxies from the TNG simulation. We then find merger probabilities on observed HSC images using the fine-tuned model. Using our merger probabilities, we examine the relationship between merger activity and environment.}
  % results heading (mandatory)
   {We find that our fine-tuned model returns an accuracy on the synthetic validation data of $\sim76\%$. This number is comparable to those of previous studies where convolutional neural networks were trained with simulation images, but with our work requiring a far smaller number of training samples. For our synthetic data, our model is able to achieve completeness and precision values of $\sim80\%$. In addition, our model is able to correctly classify both mergers and non-mergers of diverse morphologies and structures, including those at various stages and mass ratios, while distinguishing between projections and merger pairs. For the relation between galaxy mergers and environment, we find two distinct trends. Using stellar mass overdensity estimates for TNG simulations and observations using SDSS and GAMA, we find that galaxies with higher merger scores favor lower density environments on scales of 0.5 to 8 $h^{-1}$Mpc. However, below these scales in the simulations, we find that galaxies with higher merger scores favor higher density environments.}
  % conclusions heading (optional), leave it empty if necessary 
   {We fine-tuned a citizen-science trained deep representation learning model for purpose of merger galaxy classification in HSC-SSP, and make our merger probability catalogue available to the public. Using our morphology-based catalogue, we find that mergers are more prevalent in lower density environments on scales of 0.5 - 8 $h^{-1}$Mpc.}

    \titlerunning{Merger identification in HSC-SSP using deep learning}
    \authorrunning{Omori et al.}
   \maketitle
%
%-------------------------------------------------------------------

\section{Introduction}
\label{section:Intro}

   Galaxy evolution involves many processes that can affect the physical properties of the involved galaxies. Galaxy interactions and mergers are considered to be an important driver of physical phenomena and evolution in galaxies. For example, galaxy interactions and mergers can drive inflow of gas towards the centers of galaxies \citep{1989Natur.340..687H, 1992ARA&A..30..705B, 1996ApJ...464..641M,2001ASPC..249..735N,2010MNRAS.407.1529H,2018MNRAS.479.3952B}. These inflows can enhance star formation activity \citep{2008ChJAS...8...77B, 2008AJ....135.1877E,2011MNRAS.412..591P,2013MNRAS.433L..59P,2013MNRAS.430.1901H,2015MNRAS.448.1107M,2016MNRAS.462.2418S,2019MNRAS.482L..55T}, dilute central gas phase metallicities \citep{2008AJ....135.1877E,Rupke_2010, Montouri2010, 2010A&A...514A..57S, 2011MNRAS.417..580P,2012ApJ...746..108T,2016MNRAS.462.2418S,2019MNRAS.482L..55T}, trigger accretion onto supermassive blackholes \citep{1985AJ.....90..708K,1988ApJ...325...74S,Di_Matteo_2005,2010ApJ...716L.125K,2011MNRAS.418.2043E,2014MNRAS.441.1297S,2015MNRAS.451L..35E,2018PASJ...70S..37G,2019MNRAS.487.2491E}, and trigger quasars \citep{2008ApJ...674...80U}. More broadly, in the $\Lambda$-dominated cold dark matter ($\Lambda$CDM) framework for structure formation in the Universe, accretion of stellar material though galaxy mergers (ex-situ assembly) is a key process by which massive galaxies grow their stellar mass.

   Despite its importance, we do not have a full understanding of galaxy interactions and mergers. While studies have been able to quantify the role of mergers as a driver of stellar mass growth \citep{2014MNRAS.444.3986R,2016MNRAS.458.2371R}, the specific role of mergers in driving stellar mass growth, enhancement of star formation and active galactic nuclei (AGN) activity, and morphological transformations is still contentious. Even the environment in which galaxy mergers are prevalent does not have a definitive conclusion. Some dark matter halo simulations predict that mergers are more likely to happen in lower mass density regions \citep{1998MNRAS.300..146G}, however other simulation results \citep{2009MNRAS.394.1825F,2010ApJ...715..342H}, as well as observational studies \citep{2012ApJ...754...26J} do not necessarily agree with this prediction, stating that mergers occur in denser environments. A major reason for the lack of clear understanding of the role of mergers and interactions in galaxy evolution or their environments is the difficulty of precise identification of merger galaxies in observational data.
   
   There have been numerous studies employing several different methods for merger identification.
   One approach is the close-pairs method. This method searches for binary pairs in the sky using imaging and photometry or spectroscopy \citep[e.g.,][]{2004ApJ...617L...9L, 2007AJ....134...71S}.  This method has a number of issues. First, this approach misses galaxies in the post-coalescence phase of a merger, as there is only access to single-galaxy characteristics. Second, this approach requires spectroscopic redshifts, which also may be impacted by spectroscopic incompleteness. Third, merger rates in observations may be overestimated compared to those in simulations, even if the same criteria is used, due to reasons such as chance projections, as well as the difficulty of obtaining accurate  merging timescales \citep{2008MNRAS.391.1489K}. The criteria for pair selection, such as exact velocity and separation cuts, have undergone significant refinement to register fewer interlopers in merging pair samples \citep{1989ApJ...337...34Z, 1994ApJ...429L..13B, 1994ApJ...435..540C, 1995ApJ...445...37Y, 1995ApJ...454...32W, Patton_1997, 1998AJ....116.1513W}. Also, some mergers may not be detected by this method, as there may be merging galaxy pairs with a pair distance greater than the maximum projected distance adopted for a study.
   
   Other methods rely on galaxy imaging data and their morphologies. Galaxy interactions can cause disturbances to the morphology of a galaxy \citep{1972ApJ...178..623T,1977egsp.conf..401T, 1983MNRAS.205.1009N,1992ApJ...400..460H,2003ApJ...597..893N,2008ApJS..175..390H, 2014MNRAS.440L..66B}, which can be quantified into non-parametric statistics. An example of these statistics are the CAS parameters (Concentration \citep{2000AJ....119.2645B}, Asymmetry \citep{2000ApJ...529..886C}, Smoothness \citep{2003ApJS..147....1C}), Gini and $M_{20}$ \citep{2004AJ....128..163L}. An \textit{n} number of these parameters can be combined to create a criteria for merger classification in the \textit{n}-dimensional feature space \citep[e.g.,][]{2018PASJ...70S..37G, 2019MNRAS.486.3702S,2023ApJ...942...54R,2021RNAAS...5..144T, 2023MNRAS.519.4920G}. Using these statistics can reduce the dimensionality of data being used. As a result, the required number of training samples can be reduced compared to  image classification methods such as convolutional neural networks. The issue with this method is that merger-driven morphological disturbances are low surface brightness features \citep{2000ApJ...529..886C,2019MNRAS.486..390B,2021MNRAS.507..886T,10.1093/mnras/stac1962} which require high quality imaging, both in terms of depth and resolution, to identify. Modern day imaging surveys, such as the multi-tiered, wide-field, multi-band imaging survey Hyper Suprime-Cam Subaru Strategic Program (HSC-SSP, \citealt[][see Section 3.1]{2018PASJ...70S...8A}), can allow for such imaging where these features are visible. However, the issue still remains that these non-parametric statistics do not capture the complexity in high-quality imaging data provided by modern wide-field galaxy surveys, and as such, not all information from images can be extracted from these statistics and the \textit{n}-dimensional feature spaces using them.
  
   There have also been methods relying on visual inspection of galaxy images. When conducting morphological studies, visual classification by experts is largely considered the gold standard \citep{2010yCat..21860427N}. Foremost, the visual approach can incorporate domain-specific human knowledge in the classification of galaxies, i.e., human knowledge of what is visually a merger can be used to purify any classification results, particularly in classifications done by machine learning methods. Indeed, visual follow-ups by human classifiers is often employed to purify merger `candidate' samples produced by automated and quantitative approaches \citep[e.g.,][]{2022MNRAS.514.3294B, 2022A&A...661A..52P}. The visually distilled samples yield more robust scientific outcomes on merger properties. Therefore, it is clear that domain knowledge provided by human classifiers is valuable. Second, visual classifications, in principle, use all of the morphological information encoded in high-quality galaxy images, and is not restricted to a set of summary statistics \citep{2020MNRAS.492.2075B}. However, this method also has its weaknesses. First, the criteria for what is considered a merger can differ depending on the individual carrying out the visual inspection: so the same galaxy may be assigned a different label by different people. Additionally, visual inspection by humans can be very time-consuming, and not realistic for large datasets in modern day galaxy surveys. The Galaxy Zoo Project \citep[][hereinafter referred to as GZ1]{2008MNRAS.389.1179L, 2011MNRAS.410..166L} overcame these issues to an extent. GZ1 is a catalogue \citep{2010MNRAS.401.1552D} offering morphology probabilities for over 1 million Sloan Digital Sky Survey (SDSS) galaxies, with citizen scientists assigning labels for morphological features. The labels assigned in this catalogue are weighted in accordance with the `correctness' of the scientists. Merger probabilities are part of this catalogue, and have been used in many merger-related studies, such as \citet{2016MNRAS.459..720H,2017ApJ...845..145W}. However, while citizen science can be more time efficient than classifications made by expert scientists, it may not be as reliable.
   
   Recent advents in deep learning technology for image-based galaxy characterization have made the visual inspection process less time- and human-resource consuming. Convolutional neural networks (CNNs) and deep learning models have achieved performances greater than other computer imaging methods \citep{10.1145/3065386}, even surpassing the performance of some human classifications \citep{7410480}. CNNs have already been used in a number of studies, both in galaxy morphology classification as a whole \citep[e.g.,][]{2015MNRAS.450.1441D,2018MNRAS.476.3661D,2019MNRAS.484.5330J,2019Ap&SS.364...55Z,2020ApJ...895..112G,2021MNRAS.507.4425C,2022MNRAS.509.4024D, 2022MNRAS.509.3966W,2023MNRAS.520.5885C,2023PASA...40....1H}, and the specific task of galaxy merger classification, with varying levels of accuracy \citep[e.g.,][]{2019MNRAS.483.2968W, 2019A&A...626A..49P, 2019MNRAS.490.5390B,2021MNRAS.504..372B, 2020A&C....3200390C, 2020ApJ...895..115F, 2021MNRAS.506..677C,2022MNRAS.514.3294B, 2022MNRAS.511..100B, 2022A&A...661A..52P,2022ApJ...931...34F}.
   
    In this work, we investigate a particular approach of the training process in CNNs, in the form of transfer learning. Training a CNN from scratch requires a very large labeled training set, and preparing such a dataset for each classification task can be very time- and human-resource expensive, as highlighted in the issues with visual classification above. This step can be potentially streamlined and made more efficient through transfer learning, or transferring the knowledge from a previous study and adapting it to a new dataset. The approach of using transfer learning for galaxy merger identification was conducted in \citet{Ackermann_2018}, which found that transfer learning using the diverse ImageNet dataset can lead to improvements over conventional machine learning methods. 
   In \citet{2019MNRAS.484...93D}, transfer learning and fine-tuning using astronomical data was conducted. This study found that knowledge can be transferred between astronomical surveys, and that combining transfer learning and fine-tuning can boost model performance and reduce training sample size. 
   This work will combine the approaches of the above works. We will use the techniques of transfer learning and fine-tuning, through the use of the pre-trained model Zoobot \citep{Walmsley2023}. Zoobot is a pre-trained model trained on diverse astronomical images, using human knowledge in its pre-trained weights as a foundation. In \citet{2022MNRAS.513.1581W}, ring galaxies were correctly classified using a fine-tuning sample size of $\sim100$, finding that galaxy morphological classification problems can be solved through a transfer-learning and fine-tuning approach. Our approach is to use the weights of Zoobot as a foundation, and fine-tuned the model for the purpose of galaxy merger identification. The model is fine-tuned to classify HSC-SSP images, using a small sample of survey-realistic HSC-SSP images from the TNG50 cosmological magneto-hydrodynamical simulation \citep{2019MNRAS.490.3196P, 2019MNRAS.490.3234N} and corresponding ground truth merger status labels. This approach is able to construct a model combining \textbf{a)} human domain knowledge on galaxy morphology and \textbf{b)} ground truth merger labels accessible only from simulations.
   
      This work is broadly divided into two portions. In the first portion, encompassing from Section \ref{section:Method} to Section \ref{section:results}, we discuss our machine-learning based approach for classification, and evaluate the performance of our classifier. In the second portion of this work, composed of Section \ref{section:science} and Section \ref{section:discussion}, we conduct investigations on galaxies we identified using our fine-tuned model, particularly the relationship between galaxy mergers and environment. Specifically, we investigate whether mergers are found more frequently in higher density or lower density environments. We study the relationship between galaxy merger probability and their overdensities by using multiscale environmental parameters, ranging from 0.05 M$h^{-1}$Mpc to 8 $h^{-1}$Mpc, computed by \citet{2022ApJ...936..124Y}. We compare the findings of the relationship found in the observational galaxies with those in simulations, and discuss the results.

%--------------------------------------------------------------------
\section{Method}
\label{section:Method}
In this section, we describe the Zoobot deep representation model from \citep{2022MNRAS.509.3966W, 2022MNRAS.513.1581W, Walmsley2023}. We then describe our approach to fine-tuning the model using survey realistic HSC-SSP images constructed from galaxies from the TNG50 cosmological hydrodynamical simulation.

\subsection{Zoobot}
Zoobot \citep{Walmsley2023} is a publicly available pretrained model that can be fine-tuned for use in galaxy morphology classification problems. 
The initial deep learning model is trained with the methods written in \citet{2022MNRAS.509.3966W}, using data and labels from Galaxy Zoo DECaLS (hereinafter referred to as GZ DECaLS). GZ DECaLS is a project where volunteers visually classified galaxies in the deep, low-redshift images of the Dark Energy Camera Legacy Survey \citep[DECaLS,][]{2019AJ....157..168D}. Zoobot uses DECaLS imagery due to its superior depth and seeing compared to that in the imagery used in previous GZ projects. For example, in GZ 2 \citep{2010MNRAS.401.1552D}, which is often used for machine learning architecture in astronomical imaging classification \citep[e.g.,][]{2010MNRAS.406..342B, Ackermann_2018}, SDSS images are used. This imaging survey has a median $5\sigma$ point source depth of $r=22.7$ mag with a median seeing of $1\farcs4$ and a plate scale of $0\farcs396$ per pixel \citep{2000AJ....120.1579Y}. The DeCALS survey has a median $5\sigma$ point source depth of $r=23.6$ mag, and seeing better than $1\farcs3$, and a plate scale of $0\farcs262$ per pixel \citep{2019AJ....157..168D}, offering improved imaging quality. This not only allows for fainter and low surface brightness merger features to be revealed, but also is closer to the depth of the images we conduct training and make predictions on, which we explain in Section \ref{section:Data}.

The classifications made in the GZ DECaLS dataset were for galaxy features such as bars, bulges, spiral arms, and merger indicators.
A total of approximately 7.5 million classifications were given for over 310,000 galaxies in GZ DeCALS.
These classifications were then used to train a deep representation learning model. %add a bit more on representation learning
Predictions made by the trained model achieved 99\% accuracy when measured against confident volunteer classification for a variety of features, such as spiral arms, bars, and merger status.
The results of \citet{2022MNRAS.513.1581W} showed that this trained model was able to find similar galaxies and anomalies without any modification, even for tasks that it was never trained for. Further, the model can be fine-tuned for specific morphological classification tasks.

The technique of fine-tuning consists of training an initial model (usually with a large amount of data), then adapting the model to a different task (usually with a smaller amount of training data). Once the initial model is trained and representations learned, the `head' layer (i.e. the upper layer) is removed, and the weights of the remaining layers, or `base' layers, frozen. A `new head' model with outputs appropriate for the different task is added, then trained with data and labels for the new task. The characteristic of this method is that a far smaller training sample for the specific task is required compared to training a model from scratch.

\citet{2022MNRAS.513.1581W} found that when using a small training sample ($\sim$1000 samples), this fine-tuning approach can yield higher accuracies compared to training a dataset from scratch. Further, fine-tuning using a `base' model trained on generic galaxy morphology data and labels (Zoobot) yielded higher accuracies than fine-tuning a model trained with a generic terrestrial set of representations (ImageNet). Further detailed descriptions and methods used in Galaxy Zoo DeCALS and Zoobot are available in \citet{2022MNRAS.509.3966W}, \citet{2022MNRAS.513.1581W}, and \citet{Walmsley2023}.

\subsection{Fine-tuning Zoobot using simulation images}

\subsubsection{Training data}
Training a classifier, whether it be from scratch or through transfer learning, requires training data consisting of image data and corresponding ground truth labels. For this work, we require galaxy image data with labels of either merger or non-merger. 
% Many previous works using machine learning to classify galaxies used observational images as training data with labels based on visual inspection, such as GZ merger labels \citep{Ackermann_2018, 2019A&A...626A..49P} The accuracy of these labels is important to the performance of the classifier, and the issues with visual inspection highlighted in Section \ref{section:Intro} can be directly reflected by the classifier. Inaccurate or incorrect labels in the training data may result in inaccurate or incorrect classifications. For this reason, we decide to adopt training data where the ground truth labels are known.
We obtain images with ground-truth merger labels by using synthetic HSC-SSP images of galaxies from the TNG50 simulation. The use of simulations allows us access to information about a galaxy that is generally unavailable in observations, such as when the galaxy underwent or will undergo its previous or next merger, as well as properties of the merger activity itself, such as the mass ratio between the galaxies involved.

\subsubsection{IllustrisTNG50}
We use data available from simulation data to acquire galaxy samples for mergers and non-mergers. Specifically, we use a suite of large-volume cosmological magneto-hydrodynamical simulation data in the form of IllustrisTNG simulation data. 
The IllustrisTNG simulations \citep{2018MNRAS.475..676S,2018MNRAS.475..648P,2018MNRAS.477.1206N,2018MNRAS.475..624N,2018MNRAS.480.5113M}, performed with with the moving mesh code AREPO \citep{2010MNRAS.401..791S}, includes a comprehensive model for galaxy formation \citep{2017MNRAS.465.3291W,2018MNRAS.473.4077P}. This model includes treatments for stellar formation and evolution, black hole growth, magnetic fields, stellar and black hole feedback, and radiative cooling. TNG simulations track the evolution of dark matter, gas, stars, and supermassive blackholes ranging from the very early universe up to redshift $z=0$.
TNG simulations include three runs spanning a range of volume and resolution, TNG50, TNG100 and TNG300, in order of ascending volume and descending resolution. For this work we use simulation data from TNG50 \citep{2019MNRAS.490.3196P, 2019MNRAS.490.3234N}, which offers the highest resolution, with evolving $2\times2160^3$ dark matter particles and gas cells in a 50 Mpc box. 

We use survey-realistic synthetic HSC images from the TNG50 data, with the imaging to come in \citet{bottrell2023illustristng}. The galaxies from the TNG simulations go through a multiple steps to produce these synthetic images.

First, the images are forward-modelled into idealized synthetic images in HSC \textit{grizy} bands \citep{2018PASJ...70...66K} using the Monte Carlo Radiative transfer code SKIRT \citep{2020A&C....3100381C}. For each galaxy in our sample, stellar and gaseous particle data taken from its friends-of-friends (FoF) group within a spherical volume is used to run the radiative transfer simulation. The radius captured within this spherical volume is sufficiently large that extended structures, satellites, and nearby groups and clusters are included in the transfer simulations. 

SKIRT models the spectral energy distribution (SED) of stellar populations using the \citet{2003MNRAS.344.1000B} template spectra and \citet{2003PASP..115..763C} initial mass function for stellar populations older than 10 Myr, and with the MAPPINGS III SED photo-ionisation code \citep{2008ApJS..176..438G} for younger stellar populations (< 10 Myr). The MAPPINGS III library accounts for emission from HII regions, surrounding photodissociation regions, gas and dust absorptions in birth clouds around young stars, nebular and dust continuum and line emission.

Next, as TNG simulations do not explicitly track dust evolution, a dust model is required to account for the relationship between dust and gas properties. The model used follows \citet{2022MNRAS.510.3321P}, which takes into account the empirical scaling relation between the dust-to-metal mass ratio (DTM) and metallicity within gas \citep{2014A&A...563A..31R}. Following the empirical broken power law, \citet{2014A&A...563A..31R}, the metallicity in each gas cell can be converted in to a dust-to-gas mass density ratio, which then in turn can be used to compute the dust density (abundances).  As in  \citet{2020MNRAS.497.4773S} and \citet{2022MNRAS.510.3321P}, the dust abundances is set to zero for cells that are not star forming or temperatures greater than 75000 K. Dust self-absorption is not accounted for in the transfer simulations.

Finally, RealSim \citep{2019MNRAS.490.5390B} in conjunction with HSC Data Access Tools, is used for \textbf{a)} assignment of insertion location within HSC and flux calibration, \textbf{b)} spatial rebinning to HSC angular scale, and \textbf{c)} reconstruction of a HSC PSF and convolusion of the idealized image, and the final injection into HSC-SSP.
The full-color images created using these steps visually resemble those of real galaxies in the HSC-SSP (Eisert et al. in prep). Detailed descriptions on the synthetic images and how they are processed will be provided in \citet{bottrell2023illustristng}.
We use synthetic images at 3 snapshots, 78, 84, and 91, corresponding to redshifts $z=0.3$, $0.2$, and $0.1$, respectively. We also constrain the subhalo stellar mass to be  $\log{(\mathrm{M_{*}/M_{\odot}})} > 9$, which is approximately the lowest limit for stellar structures in TNG data to be well resolved.

\subsubsection{Merger and non-merger selection}
We select galaxy mergers based on the time to the closest merger event, either the most recent merger event or the next merger event.
We define a merger event to be the snapshot within a simulation galaxy's merger tree where two halos from the previous snapshot merge and become a single halo.
The observability timescale for galaxy interaction signatures in imaging data is difficult to constrain, as it can depend on a wide range of properties. These range from the method used for merger identification (pair identification, nonparametric statistics), physical properties of the interacting galaxies themselves (gas mass, pair mass ratio, dust) to the properties of the observations (wavelength, viewing angle, resolution). 
Studies have been conducted using hydrodynamical simulations \citep{2008MNRAS.391.1137L, 2010MNRAS.404..590L, 2010MNRAS.404..575L} to constrain the observability timescale for various merger identification methods and merger properties. Timescales found from these works can be as low as $0.2 - 0.4$ Gyr, and can exceed 1 Gyr, depending on the signature used, such as galaxy asymmetry or Gini - $M_{20}$ metric.

For this work, we apply a $0.5$ Gyr cut since or until the closest merger event to select a merger sample. This cutoff will allow for most merger signatures to be detected. As we would like to make the model agnostic to a diverse scope of mergers, we do not place any constraints on the physical properties of the galaxies such as gas mass or star formation rate, and include mergers of varying mass ratios: major (mass ratio < 1:4), minor (mass ratio < 1:10), and mini (mass ratio < 1:20). Mass ratios are defined comparing the maximum stellar masses of the composing galaxies of the merger pair. These restrictions leave us with 291 mergers, with 104 at snapshot 78, 111 at snapshot 84, and 76 at snapshot 91.

\begin{figure}[t]
    \centering
    \includegraphics[width=0.5\textwidth]{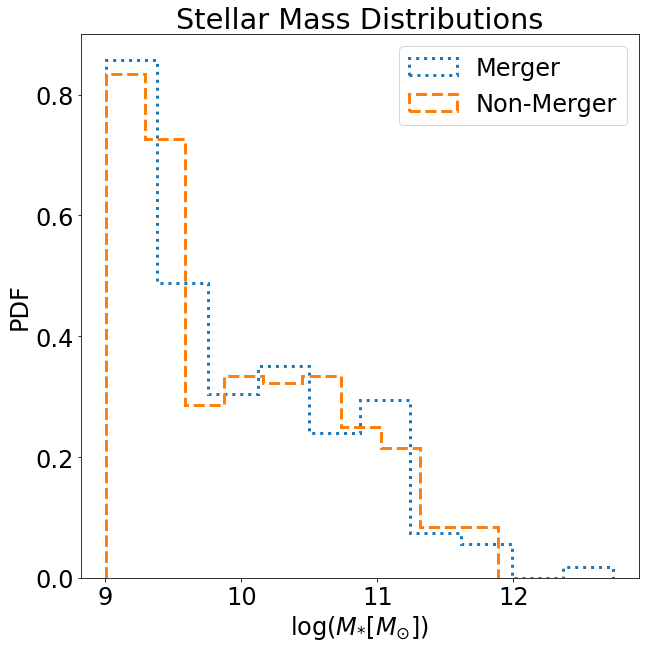}
    \caption{The stellar mass distributions for the simulated TNG50 merger and non-merger galaxies used for fine-tuning Zoobot. There are 291 each of mergers and non-mergers, with 104 at $z=0.3$. 111 at $z=0.2$, and 76 at $z=0.1$. Each merger galaxy used in the fine-tuning process has a corresponding non-merger galaxy at the same snapshot with a stellar mass within 0.1 dex.
}
    \label{fig:mstars}
\end{figure}

For non-merger selection, we adopt a cutoff so that visual merger signatures should not be visible in the images. For this work, the non-mergers have the most recent or next merger event to be $> 3$ Gyr, sufficiently greater than the observability timescales found in the works above. These cutoffs give us 1472 non-mergers. We do not use all of these non-mergers, as it is preferable that the size of classes are balanced when training models.
Further, to ensure that we do not have stellar mass biases between the merger and non-merger samples, for each merger galaxy we select a non-merger galaxy belonging to the same snapshot (redshift) with a stellar mass within $0.1$ dex. The stellar mass distributions of the merger and non-merger galaxies are shown in Fig \ref{fig:mstars}. Conducting a two-sample KS test on the merger and non-merger stellar mass distribution returns a statistic of 0.01, and a \textit{p}-value of 0.99.
The sample used for fine-tuning includes 291 mergers and non-mergers of similar stellar mass distribution, and as each galaxy in the image catalogue is processed by SKIRT along 4 lines of sight, we have $\sim$1200 synthetic HSC \textit{gri} images each for mergers (assigned with a class label of 1) and non-mergers (assigned with a class label of 0). A sample size of this order can achieve greater accuracies through fine-tuning using Zoobot as opposed to training a model from scratch, or from transfer learning using ImageNet, as shown in \citet{2022MNRAS.513.1581W}. We note that while our merger sample includes mergers at varying mass ratios and stages, they are all given the same class label. As such, the output of our model will only predict whether or not a galaxy is a merger, and will not make classifications on merger mass ratio or stage.

Cutouts are made for the $\sim2400$ galaxy images as a final pre-processing step. These cutouts encompass $10\times$ Sersic $R_{\textrm{eff}}$ of each galaxy, and are re-sized to 300 $\times$ 300 pixels for input into the model.

\subsubsection{Training procedure}
As highlighted in the previous sections, the 'head' trained on GZ DeCALS is removed, and the `base' model is frozen. Detailed architecture of the Zoobot `base' model are available in \citet{2022MNRAS.509.3966W}. We summarize the newly added 'head' layer for the merger identification task in Table \ref{table:Zoobothead}. 
\begin{table}
          % title of Table
    % is used to refer this table in the text
\centering                          % used for centering table
\resizebox{\columnwidth}{!}{
\begin{tabular}{c c c}        % centered columns (4 columns)
\hline\hline                 % inserts double horizontal lines
 Layer (type) & Output Shape & Num. Parameters \\    % table heading 
\hline                        % inserts single horizontal line
   GlobalAveragePooling2D & (None, 1280)  & 0 \\      % inserting body of the table
   Dropout & (None, 1280)  & 0 \\
   Dense & (None, 64)  & 81984 \\
   Dropout & (None, 64)  & 0 \\
   Dense & (None, 64)  & 4160 \\
   Dropout & (None, 64)  & 0 \\
   Dense & (None, 64)  & 1 \\
\hline                                   %inserts single line
\end{tabular}
}
\caption{Architecture of the new 'head' model we attach to the Zoobot `base' model. The output shape and the number of free parameters are also shown. }   
\label{table:Zoobothead}  
\end{table}

We train our new head using binary cross-entropy loss, with a maximum of 150 epochs available for training. However this maximum number of epochs may not necessarily be reached, as we follow \citet{2022MNRAS.513.1581W} and adopt an early stopping algorithm, which ends training when the validation loss stops decreasing. 
The training time is dependant on the dataset size. For the small datasets used in this work, each epoch takes about 50 seconds, with the longest training taking 48 epochs and the shortest 18 epochs.

Before we conduct tests on observational data, we first evaluate the performance of the model on simulation data alone. We split the TNG50 dataset into training, validation, and testing datasets. To prevent train/test contamination, we make sure that all 4 viewing angles of a single galaxy are contained in a single dataset, i.e. if a galaxy with viewing angle 1 is included in the training dataset, the other 3 viewing angles are also included in the training dataset, and no angles of the same galaxy are in the testing or validation datasets.
We create 10 independent training/validation/testing subsets through 10-fold cross validation so that each galaxy will be assigned a merger probability. The split for each dataset is 63\% training, 27\% validation, and 10\% testing. We record the accuracy, loss, validation accuracy and validation loss of each run.

% \subsubsection{Results on simulation data}
% \begin{figure}[h!]

%     \centering
%     \begin{subfigure}[b]{0.5\textwidth}
%         \includegraphics[width=\textwidth]{AccuracyZoobot.png}
%         \subcaption{Accuracy}
%         \label{fig:acc}
%     \end{subfigure}
% %
%     \begin{subfigure}[b]{0.5\textwidth}
%         \includegraphics[width=\textwidth]{LossZoobot.png}
%         \subcaption{Loss}
%         \label{fig:loss}
%     \end{subfigure}
% %
%     \begin{subfigure}[b]{0.5\textwidth}
%         \includegraphics[width=\textwidth]{ValidationAccuracyZoobot.png}
%         \subcaption{Validation accuracy}
%         \label{fig:valacc}
%     \end{subfigure}
%     \begin{subfigure}[b]{0.5\textwidth}
%         \includegraphics[width=\textwidth]{ValidationLossZoobot.png}
%         \subcaption{Validation loss}
%         \label{fig:valloss}
%     \end{subfigure}
    
%     \caption{Learning curves for accuracy, loss, validation accuracy, validation loss from 10 Zoobot runs. Due to the early stopping algorithm we adopt, some runs end at earlier epochs than others.}
%     \label{fig:metrics}
% \end{figure}%

%
\begin{table}
          % title of Table
    % is used to refer this table in the text
\centering                          % used for centering table
\begin{tabular}{c c c c}        % centered columns (4 columns)
\hline\hline                 % inserts double horizontal lines
Class & Precision & Recall & F1-Score \\    % table heading 
\hline                        % inserts single horizontal line % inserting body of the table
   Non-merger & 0.74 & 0.83 & 0.75 \\
    Merger & 0.80 & 0.70 & 0.80 \\
\hline                                   %inserts single line
\end{tabular}
\caption{Means of metrics of 10 individual Zoobot fine-tuning runs, assuming a complete binary class split (non-merger class (class 0): merger probability < 0.5, merger class (class 1): merger probability > 0.5). Each run split the TNG50 dataset into different 63\% training, 27\% validation, and 10\% testing datasets. The metrics are based on the validation dataset.}   
\label{table:metrics}  
\end{table}
Table \ref{table:metrics} reports the mean precision, recall, and f1-score for each class (merger and non-merger). A summary of the confusion matrices for each run are available in Appendix \ref{appendix: A}. We note a stochasticity in validation accuracy and validation loss in the confusion matrices, likely resulting from the variation in the training/validation/testing dataset splits.

The mean accuracy obtained from the 10 runs is 76\%. These results are comparable to or greater than previous works that trained CNNs from scratch using simulation images of galaxy mergers \citep[e.g.][]{2019A&A...626A..49P}, with this work using a far smaller set of training data. For example, \citet{2019A&A...626A..49P} uses $\sim$ 7000 images to train their simulation network, and achieves an accuracy of 65.2\%. We can expect a further increase in training data by incorporating mergers and non-mergers from more TNG snapshots when the images become available, which should further improve accuracies.

\begin{figure}[!hp]
\centering
\begin{subfigure}{0.49\textwidth}
\centering
  \includegraphics[width=\textwidth]{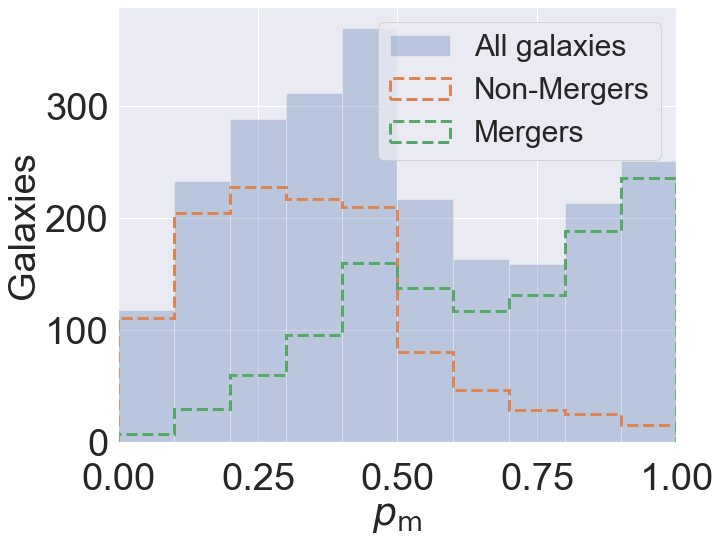}
\caption{All galaxies, mergers, and non-mergers}
\label{fig:TNGDistributionsAll}
\end{subfigure}
\begin{subfigure}{0.49\textwidth}
\centering
  \includegraphics[width=\textwidth]{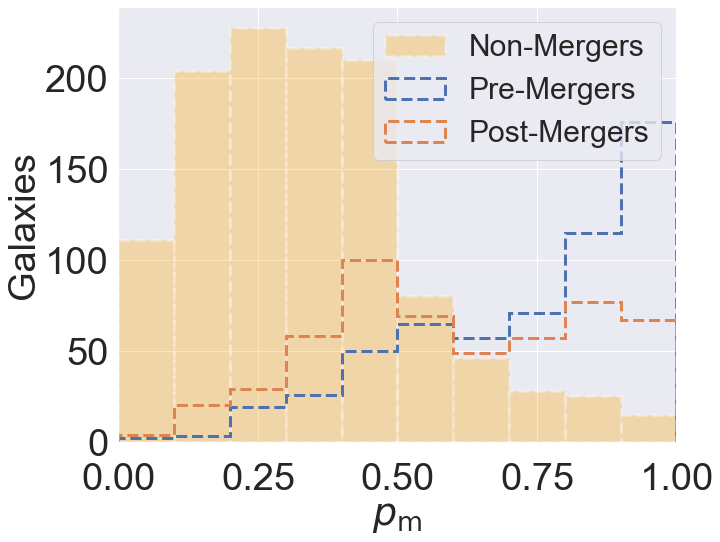}
\caption{Non-mergers, pre-mergers, and post-mergers}
\label{fig:TNGDistributionsPrePost}
\end{subfigure}
\begin{subfigure}{0.49\textwidth}
\centering
  \includegraphics[width=\textwidth]{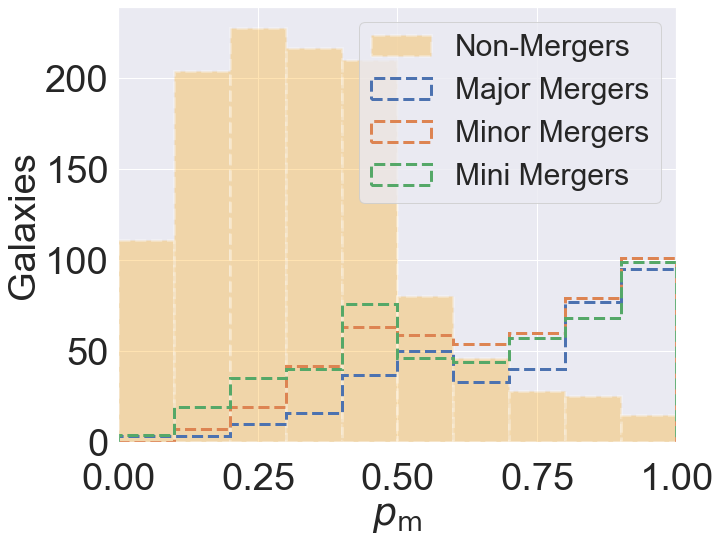}
\caption{Non-mergers, major mergers, minor mergers, and mini mergers}
\label{fig:TNGDistributionsMassRatio}
\end{subfigure}
\caption{Merger probability distributions for TNG50 synthetic galaxy images predicted using our fine-tuned model. We used 10-fold cross-validation during training for 10 independent testing samples to find probabilities for all galaxies., i.e. each galaxy was given one probability during the process. Each subfigure shows the merger probability distributions for different types of ground-truth mergers. a) All galaxies, mergers, and non-mergers. b) Non-mergers, pre-mergers (0.5 Gyr > time until merger > 0 Gyr), and post-mergers (0.5 Gyr > time since merger > 0 Gyr). c) Non-mergers, major mergers (mass ratio > 1:4), minor mergers (1:4 > mass ratio > 1:10), mini mergers (1:10 > mass ratio > 1:20).}
\label{fig:TNGDistributions}
\end{figure}

Making predictions on galaxies returns a merger probability between 0 and 1, with 0 indicating a non-merger galaxy and 1 indicating a merger galaxy , independent of merger mass ratio or merger stage. Figure \ref{fig:TNGDistributions} shows histograms of these probabilities for the 10 runs. In Fig. \ref{fig:TNGDistributionsAll} we see that the probabilities are peaked in a range between 0.4 - 0.5, indicating that many galaxies have unclear classifications, and not as many galaxies are `confidently' labeled mergers or non-mergers. However, we find that more mergers are given a probability > 0.5, and more non-mergers are given a probability < 0.5. We further investigate what type of mergers are given unclear merger probabilities. Figure \ref{fig:TNGDistributionsPrePost} shows the merger probability distributions on ground truth pre-merger (within 0.5 Gyr until the merger event) and post-merger (within 0.5 Gyr since the merger event) galaxy images. We find that while the model most frequently gives pre-mergers  a probability > 0.8, post-mergers are found to be most frequently given a probability between 0.4 - 0.5. Figure  \ref{fig:TNGDistributionsMassRatio} shows the merger probability distributions on ground truth major merger (mass ratio > 1:4), minor merger (1:4 > mass ratio > 1:10), and mini merger (1:10 > mass ratio > 1:20) images. We find that while the model is able to give high merger probabilities to mergers of all mass ratios, more minor and mini mergers are given lower probabilities (0.5 < merger probability < 0.8) compared to major mergers. As such, the galaxies given unclear merger probabilities are likely to be minor and mini post-mergers.

% \begin{figure}[t]
%     \centering
%     \includegraphics[width=0.5\textwidth]{TNGDistributions.png}
%     \caption{Merger prediction distribution for TNG50 synthetic galaxy images predicted using our fine-tuned model. We used 10-fold cross-validation during training for 10 independent testing samples to find probabilities for all galaxies., i.e. each galaxy was given one probability during the process.}
%     \label{fig:TNGDistributions}
% \end{figure}

\begin{figure}[h!]
    \centering
    \includegraphics[width=0.5\textwidth]{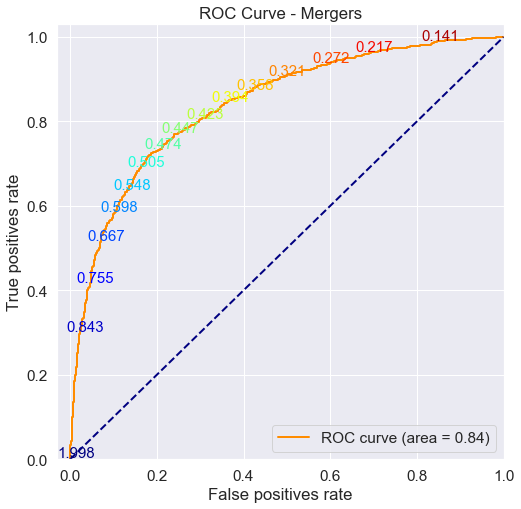}
    \caption{The trained model's ROC curve. The numbers overlayed on the curve indicate thresholds where the false positive and true positive rates along the curve are recorded. The dotted diagonal line is the performance of a completely random model which outputs a random probability for any image. An ROC curve is preferred to be above this dotted diagonal, and the curve generated from our model is above this line. We also find an AUC value of 0.84.}
    \label{fig:roccurve}
\end{figure}

We next evaluate the model's performance at various thresholds using an ROC curve. The ROC curve plots the true positive ($\frac{TP}{TP+FN}$) against the false positive ($\frac{FP}{FP+TN}$) rates at all probability thresholds between 0 and 1. Figure \ref{fig:roccurve} shows the ROC curve. The area under the ROC curve (AUC) measures the ability of the model, returning a value between 0 and 1. This value is the probability that a randomly selected merger has a merger probability greater than that of a randomly selected non-merger. An AUC of 1 indicates a perfect classifier. Our AUC is 0.84.

\begin{figure*}
        \centering
        \begin{subfigure}[t]{0.48\textwidth}
            \centering
            \includegraphics[width=\textwidth]{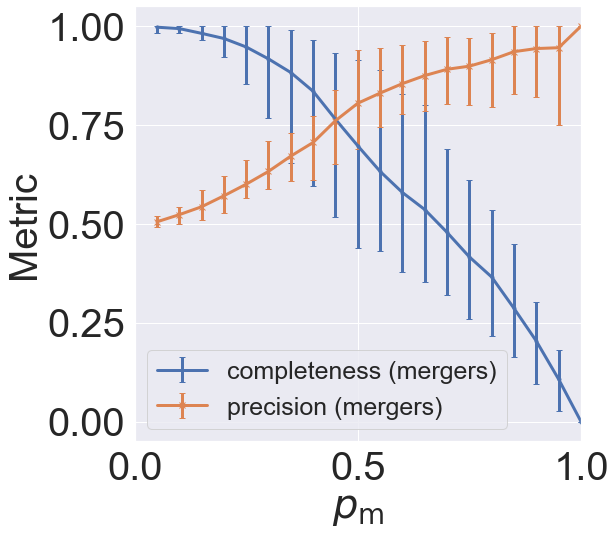}
            \caption[Network2]%
            {{\small The trained model's performance, with mean completeness and precision as a function of merger probability. The blue curve indicates completeness, and the orange curve indicates precision. Vertical error bars indicate the maximum and minimum values for the metric from the 10 runs at each probability bin.}}    
            \label{fig:completenessprecision}
        \end{subfigure}
        \hfill
        \centering
        \begin{subfigure}[t]{0.48\textwidth}
            \centering
            \includegraphics[width=\textwidth]{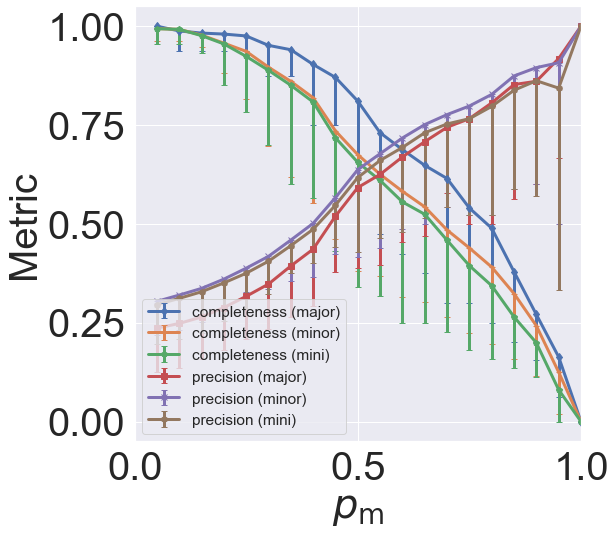}
            \caption[Network2]%
            {{\small Same as Fig. \ref{fig:completenessprecision} but metrics are split by merger mass ratio (major, minor, mini). We find mini and minor mergers are predicted at an equivalent, if not greater precision compared to major mergers. We also find that the completeness at any fixed merger probability decreases with decreasing merger mass ratio.}}    
            \label{fig:completenessprecisionmassratios}
        \end{subfigure}
        \hfill
        \caption[ The average and standard deviation of critical parameters ]
        {\small Precision and completeness curves of the TNG images. Figure \ref{fig:completenessprecision} shows the curves assuming a complete binary split, and Fig. \ref{fig:completenessprecisionmassratios} investigates the curves taking into consideration differing merger mass ratios.} 
        \label{fig:metriccurves}
\end{figure*}

% \begin{figure*}[h!]
%     \centering
%         \begin{subfigure}[b]{0.450\textwidth}
%             \centering
%             \includegraphics[width=\textwidth]{binaryprecisioncompleteness1226.png}
%             \caption[Network2]%
%             {{\small The trained model's performance, with mean completeness and precision as a function of merger probability. The blue curve indicates completeness, and the orange curve indicates precision. Vertical error bars indicate the maximum and minimum values for the metric from the 10 runs at each probability bin.}}    
%             \label{fig:completenessprecision}
%         \end{subfigure}
%         \hfill
        
%         \begin{subfigure}[b]{0.450\textwidth}
%         \centering
%         \includegraphics[width=\textwidth]{classseparateprecisioncompleteness1226.png}
%          \caption[Network2]%
%             {{\small Same as Fig. \ref{fig:completenessprecision} but metrics are split by merger mass ratio (major, minor, mini). We find mini and minor mergers are predicted at an equivalent, if not greater precision compared to major mergers.}} 
%             \label{fig:completenessprecisionmassratios}
%          \end{subfigure}
%             \centering
%     \caption{Precision and completeness curves of the TNG images }

% \end{figure*}

We further show the model's performance as a function of merger probability in Fig. \ref{fig:completenessprecision}, in the form of mean completeness and precision curves of the 10 runs. The completeness is obtained by the dividing the number of ground truth mergers with a greater merger probability than the probability bin, by the total number of ground truth mergers in the testing dataset:
\begin{multline}\label{eq:1}
    \textrm{Completeness}(p_{\textrm{merg}})\\
    = \frac{(\textrm{Num. GT Mergers}>p_{\textrm{merg}}) - \textrm{Total GT Mergers}}{\textrm{Total GT Mergers}}
\end{multline}

where GT means ground truth. The precision is obtained by the dividing the number of ground truth mergers with a greater merger probability than the probability bin by the total number of objects with a greater merger probability than the probability bin:
\begin{multline}\label{eq:2}
    \textrm{Precision}(p_{\textrm{merg}})\\
    =\frac{\textrm{Num. GT Mergers}>p_{\textrm{merg}}}{\textrm{Num. Objects}>p_{\textrm{merg}}} 
\end{multline}

The model has values of 80\% for both completeness and precision for mergers on the testing dataset if we adopt a complete binary split, that is, any galaxy with a merger probability $>0.5$ is classified as a merger, and any $<0.5$ is classified as a non-merger. This accuracy can be considered a reasonable result ($>80\%$). However, if we adopt a merger/non-merger split based on `confident' predictions, i.e., merger probability $>0.8$, our mean accuracy increases to 91\%. 

We further investigate the role of mass ratio in the completeness and precision curves. Fig. \ref{fig:completenessprecisionmassratios} shows the mean precision and completeness as a function of the same merger probabilities, however this time separating the major, minor, and mini mergers. We find that the mean precision at a binary split is 59\% for major mergers, 63\% for minor mergers, and 62\% for mini mergers. These metrics increase to 80\% for major mergers, 82\% for minor mergers, and 79\% for mini mergers for `confident' merger probabilities. The precision values are lower when the mergers are split by mass ratio as opposed to a 2-class split due to the method of computing the metrics. There are an equal number of overall mergers and non-mergers, however the respective number of major, minor, and mini mergers are lower than the number of non-mergers in each test dataset. As a result while the numerator in Equation \ref{eq:2} accounts only for mergers of the labeled mass ratio, the denominator includes all non-mergers and mergers regardless of mass ratio. As such, our findings with respect to the precisions shown in Fig. \ref{fig:completenessprecisionmassratios} are more relevant in a qualitative manner rather than quantitative. 
Noting the gradients for the metric curves, we find that the completeness curve for major mergers drops off most gradually, indicating that our model is most confident in classifying major mergers. 
We also note that the precision for mini mergers (mass ratio < 1:20) is the highest between the 3 mass ratios at a binary split, and remains the highest at the `confident' probabilities, indicating that the model is able to predict mini mergers at an equivalent or greater precision compared to major and minor mergers. This precision will allow for studies using large, precise samples of sub-major mergers, which have adverse effects on galaxy properties such as size growth \citep{2013MNRAS.431..767B, 2014ApJ...790L..33L, 2018MNRAS.480.2266M,bottrell2023illustristng}.

\begin{figure}[h!]

\end{figure}

We provide some examples of true positive, true negative, false positive and false negative galaxy classifications from the 10 runs in Appendix \ref{appendix: B}, in Figures \ref{fig:TruePos} - \ref{fig:FalseNeg}, respectively, adopting a completely binary split. With this split, the model seems to be able to identify mergers of diverse morphologies, ranging from interacting pairs to merger remnants with visual signatures.
We also find that the model is able to identify non-mergers of diverse morphologies, including projections, overlaps, and isolated galaxies, as shown in the varying appearances of true negative predictions in Fig. \ref{fig:TrueNeg}. 
The model seems to have mixed results on galaxy projections, as there are examples in both Fig. \ref{fig:FalsePos} (false positives) and Fig. \ref{fig:TrueNeg} (true negatives). For incorrectly classified mergers (false negatives), many of the mergers with low probabilities (merger probability $<0.3$) are those with large mass ratios ($\mu<0.25$) whose times since or until the nearest merger event are close to the selection threshold of 0.5 Gyr ($>0.3$ Gyr), meaning that potential merger indicators may not be visible, for example 84-569599\textunderscore v0 (merger probability 0.05) in row 4, column 5 of Fig. \ref{fig:FalseNeg} is a mini merger that is $0.31$ Gyr until its merger event, and look visually very similar to non-mergers. Many of these galaxies would also likely not be classified as mergers by human-based visual classification methods. However, we also note there are also misclassifications with major mergers, for example 84-577873\textunderscore v0 (merger probability 0.40) in row 3, column 4 of Fig. \ref{fig:FalseNeg} is a major merger that is $0.49$ Gyr until its merger event, which would likely be labeled as a merger by human-based methods.
For misclassified non-mergers (false positives), galaxies with high merger probabilities (>0.8) are likely to be classified as merging by human-based methods, as they show merger-like disturbances, and may also have projections. As such, our machine-learning based approach may encounter similar issues as previous human-based visual approaches.
 Nevertheless, even with its misclassifications, we find that the model is able to correctly classify a diverse range of mergers and non-mergers, which should be useful for merger galaxy sciences.

%input type of merger and timescale for each figure.
%Find diverse non-mergers

\section{Predictions on observations}
\label{section:Data}

In this section, we describe the work we do to make predictions on observational images using the fine-tuned model.
For the work conducted in this section, we trained a new head model using a 70\% training and 30\% validation split of the simulation dataset used in the previous section, and no further split of testing data, and attached it to the original Zoobot base model. Armed with our fine-tuned Zoobot model, we apply it to SDSS and GAMA spectroscopically confirmed galaxies in the HSC-SSP public data release 3 \citep{2022PASJ...74..247A}.The training with this split lasted 37 epochs, with a duration of 23 minutes using GPU (NVIDIA Quadro P400).

\subsection{Data; The Subaru Hyper Suprime-Cam Subaru Strategic Program}

The HSC-SSP is a multi-tiered, wide-field, multi-band imaging survey on the Subaru 8.2 m telescope on Maunakea in Hawaii. Detailed information about the survey, its instrumentation, and its techniques are available in \citet{2018PASJ...70S...8A} and relevant papers \citep{2018PASJ...70S...5B,2018PASJ...70S...1M,2018PASJ...70S...2K,2018PASJ...70S...3F}.
We use data from HSC-SSP due to its wide field of observation and exceptional ground-based depth and resolution. In its widest component (Wide layer), HSC-SSP covers about 600 deg$^2$ of the sky in five broad band filters (\textit{grizy}), with observations from 330 nights coadded. The depth of the HSC survey is $r\approx26$ mags ($5\sigma$, point source) for the Wide layer. The coverage and depth of HSC-SSP will give us access to high quality imaging data of several million galaxies. We plan to make merger probabilities for all of HSC-SSP Wide in future works, a catalogue of which will be made publicly available for the benefit of the galaxy astronomy community. This merger probability catalogue is expected to be one of the largest catalogues of its kind.

For the work conducted in Section \ref{section:science}, we use a subsample of galaxies from the HSC-SSP Wide internal data release S21A catalogue to conduct predictions on with our trained model. To select a galaxy sample, we cross-match HSC-SSP S21A galaxies to SDSS Data Release 17 \citep{2022ApJS..259...35A} and Galaxy And Mass Assembly Data Release 4 \citep[GAMA DR4,][]{2022MNRAS.513..439D} galaxies within 1 sky arcsec. We are left with 145,544 matches in SDSS and 156,604 matches in GAMA, for a total of 302,148 galaxies.
All galaxies matched have spec-\textit{z} measurements from their respective catalogues. We only use spectroscopic redshifts, due to photometric redshift errors. The galaxies have a magnitude limit of  \textit{r} < 17.7 mag for SDSS galaxies and \textit{r} < 19.8 for GAMA galaxies. The galaxies lie within a redshift range of $z=0.01 - 0.35$ and have $M_{*}=3\times10^{9} -3\times10^{11}M_{\odot}$. $M_{*}$ values are obtained following \citet{2012MNRAS.421..314C} for the SDSS galaxies and its own catalogue for the GAMA galaxies. The SDSS galaxy stellar masses, using a principal component analysis method on SDSS spectra. First, a model spectra is created based on \citet{2003MNRAS.344.1000B} stellar population synthesis model. Next, principal components are identified from the model library. Finally, the SDSS spectra are fitted to the model and physical properties estimated. The masses obtained are consistent with the GALEX-SDSS-WISE Catalog \citep{2016ApJS..227....2S, 2018ApJ...859...11S}. The GAMA galaxy stellar masses are obtained using the SED fitting code MAGPHYS \citep{2018MNRAS.475.2891D}. MAGPHYS also uses a library based on \citet{2003MNRAS.344.1000B}. Sets of optical and infrared spectra are regressed towards flux measurements and errors to find a best-fit SED and physical parameters, including stellar mass.

Each galaxy in the catalogue also has several environmental parameters related to its local mass density, studied in \citet{2022ApJ...936..124Y}.
The stellar mass overdensities within radii of 0.5, 1, and 8 $h^{-1}$Mpc, as well as within the radii determined by the projected distance to the 5th nearest neighbor are calculated.
Only galaxies within $|\Delta v|<1000 \mathrm{km s^{-1}}$ relative to the primary source are considered in this calculation. This cutoff prevents unrelated foreground or background galaxies from being included.
The densities are normalized by the median densities of all galaxies within a given mass range and redshift bin, making them overdensities relative to the median in the redshift bin. Details on how the densities are calculated are written in \citet{2022ApJ...936..124Y} and the papers referenced within.

% \begin{figure}[h!]

%     \centering
%     \begin{subfigure}[b]{0.4\textwidth}
%         \includegraphics[width=\textwidth]{78-284883_v0.png}
%         \subcaption{HSC image cutout}
%         \label{fig:HSC image}
%     \end{subfigure}
% %
%     \begin{subfigure}[b]{0.4\textwidth}
%         \includegraphics[width=\textwidth]{1237648674510733463.png}
%         \subcaption{TNG synthetic image}
%         \label{fig:TNG Image}
%     \end{subfigure}
% %

%     \caption{An image comparison between and HSC image cutout (upper) and a synthetic TNG image (lower), with dimensions matched to $300 \times 300$ pixels.}
%     \label{fig:HSCTNGComp}
% \end{figure}%

We make predictions on \textit{gri} images for each HSC galaxy, with the images having the same dimensions as the synthetic HSC images of the TNG50 galaxies - cutouts encompassing $10\times R_{\textrm{eff}}$, re-sized to 300 $\times$ 300 pixels. For each input image, our model will output a merger probability between 0 and 1, with 0 indicating non-merger and 1 indicating a merger, again independent of merger mass ratio or merger stage.

% \begin{figure}[h!]

%     \centering
%     \begin{subfigure}[b]{0.3\textwidth}
%         \includegraphics[width=\textwidth]{1237679322861666457.png}
%         \subcaption{Cutout image used in classification.}
%         \label{fig:shredding}
%     \end{subfigure}
% %
%     \begin{subfigure}[b]{0.3\textwidth}
%         \includegraphics[width=\textwidth]{1237679322861666457full.png}
%         \subcaption{Non-cropped image.}
%         \label{fig:shreddingfull}
%     \end{subfigure}
% %

%     \caption{Example of "bright galaxy shredding" occuring in HSC galaxy identifications. This object, with the image inputted in the model shown in \ref{fig:shredding}, is identified, even after cross-match between spectroscopic detections (SDSS) and HSC, to be a galaxy with an effective radius of 0.77 arcsec. However, the non-cropped image, shown in \ref{fig:shreddingfull}, finds that the object is the spiral arm of a larger galaxy with a separate identification, and deblended by this "shredding" issue.}
%     \label{fig:shreddingexample}
% \end{figure}%

We note that some HSC galaxy identifications are susceptible to "bright galaxy shredding", discussed in \citet{2018PASJ...70S...8A}. Bright $(i<19)$ galaxies, especially of late-type, are deblended into multiple objects, with the debelending seen even after cross-matching with a spectroscopic catalogue. An inspection of the non-cropped image of objects affected by shredding reveals that that it is part of the spiral arm of a larger galaxy. As much as 15\% of bright galaxies in HSC-SSP suffer from shredding, so it is expected that there are galaxies suffering from similar effects in our classified samples. However, these issues are expected to be resolved when making predictions on future HSC-SSP public data releases. For this work, as we are using the SDSS/GAMA coordinates and their petrosian radii to make our cutouts, we should not be as affected by shredding compared to HSC data.

\section{Results}
\label{section:results}
\subsection{Prediction results}

\begin{table*}
          % title of Table
    % is used to refer this table in the text
\centering                          % used for centering table
\begin{center}
\begin{tabular}{l c c c c c r}        % centered columns (4 columns)
\hline\hline                 % inserts double horizontal lines
HSC ID & ra & dec & z & $\log M_*$ & $R_{\mathrm{eff}}$ & $p_\mathrm{m}$ \\    % table heading 
\hline                        % inserts single horizontal line % inserting body of the table
   40976904286657178 & 135.21693 & -0.21783 & 0.13429 & 9.94 & 0.9441 & 0.775 \\
   40976904286655515 & 135.22422 & -0.25286 & 0.26773 & 10.69 & 1.3761 & 0.748 \\
   40976899991693591 & 135.15696 & -0.35011 & 0.25953 & 10.82 & 0.9324 & 0.373 \\
   40976899991692475 & 135.17823 & -0.37379 & 0.19888 & 10.50 & 0.6889 & 0.361 \\
   40976899991691083 & 135.20912 & -0.40034 & 0.26186 & 10.74 & 1.0056 & 0.525 \\
\hline                                   %inserts single line
\end{tabular}
\end{center}
\caption{The first 5 rows of the merger probability catalogue, which will be made publicly available. The columns are, from left to right: HSC ID, right ascension, declination, redshift, log stellar mass, effective radii in arcseconds, merger probability.}   
\label{table:predtable}  
\end{table*}

\begin{figure}[h!]

    \centering
    \begin{subfigure}[b]{0.45\textwidth}
        \includegraphics[width=\textwidth]{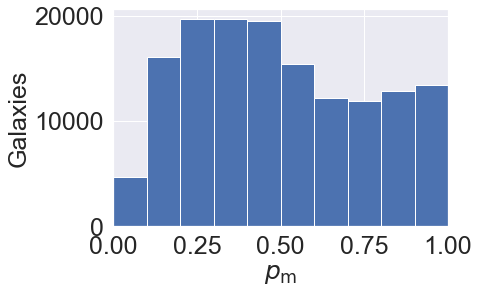}
        \subcaption{ Merger probability distributions of HSC galaxies cross-matched with SDSS DR17.}
        \label{fig:SDSS Distributions}
    \end{subfigure}
    \begin{subfigure}[b]{0.45\textwidth}
        \includegraphics[width=\textwidth]{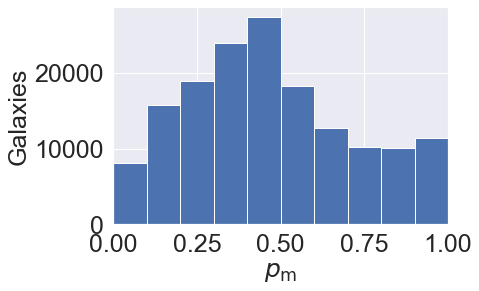}
        \subcaption{Merger probability distributions of HSC galaxies cross-matched with GAMA DR4.}
        \label{fig:GAMA distributions}
    \end{subfigure}

    \caption{Merger probability distributions for HSC-SDSS (upper) and HSC-GAMA (lower) cross-matched galaxies predicted using our fine-tuned model. Note that in both cases, more galaxies lie in an unclear range (merger probability $0.2 - 0.5$) for non-mergers than `confident' non-mergers (merger probability $0 - 0.1$). However, we find that many of the mergers identified are confident (merger probability $>0.8$).}
    \label{fig:HSCSDSSGAMA}
\end{figure}%

We plot a histogram of the merger probabilities of the SDSS and GAMA galaxies in Fig. \ref{fig:HSCSDSSGAMA}. We see that the outputted probabilities are diverse in range, with the most galaxies being assigned `unconfident' labels, with the peak being between 0.1 - 0.5 for the SDSS galaxies and 0.3 - 0.5 for the GAMA galaxies. This location of the peak is similar to that found in the simulation predictions in the previous Section, but differs from the probabilities found by transfer learning in \citet{Ackermann_2018, 2020ApJ...895..115F}, where very clear probabilities both for mergers and non-mergers were favored. 
A possible explanation for the difference in distribution between our results and previous works, i.e., the peak in the 0.1 - 0.6 merger probability range, is the inclusion of the substantial number of mini mergers in our training sample. We will be able to examine this upon the completion of the images in \citet{bottrell2023illustristng}, as we will have many more major and minor mergers to fine-tune our model with. Many galaxies with similar appearances as the true mini mergers used in the fine-tuning process could be assigned `unconfident' merger probablities in this range, whether they be merging or non-merging in ground truth. However, we also note the secondary peak at higher merger probability ($>0.9$), meaning many mergers are confidently identified. This is consistent with the above mentioned works, and also allows for a large `confident' merger sample for science purposes.

We show examples of galaxies within various merger probability bins in our predicted samples in Appendix \ref{appendix: C}, in Figures \ref{fig:GAMA03} through \ref{fig:GAMA08}. We find that in the low merger probability bins (merger probabilities $0 - 0.3$, Fig. \ref{fig:GAMA03}), our model correctly identifies likely stellar overlaps and projections as non-mergers, as well as a diverse appearance of non-mergers, similar to the correctly identified non-mergers in the simulation data. On the high merger probability end (merger probabilities $>0.8$), similar to as seen in the simulation data, the model can predict mergers with diverse appearances, including clear interacting pairs and late or early stage mergers with disturbances.

\begin{figure*}[h!]
    \centering
    \includegraphics[width=0.8\textwidth]{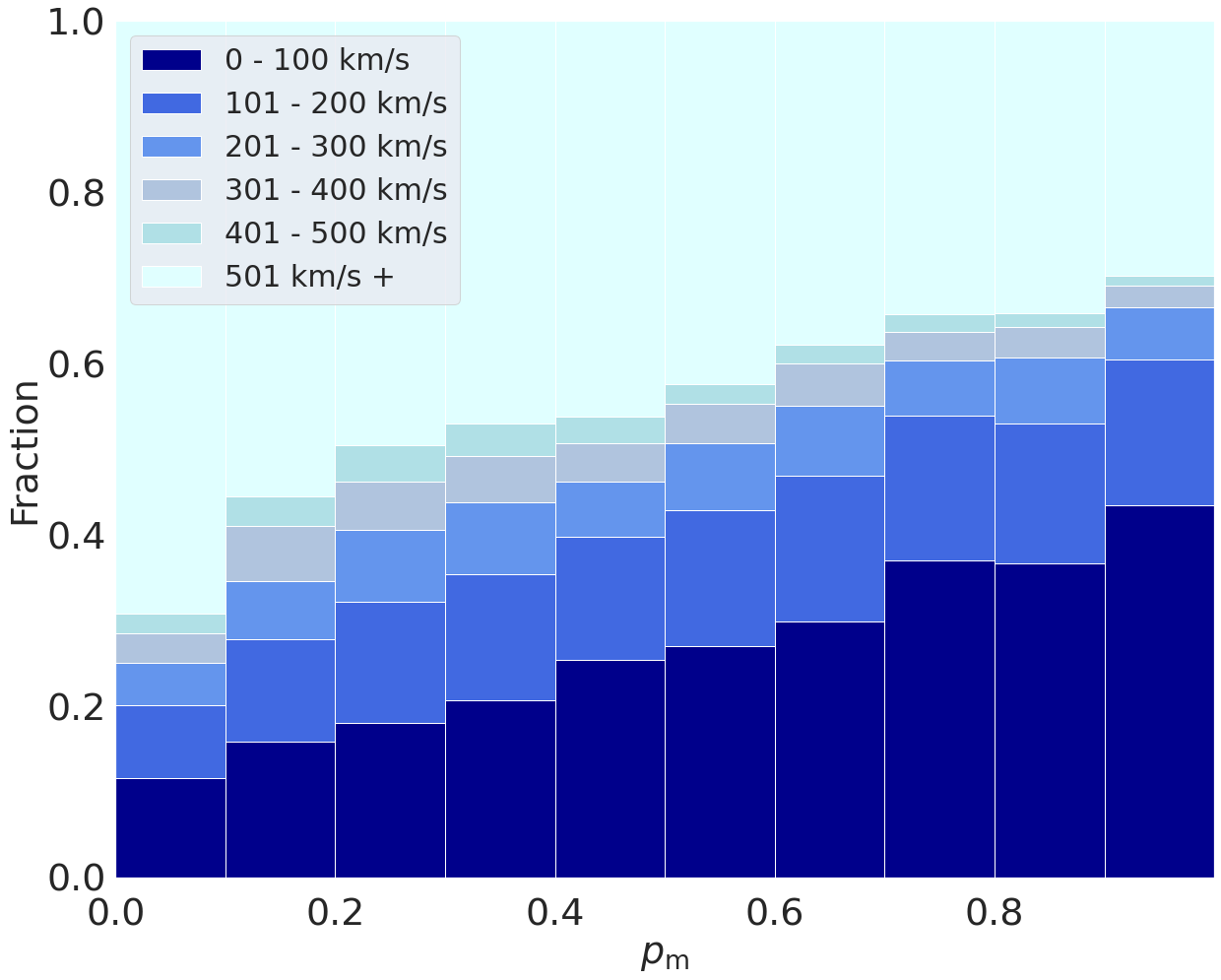}
    \caption{Merger probability distributions for galaxies with a spectroscopically identified pair within a 30 kpc radius aperture, binned by velocity difference of the pair. The vertical axis represents the fraction of galaxies in each merger probability bin belonging to each velocity difference bin. We find that the fraction of smaller velocity difference galaxy pairs increases monotonically with merger probability bin.}
    \label{fig:pairseparation}
\end{figure*}

As further validation, we confirm our model's ability to differentiate between projections and physically connected pairs in observational images. For each galaxy within our two samples that has a neighbor within a 30 kpc physical aperture, we calculate the line-of-sight velocity offset between the galaxy and its neighbor by $\Delta v = c \Delta z / (1+z_{\textrm{target}})$. We plot the merger probability distribution of these galaxies, binned by the line-of-sight velocity offset with its neighbor, or in the case of galaxies with multiple neighbors the minimum line-of-sight velocity offset, in Fig. \ref{fig:pairseparation}. We find that pairs with offset $\Delta v < 500 \mathrm{km s^{-1}}$ makes up the greatest fraction of galaxies pairs with merger probability > 0.8, and as the velocity offset increases, a greater distribution of galaxies are given unclear to lower merger probabilities (merger probability < 0.5). Of the galaxies with both high merger probability and offset, while a fraction of such may be projections, there are also likely to be true mergers that coincidentally have a projection. In particular, there may be post-merger galaxies that are not a pair but still have high merger probability. Based on our simulation results in Fig. \ref{fig:TNGDistributionsPrePost}, our model identifies 1 post-merger galaxy for every 2 pre-merger galaxies. Following these results, a fraction of galaxies with $\Delta v > 500 \mathrm{km s^{-1}}$ and high merger probabilities in Fig. \ref{fig:pairseparation} are likely to be post-mergers.
.

\subsection{Merger sample selection for science}

The results from simulation data show that a complete binary split, i.e., merger probability $>0.5$ is a merger and $<0.5$ is a non-merger, gives a precision that is $\sim 80\%$. However, the merger probability distribution for both observation and simulation results show that many galaxies lie in a range between merger probability $0.2 - 0.6$. As such, while the merger probabilities we found can be sufficiently useful to investigate trends between merger probability and physical properties, conducting studies adopting a complete binary split may suffer from contamination of `unconfident' galaxies. We can see from Fig. \ref{fig:GAMA0508} that there are many unclear galaxies in the probability range $0.5 - 0.8$ that may or may not be real mergers, thus using galaxies in this range may contaminate both merger and non-merger samples.

\begin{figure*}[h!]
    \centering
    \includegraphics[width=0.8\textwidth]{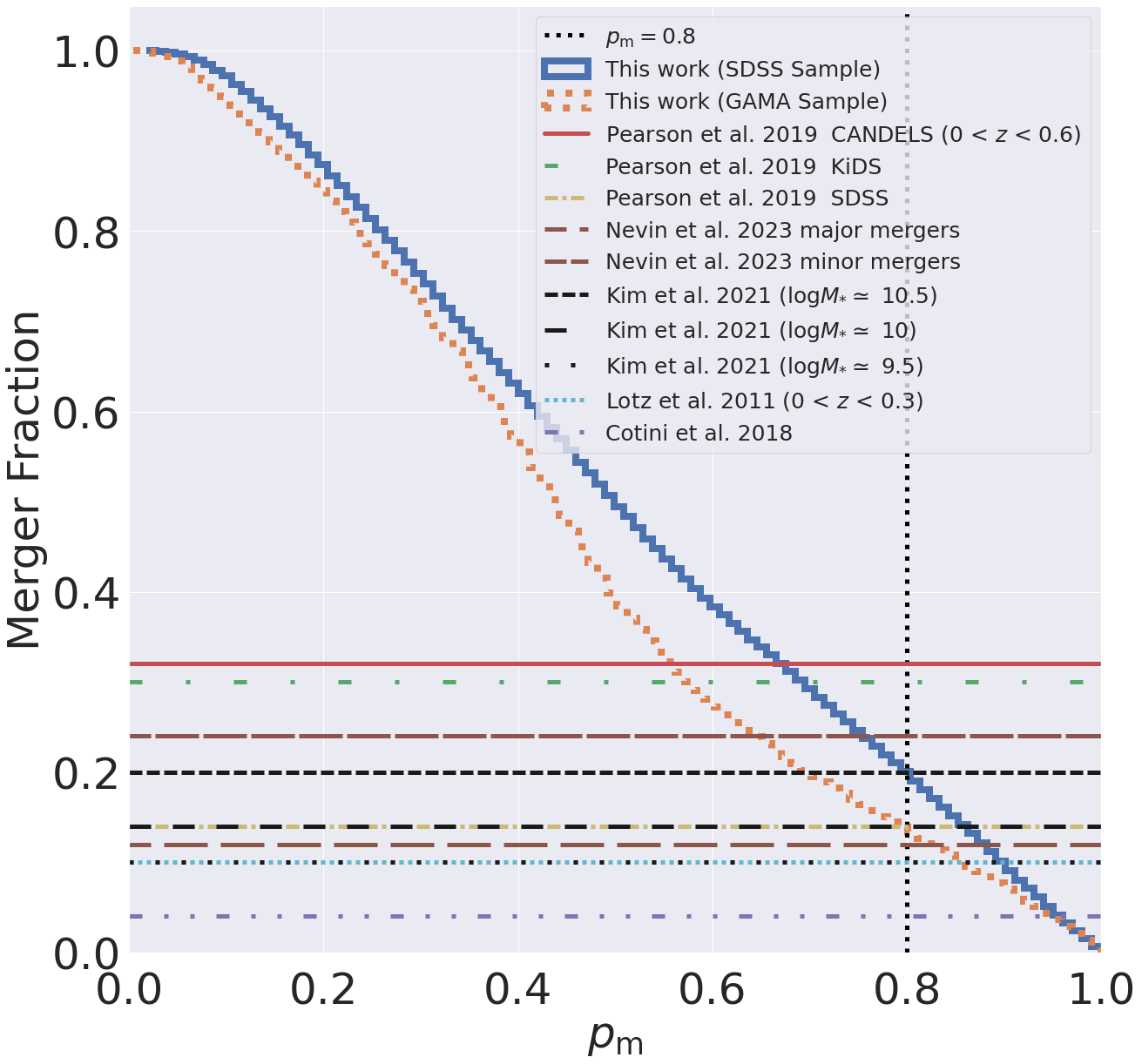}
    \caption{Merger fraction as a function of merger probability for our two samples. The black vertical dotted line indicates the `confident' merger probability threshold of 0.8. We also plot the merger fractions found in \citet[][red, green, and yellow lines]{2019A&A...631A..51P}, \citet[][brown lines]{Nevin_2023}, \citet[][black lines]{2021MNRAS.507.3113K}, \citet[][light blue dotted line]{2011ApJ...742..103L}, and \citet[][purple line]{2013MNRAS.431.2661C}.  We find that our merger fraction matches those of previous studies at increased merger probabilities.}
    \label{fig:mergprobdistribution}
\end{figure*}

We plot the merger fraction of our two samples as a function of merger probability in Fig. \ref{fig:mergprobdistribution}. We compare our results with previous works calculating the merger fraction in similar redshift ranges, both in observations and simulations \citep{2011ApJ...742..103L, 2013MNRAS.431.2661C, 2019A&A...631A..51P, 2021MNRAS.507.3113K, Nevin_2023}. We find that our merger fractions, in most cases, become statistically consistent with these works if we consider `confident' classifications (merger probability $>0.8$) as our threshold, with the threshold indicated by the dotted vertical line. However, we note that despite the statistical consistency there are likely to be contaminants regardless of threshold.

We will not set a definite, arbitrary threshold to define a merger in the catalogue to be released, instead we will just provide the merger probabilities for every galaxy, as shown in Table \ref{table:predtable}, and users can determine their thresholds. However we recommend that a `confident' threshold (merger probability $>0.8$) is used to determine a merger sample.

% \begin{figure*}[h!]
%     \centering
%     \hspace*{-3.5cm}\includegraphics[trim=0cm 3cm 0cm 0cm,scale=0.65]{SDSS03.pdf}
%     \caption{20 randomly drawn examples of SDSS galaxies with a merger probability $<0.3$, with merger probabilities in descending order. The merger probabilities are indicated in the image and the SDSS ID below the image. We find a diverse range of appearances of non-merger galaxies, including possible projections and overlaps, given low merger probabilities using our model.}
%     \label{fig:SDSS03}
% \end{figure*}

% \begin{figure*}[h!]
%     \centering
%     \hspace*{-3.5cm}\includegraphics[trim=0cm 3cm 0cm 0cm,scale=0.65]{SDSS0305.pdf}
%     \caption{The same as Fig. \ref{fig:SDSS03} but for merger probability $>0.3$ and $<0.5$.}
%     \label{fig:SDSS0305}
% \end{figure*}

% \begin{figure*}[h!]
%     \centering
%     \hspace*{-3.5cm}\includegraphics[trim=0cm 3cm 0cm 0cm,scale=0.65]{SDSS0508.pdf}
%     \caption{The same as Fig. \ref{fig:SDSS03} but for merger probability $>0.5$ and $<0.8$.}
%     \label{fig:SDSS0508}
% \end{figure*}

% \begin{figure*}[h!]
%     \centering
%     \hspace*{-3.5cm}\includegraphics[trim=0cm 3cm 0cm 0cm,scale=0.65]{SDSS08.pdf}
%     \caption{The same as Fig. \ref{fig:SDSS03} but for merger probability $>0.8$. We find a diverse range of appearances of galaxies predicted with high merger probabilities using our model.}
%     \label{fig:SDSS008}
% \end{figure*}

\section{Merger galaxy properties}
\label{section:science}
In this section, we use merger probabilities obtained using our model to investigate the relationship between galaxy mergers and local galaxy environments. We evaluate the merger probability distribution in differing environmental density bins for the various environmental parameters. We also investigate the relationship between galaxy merging and environment in TNG simulations, and look for any agreements between observations and simulations.

\subsection{Environmental overdensities as a function of merger probability}

\begin{figure*}
        \vspace*{-2cm}
        \centering
        \begin{subfigure}[b]{0.475\textwidth}
            \centering
            \includegraphics[width=\textwidth]{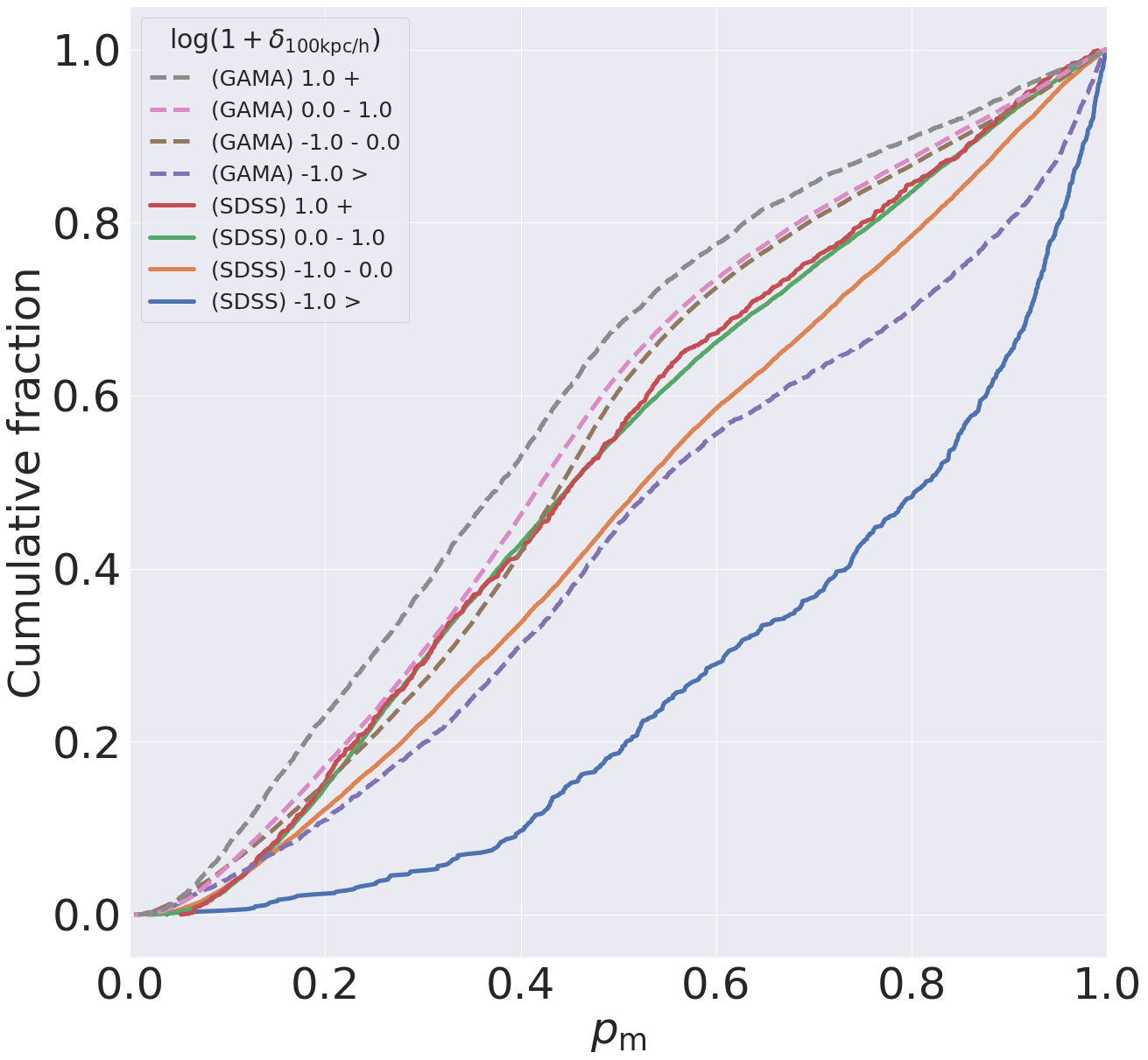}
            \caption[Network2]%
            {{\small $0.1h^{-1}$ Mpc stellar mass overdensity $\log(1+\delta_{0.1\textrm{Mpc}/h})$\\ KS test (SDSS) statistic: 0.394 %p-value: $2.710\times10^{-50}$ \\
            \\KS test (GAMA) statistic: 0.238}}% p-value: $6.429\times10^{-109}$}}    
            \label{fig:100kpc}
        \end{subfigure}
        \begin{subfigure}[b]{0.475\textwidth}
            \centering
            \includegraphics[width=\textwidth]{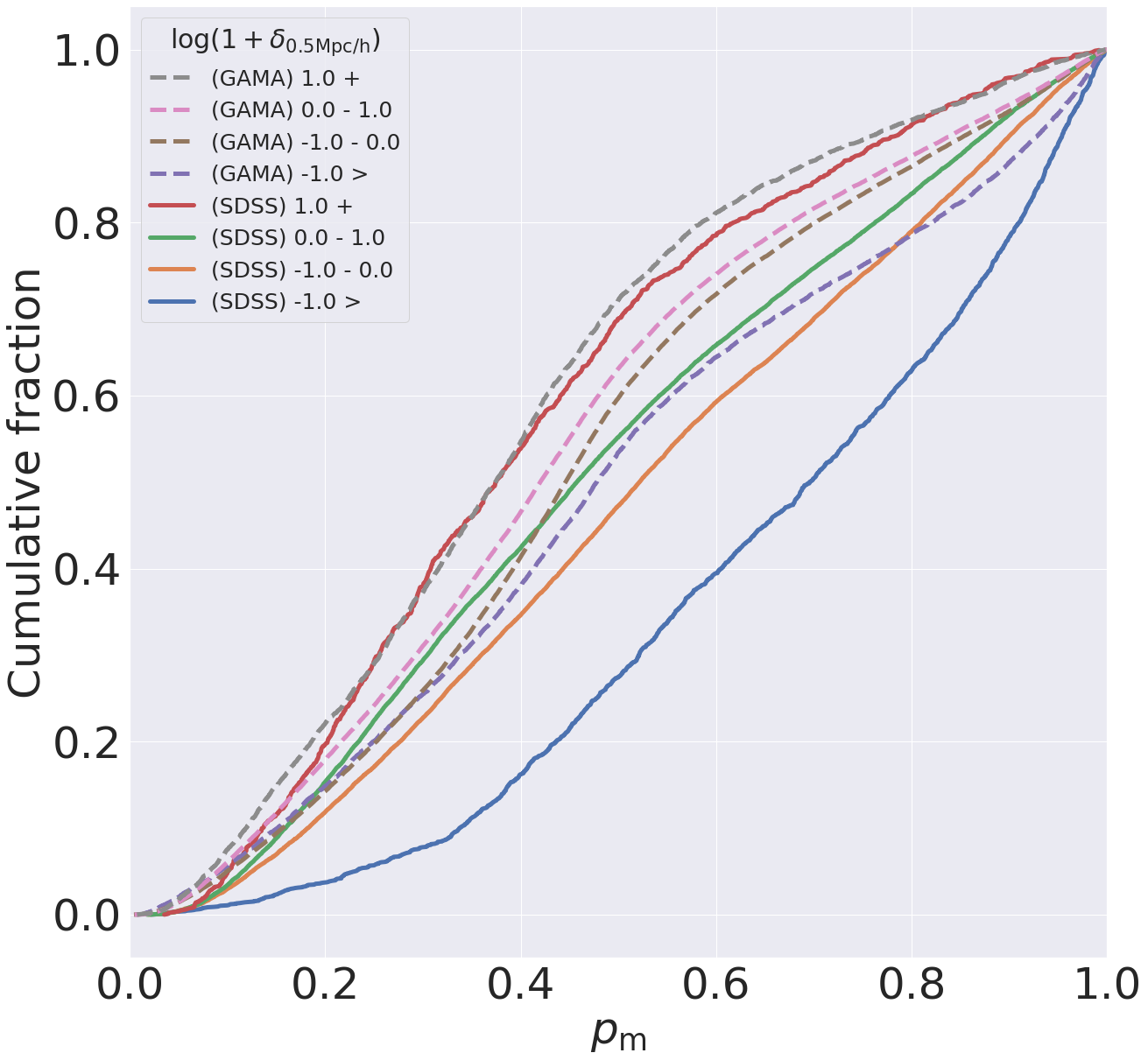}
            \caption[Network2]%
            {{\small $0.5h^{-1}$ Mpc stellar mass overdensity $\log(1+\delta_{0.5\textrm{Mpc}/h})$\\ KS test (SDSS) statistic: 0.417 %p-value: $1.233\times10^{-118}$ 
            \\ KS test (GAMA) statistic: 0.184}}% p-value: $8.847\times10^{-105}$}}    
            \label{fig:500kpc}
        \end{subfigure}
        \vskip\baselineskip
        \begin{subfigure}[b]{0.475\textwidth}  
            \centering 
            \includegraphics[width=\textwidth]{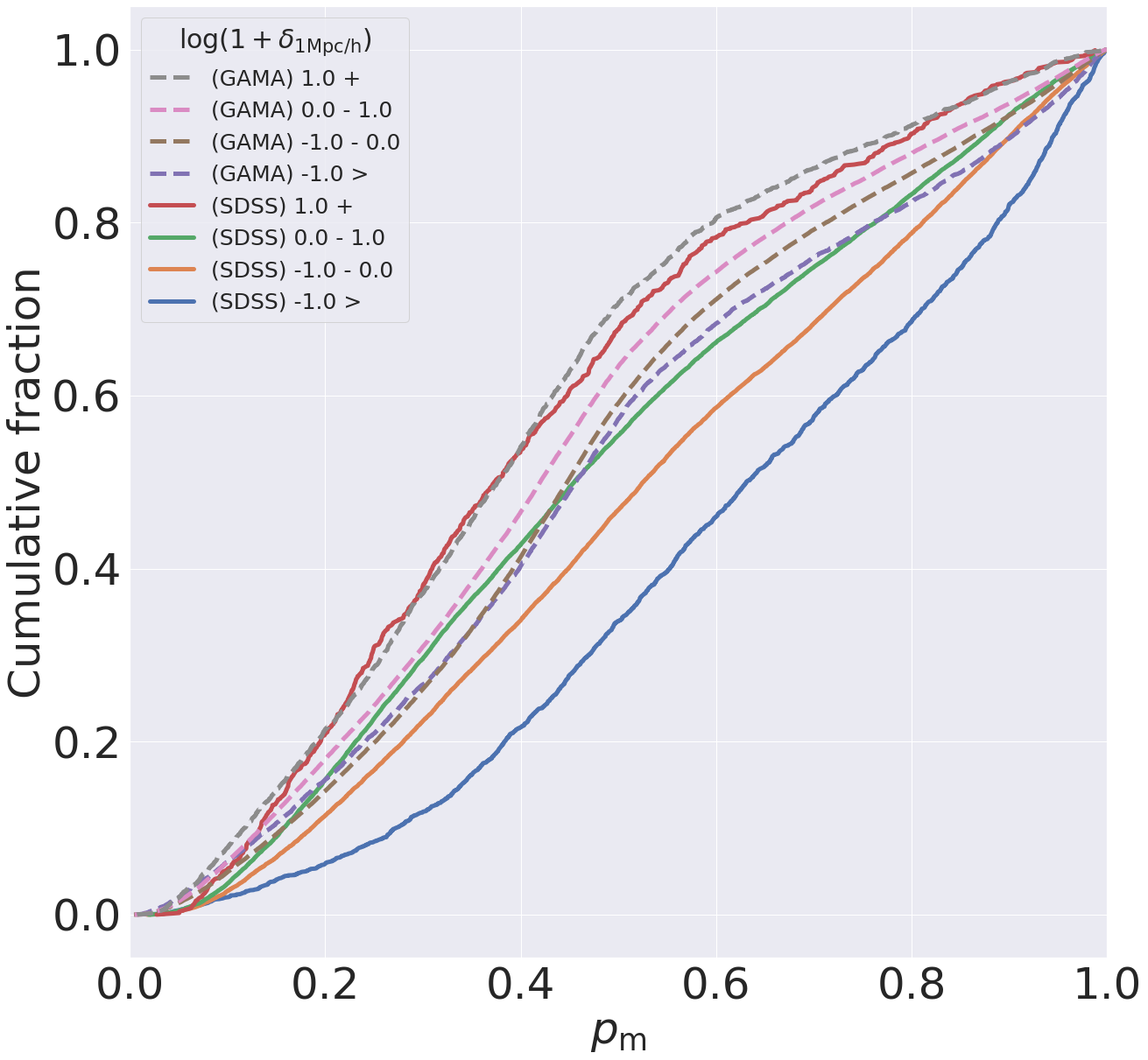}
            \caption[]%
            {{\small $1h^{-1}$ Mpc stellar mass overdensity $\log(1+\delta_{1\textrm{Mpc}/h})$\\ KS test (SDSS) statistic: 0.341 %p-value: $2.262\times10^{-75}$ 
            \\ KS test (GAMA) statistic: 0.143}}% p-value: $6.242\times10^{-44}$}}  
            \label{fig:1Mpc}
        \end{subfigure}
        \hfill
        \begin{subfigure}[b]{0.475\textwidth}   
            \centering 
            \includegraphics[width=\textwidth]{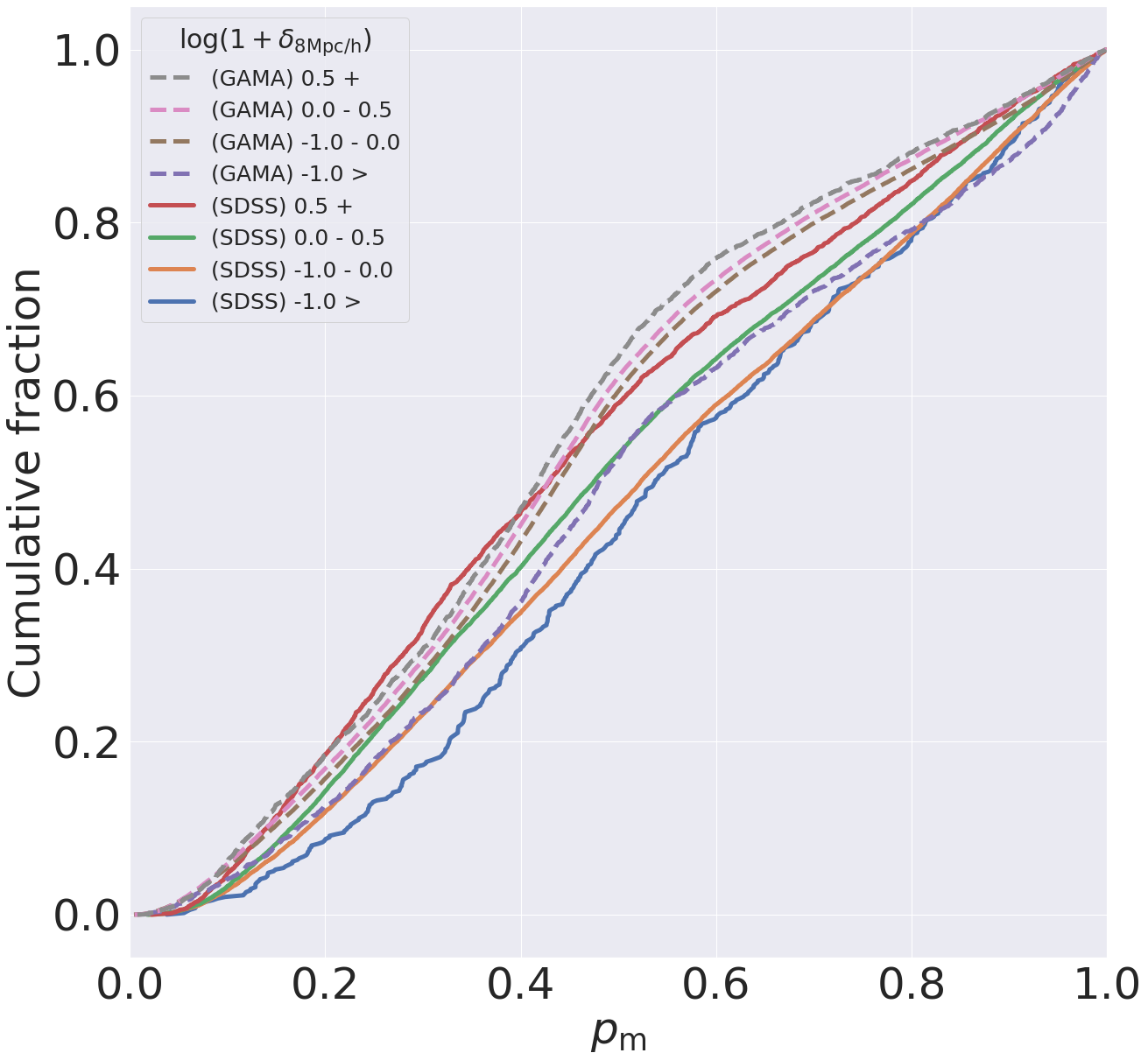}
            \caption[]%
            {{\small $8h^{-1}$ Mpc stellar mass overdensity $\log(1+\delta_{8\textrm{Mpc}/h})$\\ KS test (SDSS) statistic: 0.183 %p-value: $1.617\times10^{-12}$ 
            \\ KS test (GAMA) statistic: 0.127}}% p-value: $1.591\times10^{-22}$}}    
            \label{fig:8Mpc}
        \end{subfigure}
        \vskip\baselineskip
        \begin{subfigure}[b]{0.475\textwidth}   
            \centering 
            \includegraphics[width=\textwidth]{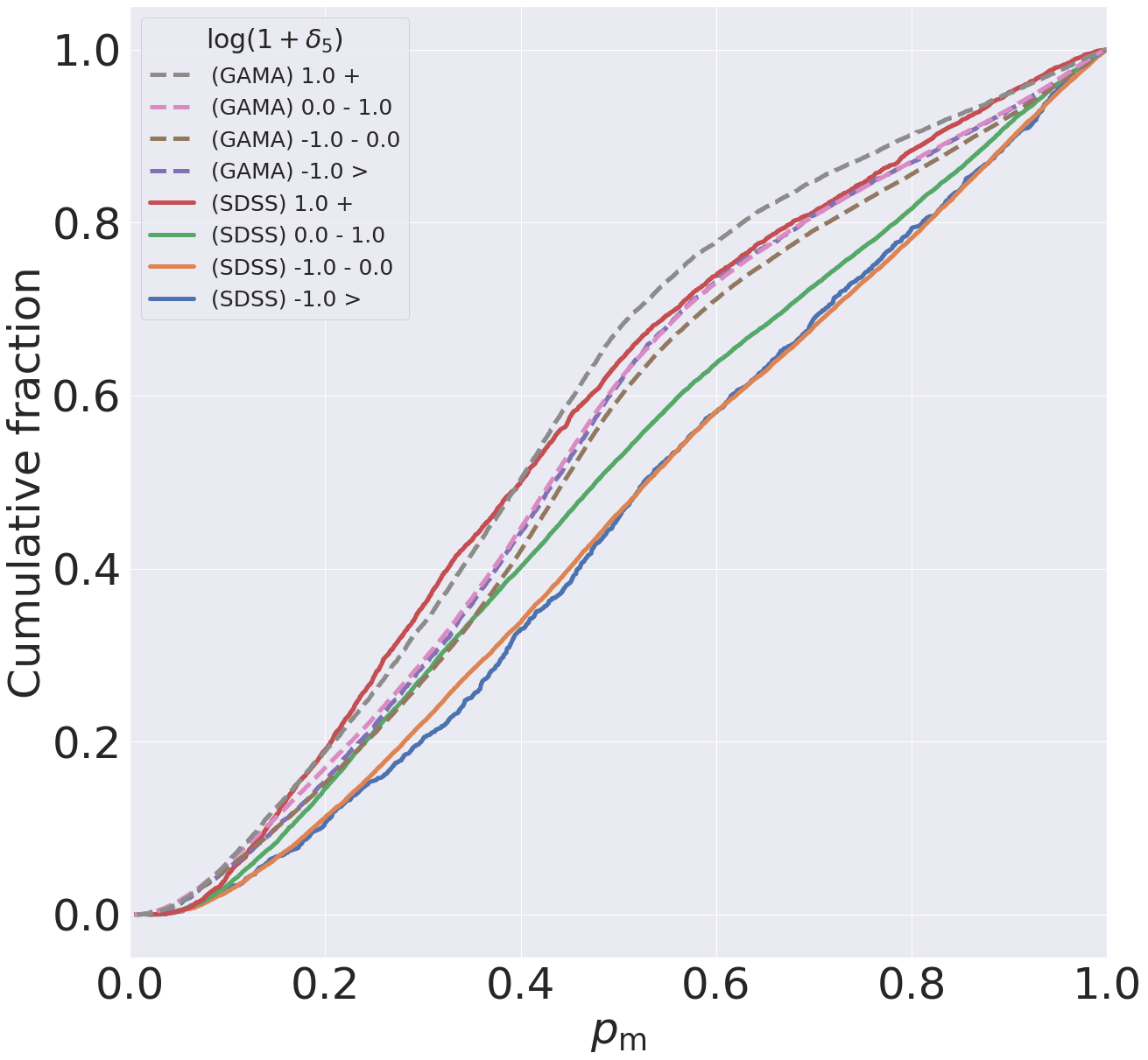}
            \caption[]%
            {{\small 5th nearest neighbor overdensity $\log(1+\delta_{5})$\\ KS test (SDSS) statistic: 0.191 %p-value: $1.564\times10^{-31}$
            \\ KS test (GAMA) statistic: 0.069}}% p-value: $1.148\times10^{-19}$}}    
            \label{fig:5nearest}
        \end{subfigure}
        \caption[ The average and standard deviation of critical parameters ]
        {\small Cumulative distribution curves of merger probabilities of HSC galaxies cross-matched with GAMA (dotted lines) and SDSS (solid lines) predicted by our fine-tuned model. Each curve represents a different environmental density bin. The gradients of these curves show that in lower density environments, there are more higher merger probability galaxies, and in higher density environments, there are more galaxies with lower merger probabilities. This trend holds true for each parameter investigated. The trend also holds qualitatively between the SDSS and GAMA samples. We also indicate the KS test statistic between the densest and least dense regions in each figure as reference.}
        \label{fig:envcumul}
\end{figure*}

Figure \ref{fig:envcumul} shows the cumulative distributions of merger probability in bins of various environment metrics described in Section \ref{section:Data}. The distributions are split into separate density bins, depending on the scale of the parameter used.

The upper left panel of Fig. \ref{fig:envcumul} shows the sensitivity of merger probability to environmental overdensities at 0.5 $h^{-1}$Mpc scales (1 $h^{-1}$Mpc aperture, centered on the target galaxy). The upper right panel shows the same at 1 $h^{-1}$Mpc scales (2 $h^{-1}$Mpc aperture). The bottom left panel shows the same at 8 $h^{-1}$Mpc scales (16 $h^{-1}$Mpc aperture). The bottom right shows the sensitivity of merger probability to environmental overdensities within the radii of the target galaxy's 5th nearest neighbor. From blue to red, the curves show the cumulative distributions of merger probability in increasingly dense environments.

We find a clear difference in the distribution curves and histograms between the lowest density $(\log(1+\delta_{x})<-1.0)$ and highest density $(\log(1+\delta_{x})>1.0)$ environments, with a similar trend holding across all 4 environmental parameters investigated. In each panel, we find that the lowest density environments contain the largest number of galaxies with high merger probability, as seen by the blue curve in each panel. This is the most pronounced in the blue curves in the upper left (0.5 $h^{-1}$Mpc) and upper right (1 $h^{-1}$Mpc) panels, but still qualitatively hold true in the bottom two panels (8 $h^{-1}$Mpc and 5th nearest neighbor scale). 

Conversely, we find that higher density environments contain the largest number of galaxies with low merger probability, as seen by the red curves. 

In higher density environments $(\log(1+\delta_{x})>0.0)$, we find that galaxies with a lower merger probability tend to be in this environment, as seen by the red and green curves. For lower density environments $(\log(1+\delta_{x})<0.0)$, we find that the opposite holds true, that higher merger probability galaxies tend to be in lower density environments, shown by the blue and orange curves.
We also find that these trends, in general, qualitatively hold true regardless of stellar mass of the target galaxy, as shown in Appendix \ref{appendix: D}. In most figures, the most mass overdense regions have more galaxies with low merger probability, and high merger probability galaxies are more likely to lie in mass underdense regions. We note that mass overdense and underdense do not necessarily refer to group members and non-group members, and we find no differences when investigating these qualitative trends separately for group and non-group members. Further, as the stellar mass overdensities have a strong correlation with halo mass \citep{2022ApJ...936..124Y}, these trends also appear at differing halo mass scales.

These results are consistent with the findings of works such as \citet{1998MNRAS.300..146G}, \citet{2010ApJ...718.1158L} and \citet{2012A&A...539A..46A}. These works, with environments computed at similar scales, suggest that mergers and merging pairs are more likely to happen in less dense environments. In addition, spectroscopic pair matching methods with strict spectroscopic cuts are more likely to produce results similar to our findings \citep{2010MNRAS.407.1514E}.

However, we also note that these results contradict with the suggestion that galaxy interactions are associated with intermediate to higher density regions, where galaxies have close companions and neighbors \citep{2010MNRAS.401.1552D, 2012ApJ...754...26J}. For example, \citet{2010MNRAS.401.1552D} found that at the $\log(1+\delta_{2\mathrm{h^{-1}Mpc}})$ scale, even though both mergers and non-mergers peak in an intermediate environment, mergers occupy a slightly denser environment, which differs from our findings. We will discuss possible reasons in Section \ref{section:discussion}.
% We also note that there are studies where mergers samples are made using visual identification techniques, where galaxies with many visual neighbors, i.e., galaxies in group to high density environments, may be considered mergers, and the difference in merger sample selection may result in contradicting conclusions.

\subsection{Comparison with simulation data}

\begin{figure*}
        \centering
        \begin{subfigure}[b]{0.475\textwidth}
            \centering
            \includegraphics[width=\textwidth]{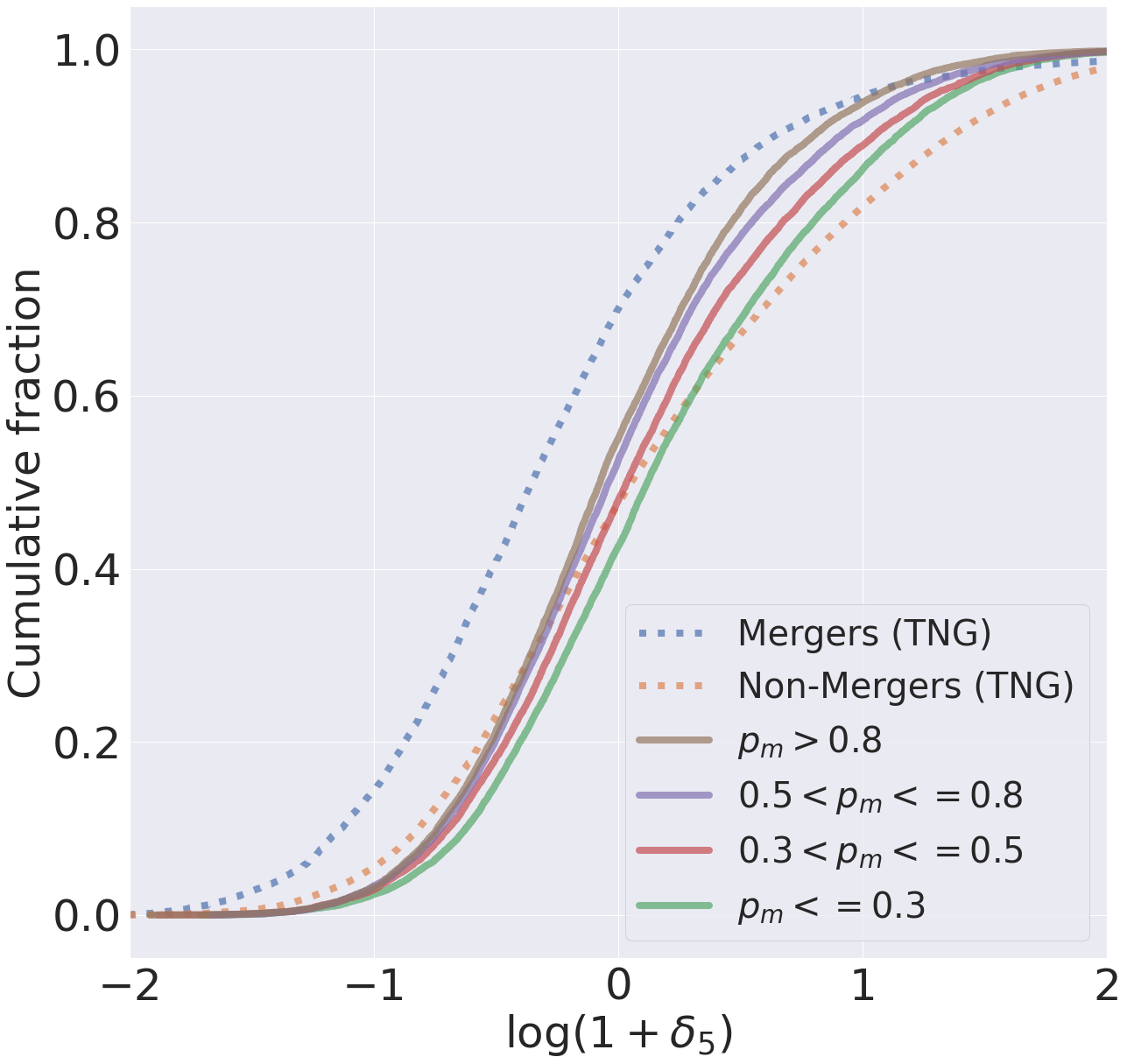}
            \caption[Network2]%
            {{\small Major companions\\ KS test (Simulation) statistic: 0.226 %p-value: 0
            \\ KS test (Observation) statistic: 0.131}}% p-value: $1.273\times10^{-78}$}}}    
            \label{fig:5maj}
        \end{subfigure}
        \hfill
        \centering
        \begin{subfigure}[b]{0.475\textwidth}
            \centering
            \includegraphics[width=\textwidth]{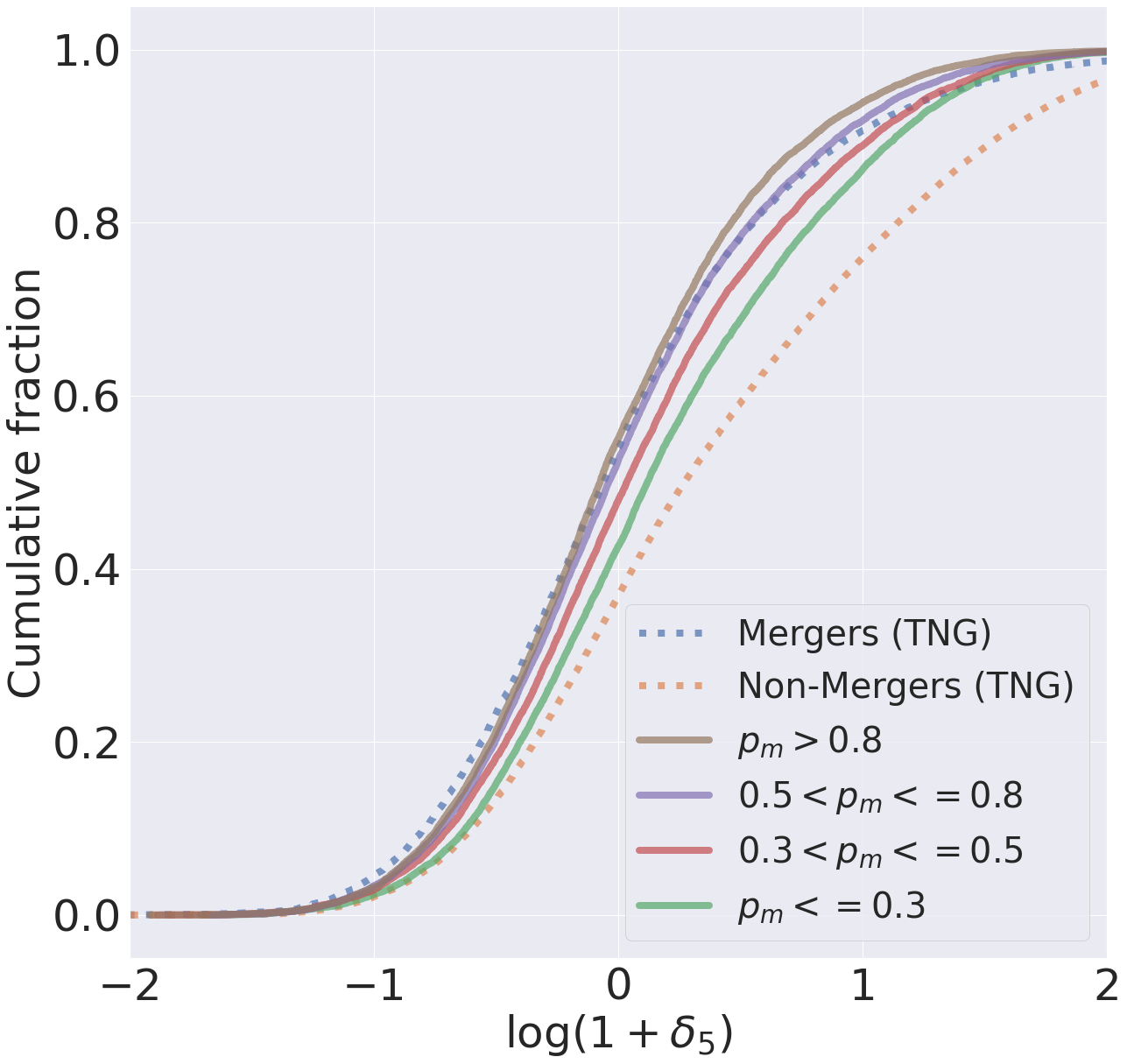}
            \caption[Network2]%
            {{\small Major and minor companions\\ KS test (Simulation) statistic: 0.194 %p-value: 0
            \\ KS test (Observation) statistic: 0.131}}% p-value: $1.273\times10^{-78}$}}}    
            \label{fig:5min}
        \end{subfigure}
        \hfill
        % \begin{subfigure}[b]{0.350\textwidth}  
        %     \centering 
        %     \includegraphics[width=\textwidth]{5thnearestmini.png}
        %     \caption[]%
        %     {{\small mini Neighbors}}    
        %     \label{fig:5mini}
        % \end{subfigure}
        \caption[ The average and standard deviation of critical parameters ]
        {\small Environment distribution cumulative histograms of the environmental densities within a spherical volume including the 5th nearest neighbor to the target galaxy, for TNG50 and TNG100 mergers and non-mergers. From left to right: a) environmental mass densities taking into account companions with mass ratio >1:4 (major companions), and b) environmental mass densities taking into account companions with mass ratio > 1:10 (major and minor companions). We also make the histograms for the observational data visible as reference where available, indicated by the dotted lines.KS test statistics between simulation mergers and non-mergers, as well as between confident mergers (merger probability >0.8) and non-mergers (merger probability <0.3) are indicated.}
        \label{fig:5envhist}
\end{figure*}
\begin{figure*}
        \centering
        \begin{subfigure}[b]{0.475\textwidth}
            \centering
            \includegraphics[width=\textwidth]{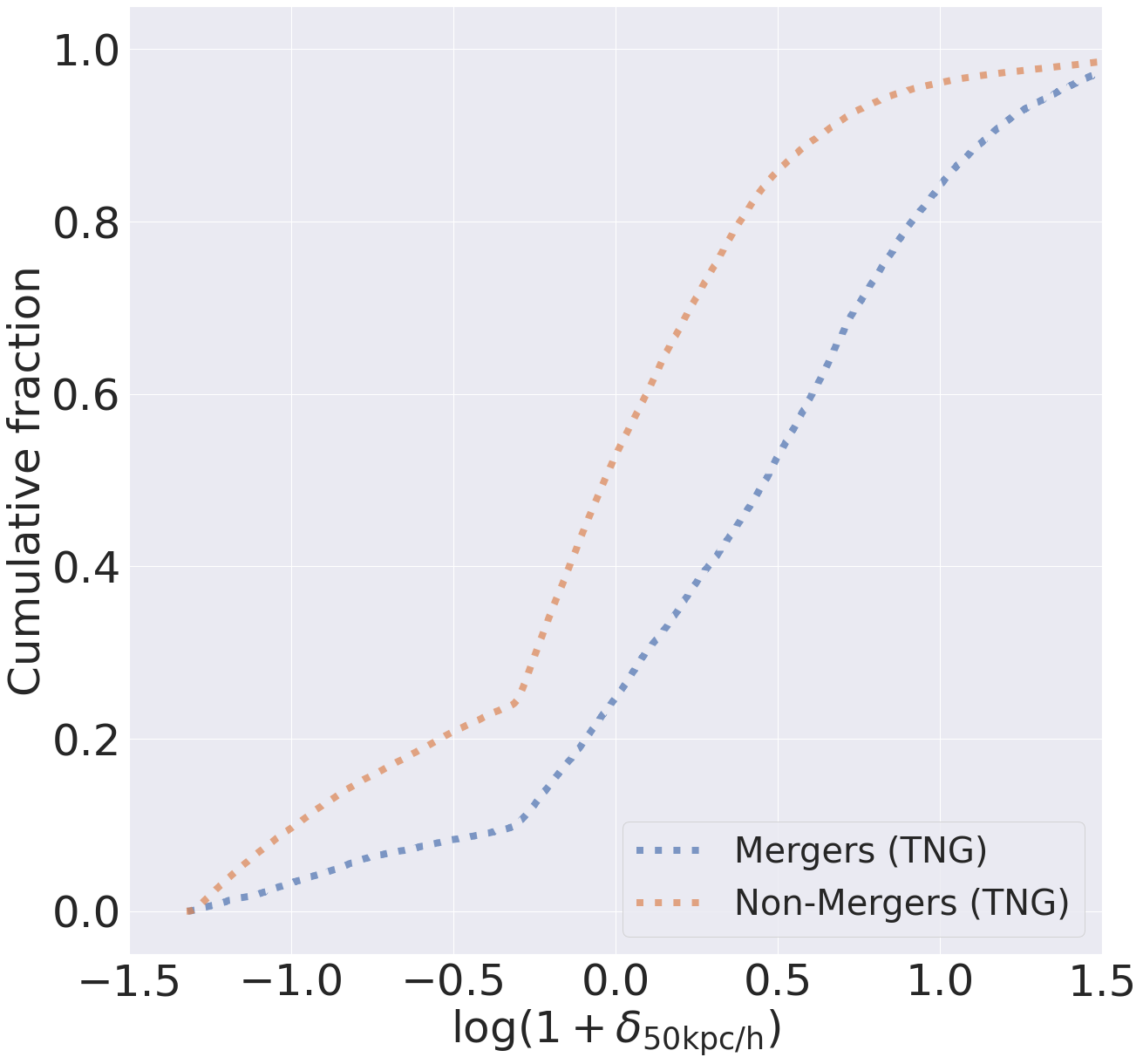}
            \caption[Network2]%
            {{\small Major companions\\ KS test (Simulation) statistic: 0.348 %p-value: 0
            }}  
            \label{fig:50kmaj}
        \end{subfigure}
        \hfill
        \centering
        \begin{subfigure}[b]{0.475\textwidth}
            \centering
            \includegraphics[width=\textwidth]{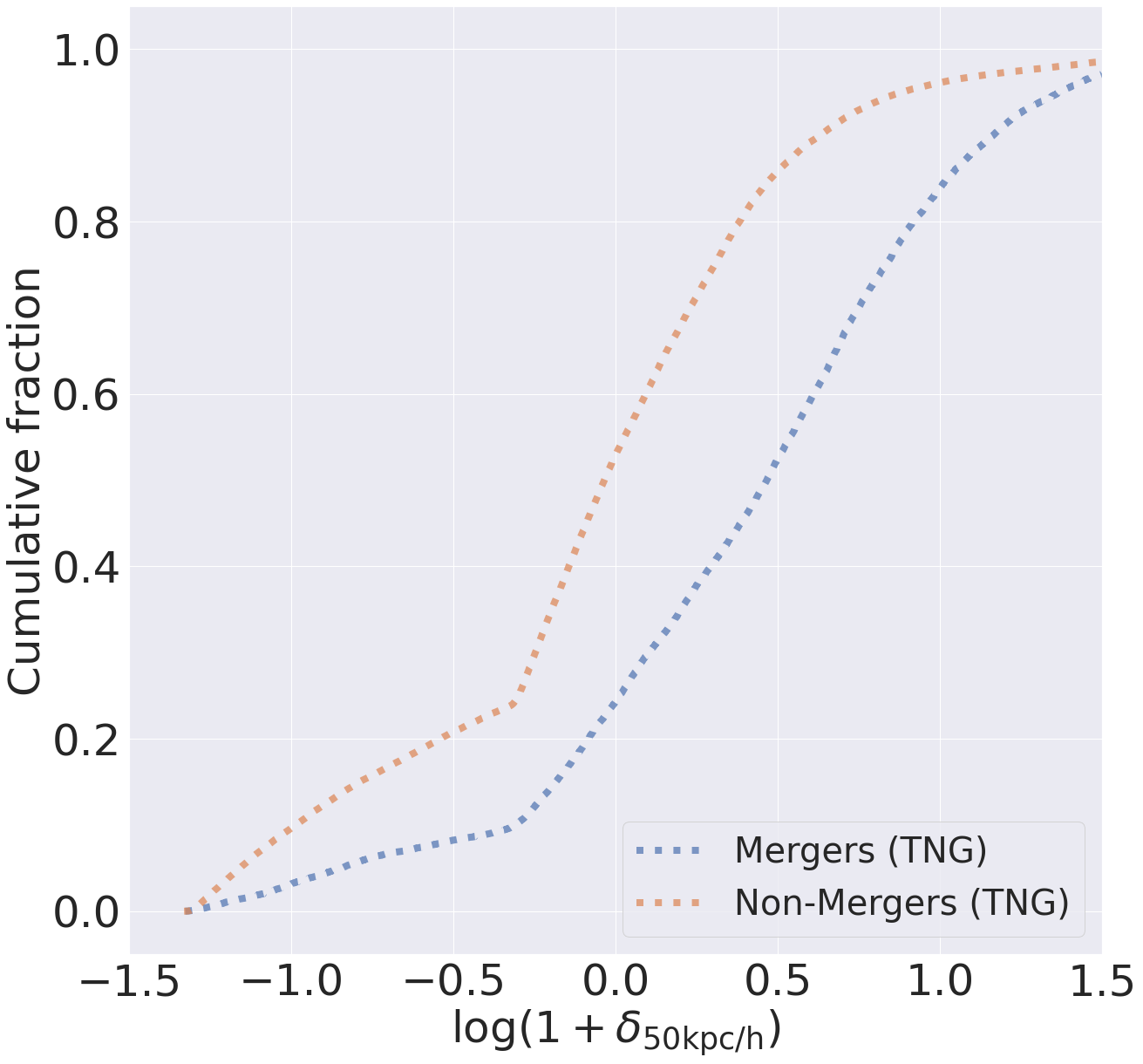}
            \caption[Network2]%
            {{\small Major and minor companions\\ KS test (Simulation) statistic: 0.353% p-value: 0
            }}
            \label{fig:50kmin}
        \end{subfigure}
        \hfill
        % \begin{subfigure}[b]{0.350\textwidth}  
        %     \centering 
        %     \includegraphics[width=\textwidth]{50kpcmini.png}
        %     \caption[]%
        %     {{\small mini Neighbors}}       
        %     \label{fig:50kmini}
        % \end{subfigure}
        \caption[ The average and standard deviation of critical parameters ]
        {\small Same as Fig. \ref{fig:5envhist}, but for stellar mass overdensities within a 50 kpc radii spherical volume.} 
        \label{fig:50kpcenvhist}
\end{figure*}
\begin{figure*}
        \centering
        \begin{subfigure}[b]{0.475\textwidth}
            \centering
            \includegraphics[width=\textwidth]{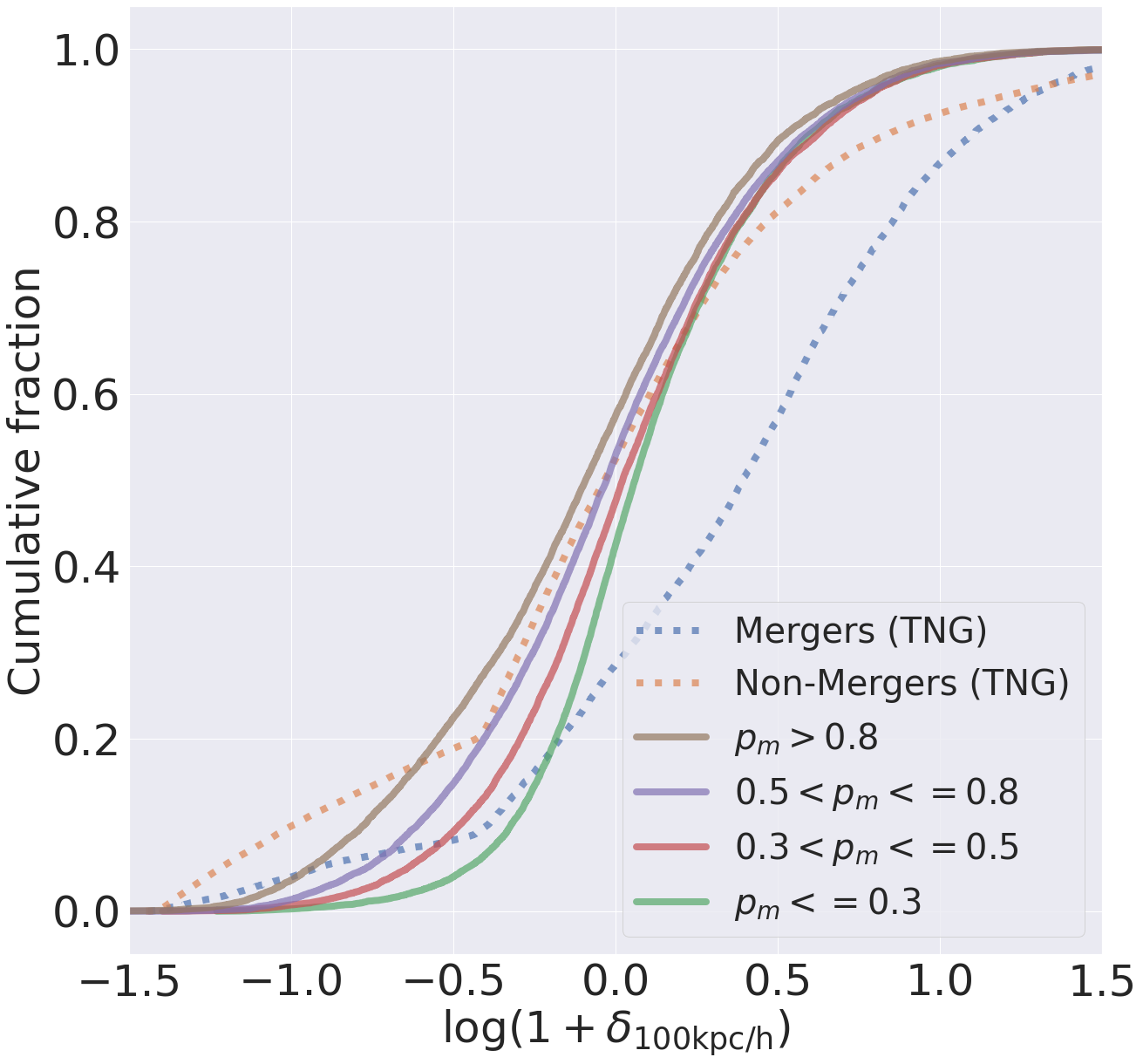}
            \caption[Network2]%
            {{\small Major companions\\ KS test (Simulation) statistic: 0.285% p-value: 0
            \\ KS test (Observation) statistic: 0.231}}% p-value: $7.806\times10^{-244}$}}}    
            \label{fig:100kmaj}
        \end{subfigure}
        \hfill
        \centering
        \begin{subfigure}[b]{0.475\textwidth}
            \centering
            \includegraphics[width=\textwidth]{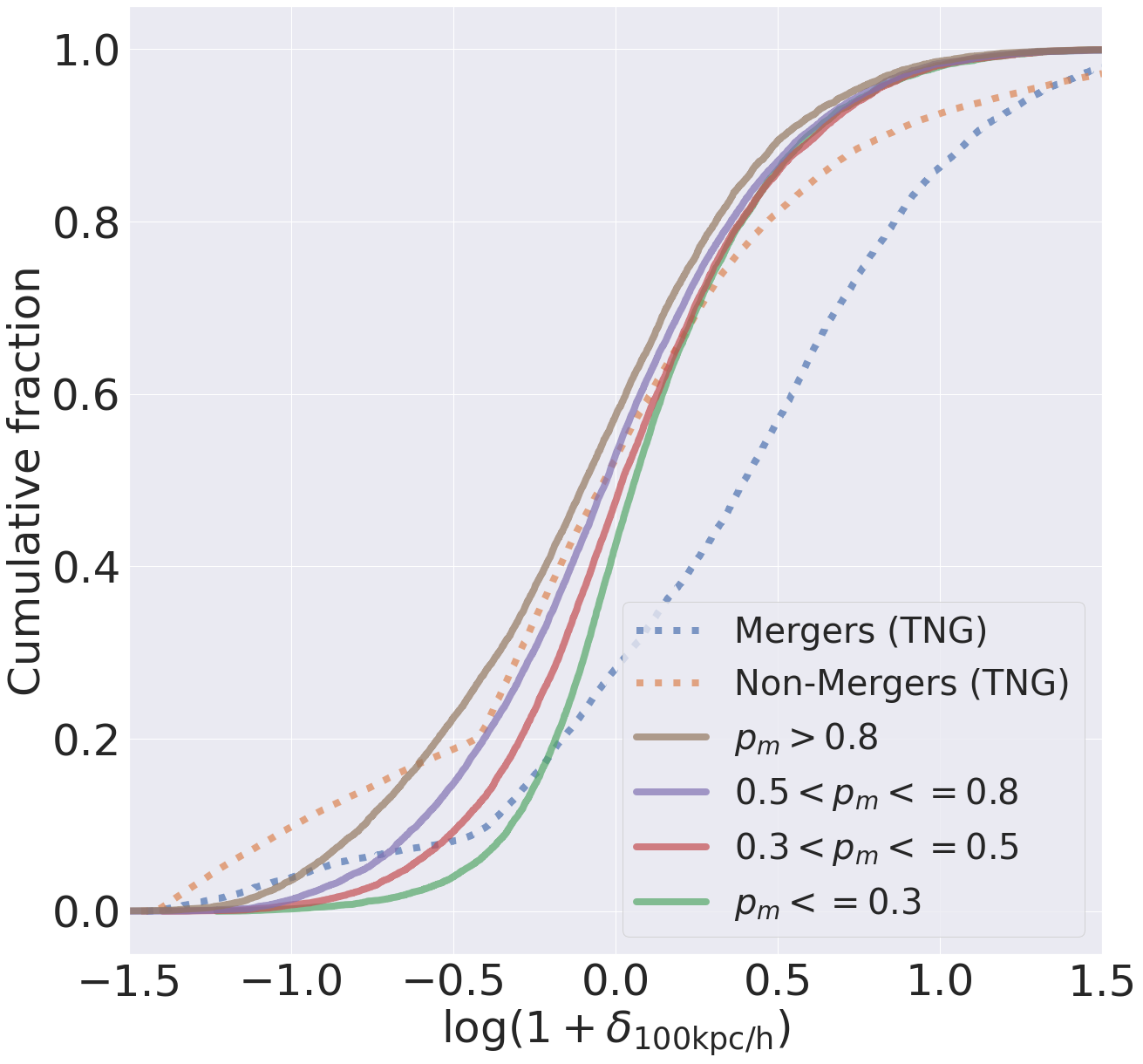}
            \caption[Network2]%
            {{\small Major and minor companions\\ KS test (Simulation) statistic: 0.260% p-value: 0
            \\ KS test (Observation) statistic: 0.231}}% p-value: $7.806\times10^{-244}$}}}   
            \label{fig:100kmin}
        \end{subfigure}
        \hfill
        % \begin{subfigure}[b]{0.350\textwidth}  
        %     \centering 
        %     \includegraphics[width=\textwidth]{100kpcmini.png}
        %     \caption[]%
        %     {{\small mini Neighbors}}      
        %     \label{fig:100kmini}
        % \end{subfigure}
        \caption[ The average and standard deviation of critical parameters ]
        {\small Same as Fig. \ref{fig:5envhist}, but for stellar mass overdensities within a 100 kpc radii spherical volume.} 
        \label{fig:100kpcenvhist}
\end{figure*}
\begin{figure*}
        \centering
        \begin{subfigure}[b]{0.475\textwidth}
            \centering
            \includegraphics[width=\textwidth]{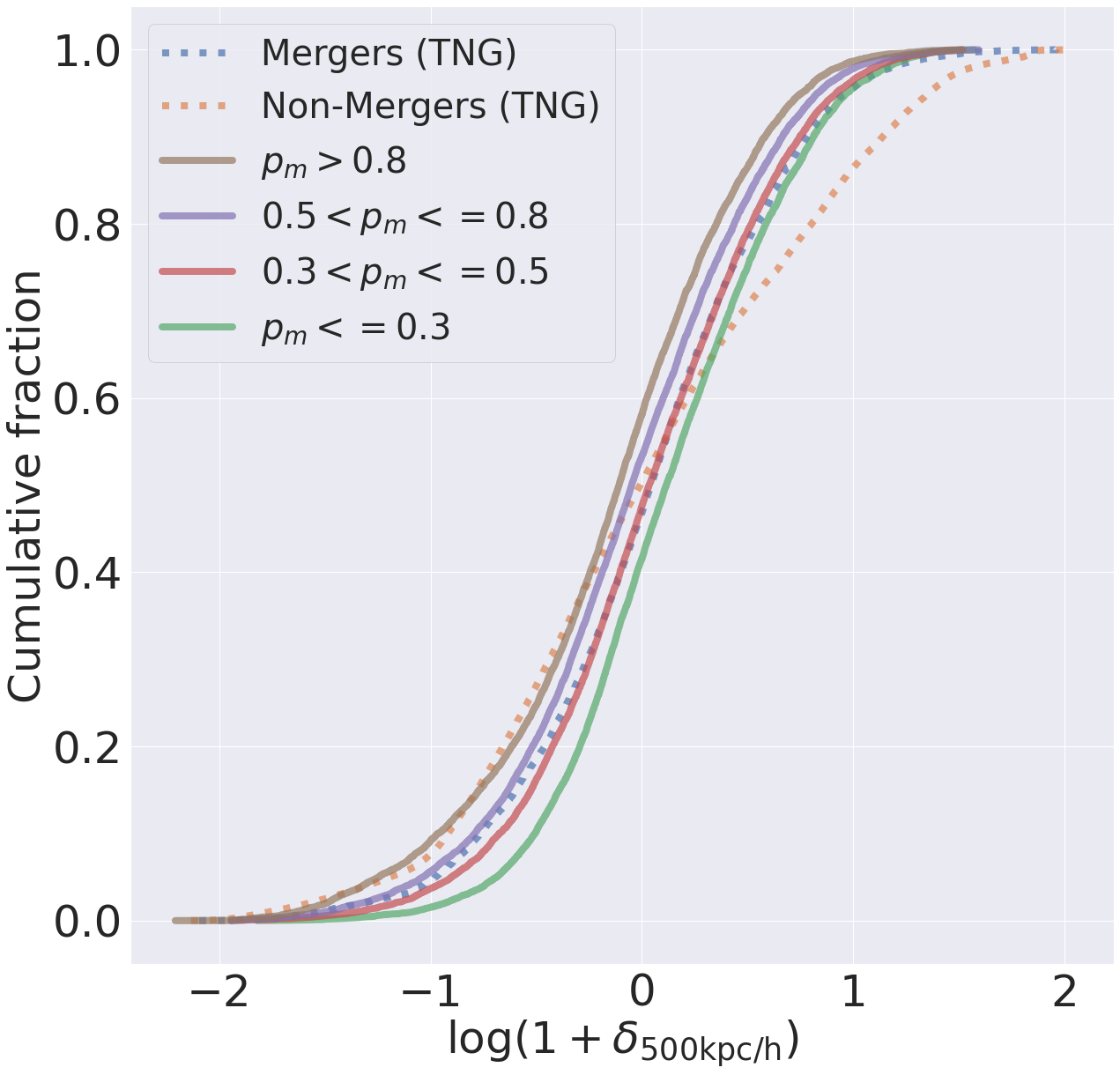}
            \caption[Network2]%
            {{\small Major companions\\ KS test (Simulation) statistic: 0.108% p-value: $8.746\times10^{-222}$
            \\ KS test (Observation) statistic: 0.171}}% p-value: $4.115\times10^{-133}$}}}    
            \label{fig:500kmaj}
        \end{subfigure}
        \hfill
        \centering
        \begin{subfigure}[b]{0.475\textwidth}
            \centering
            \includegraphics[width=\textwidth]{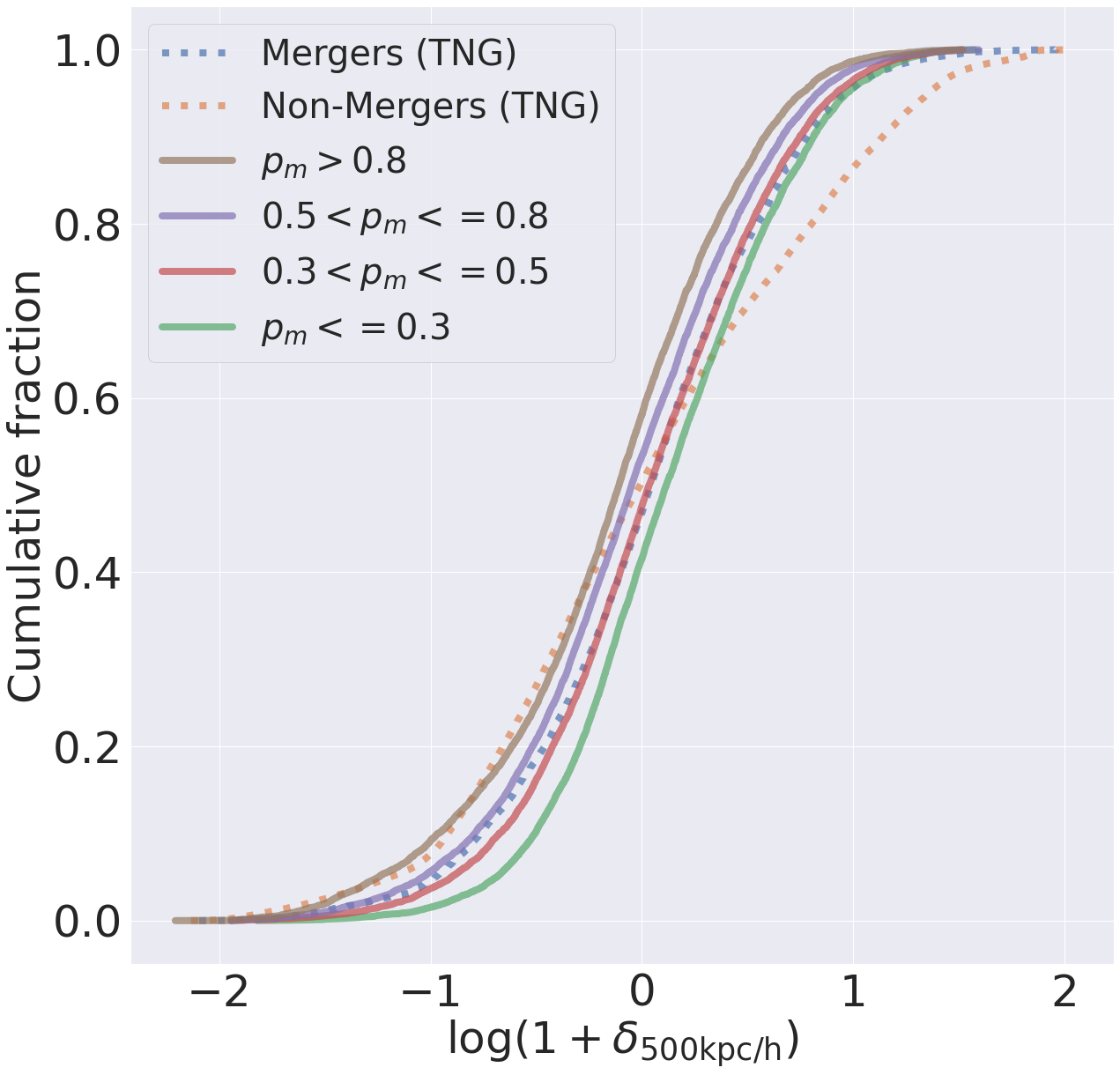}
            \caption[Network2]%
            {{\small Major and minor companions\\ KS test (Simulation) statistic: 0.100% p-value: $1.739\times10^{-190}$
            \\ KS test (Observation) statistic: 0.171}}% p-value: $4.115\times10^{-133}$}}}    
            \label{fig:500kmin}
        \end{subfigure}
        \hfill
        % \begin{subfigure}[b]{0.350\textwidth}  
        %     \centering 
        %     \includegraphics[width=\textwidth]{500kpcmini.png}
        %     \caption[]%
        %     {{\small mini Neighbors}}       
        %     \label{fig:500kmini}
        % \end{subfigure}
        \caption[ The average and standard deviation of critical parameters ]
        {\small Same as Fig. \ref{fig:5envhist}, but for stellar mass overdensities within a 500 kpc radii spherical volume. Observational data is available for this parameter.} 
        \label{fig:500kpcenvhist}
\end{figure*}
\begin{figure*}
        \centering
        \begin{subfigure}[b]{0.475\textwidth}
            \centering
            \includegraphics[width=\textwidth]{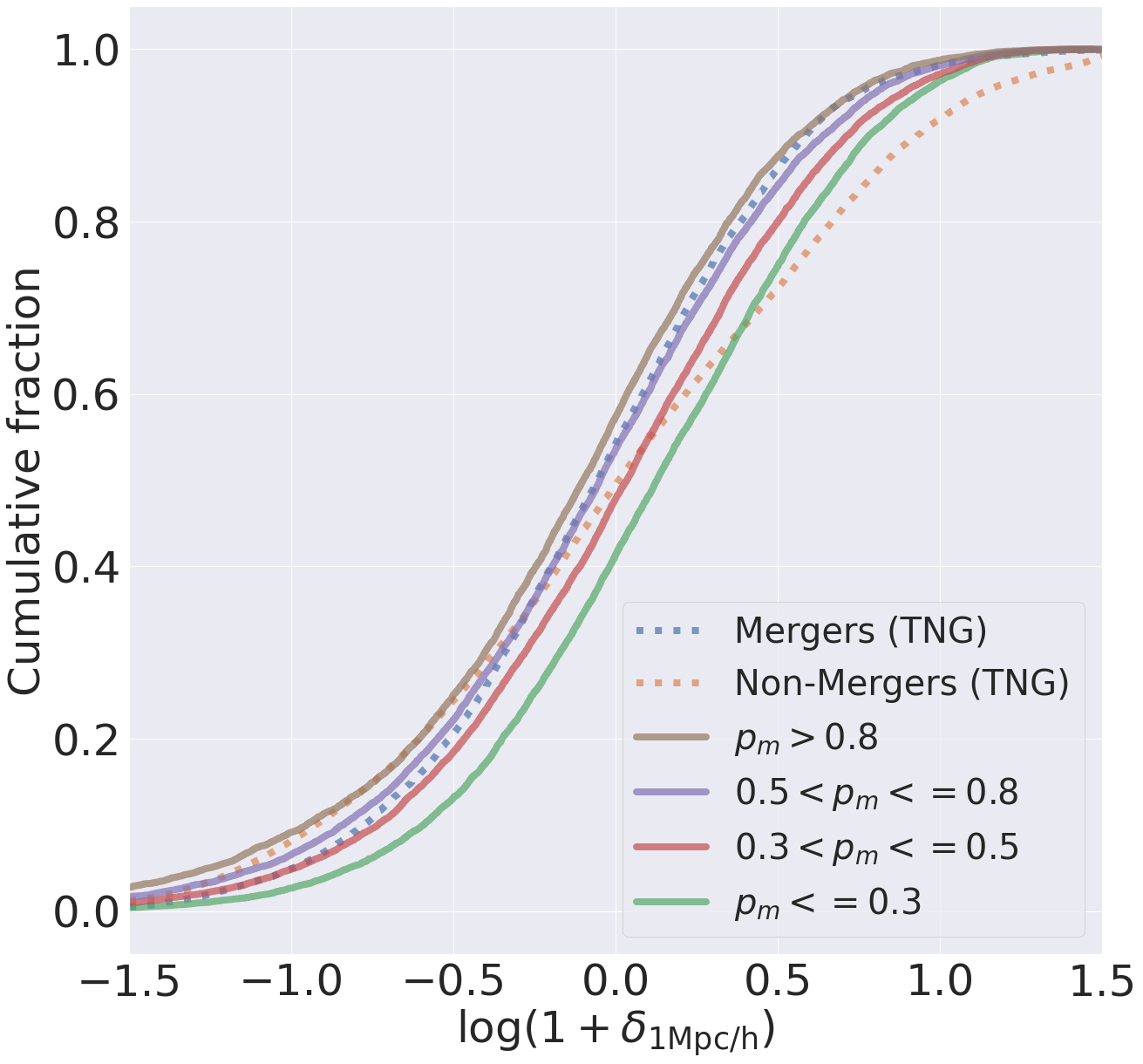}
            \caption[Network2]%
            {{\small Major companions\\ KS test (Simulation) statistic: 0.140% p-value:0
            \\ KS test (Observation) statistic: 0.168}}% p-value: $1.737\times10^{-128}$}}}    
            \label{fig:1Mmaj}
        \end{subfigure}
        \hfill
        \centering
        \begin{subfigure}[b]{0.475\textwidth}
            \centering
            \includegraphics[width=\textwidth]{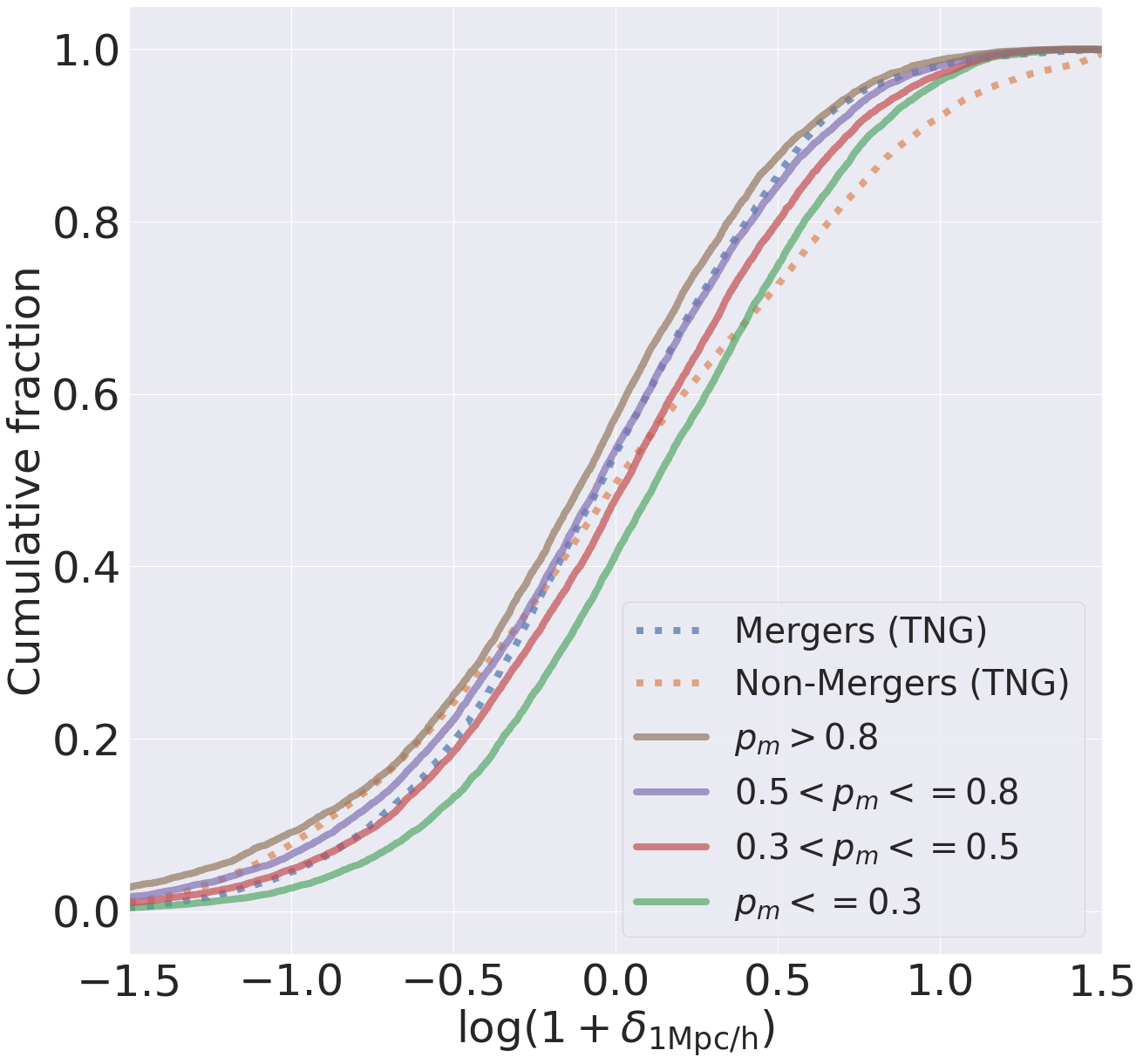}
            \caption[Network2]%
            {{\small Major and minor companions\\ KS test (Simulation) statistic: 0.129 %p-value: $1.179\times10^{-316}$
            \\ KS test (Observation) statistic: 0.168}}% p-value: $1.737\times10^{-128}$}}}    
            \label{fig:1Mmin}
        \end{subfigure}
        \hfill
        % \begin{subfigure}[b]{0.350\textwidth}  
        %     \centering 
        %     \includegraphics[width=\textwidth]{1Mpcmini.png}
        %     \caption[]%
        %     {{\small mini Neighbors}}      
        %     \label{fig:1Mmini}
        % \end{subfigure}
        \caption[ The average and standard deviation of critical parameters ]
        {\small Same as Fig. \ref{fig:5envhist}, but for stellar mass overdensities within a 1 Mpc radii spherical volume. Observational data is available for this parameter.} 
        \label{fig:1Mpcenvhist}
\end{figure*}
\begin{figure*}
        \centering
        \begin{subfigure}[b]{0.475\textwidth}
            \centering
            \includegraphics[width=\textwidth]{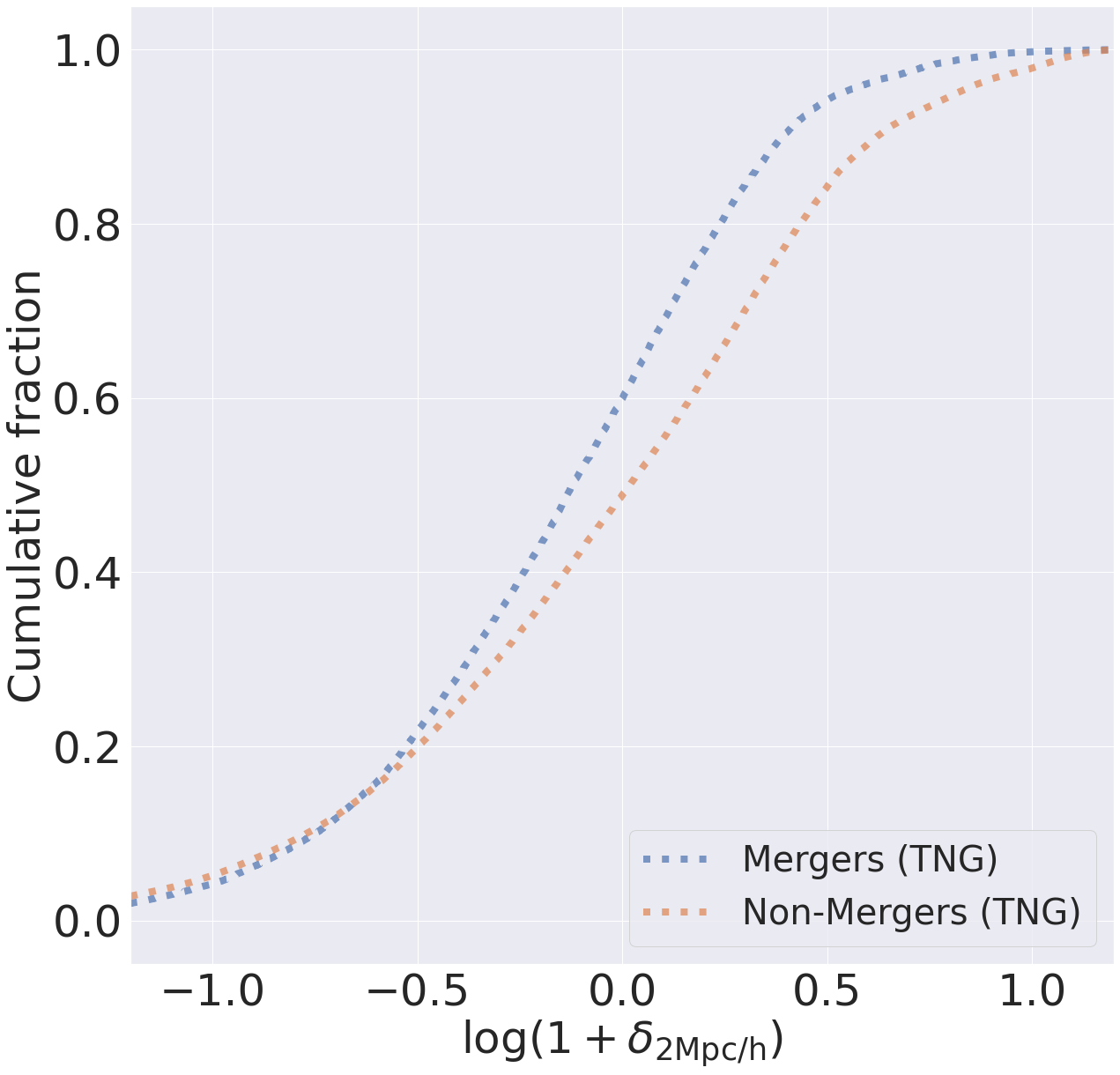}
            \caption[Network2]%
            {{\small Major companions\\ KS test (Simulation) statistic: 0.147 %p-value: 0
            }}    
            \label{fig:2Mmaj}
        \end{subfigure}
        \hfill
        \centering
        \begin{subfigure}[b]{0.475\textwidth}
            \centering
            \includegraphics[width=\textwidth]{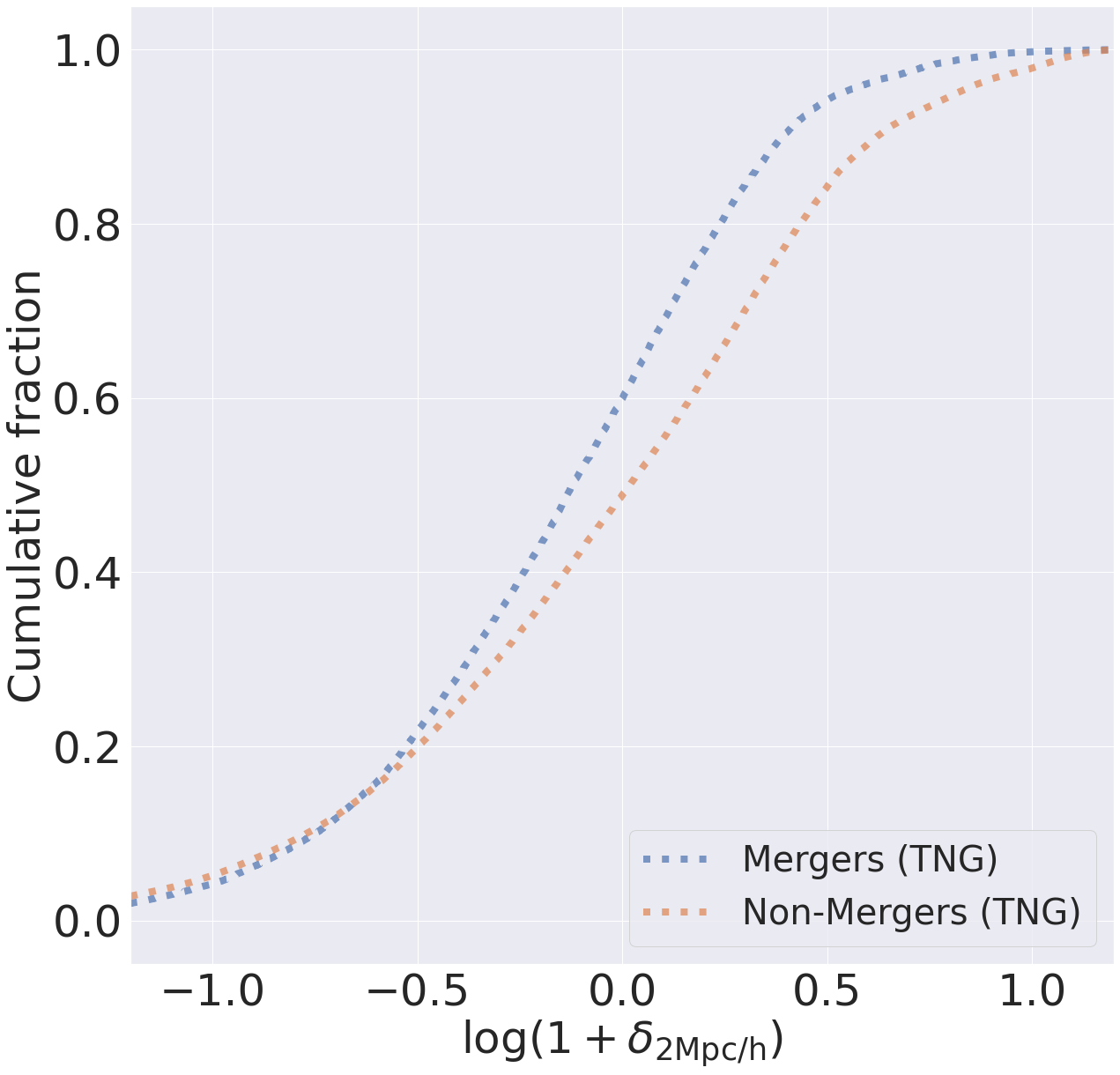}
            \caption[Network2]%
            {{\small Major and minor companions\\ KS test (Simulation) statistic: 0.131}% p-value: 0
            }
            \label{fig:2Mmin}
        \end{subfigure}
        \hfill
        % \begin{subfigure}[b]{0.350\textwidth}  
        %     \centering 
        %     \includegraphics[width=\textwidth]{2Mpcmini.png}
        %     \caption[]%
        %     {{\small mini Neighbors}}      
        %     \label{fig:2Mmini}
        % \end{subfigure}
        \caption[ The average and standard deviation of critical parameters ]
         {\small Same as Fig. \ref{fig:5envhist}, but for stellar mass overdensities within a 2 Mpc radii spherical volume.} 
        \label{fig:2Mpcenvhist}
\end{figure*}
\begin{figure*}
        \centering
        \begin{subfigure}[b]{0.475\textwidth}
            \centering
            \includegraphics[width=\textwidth]{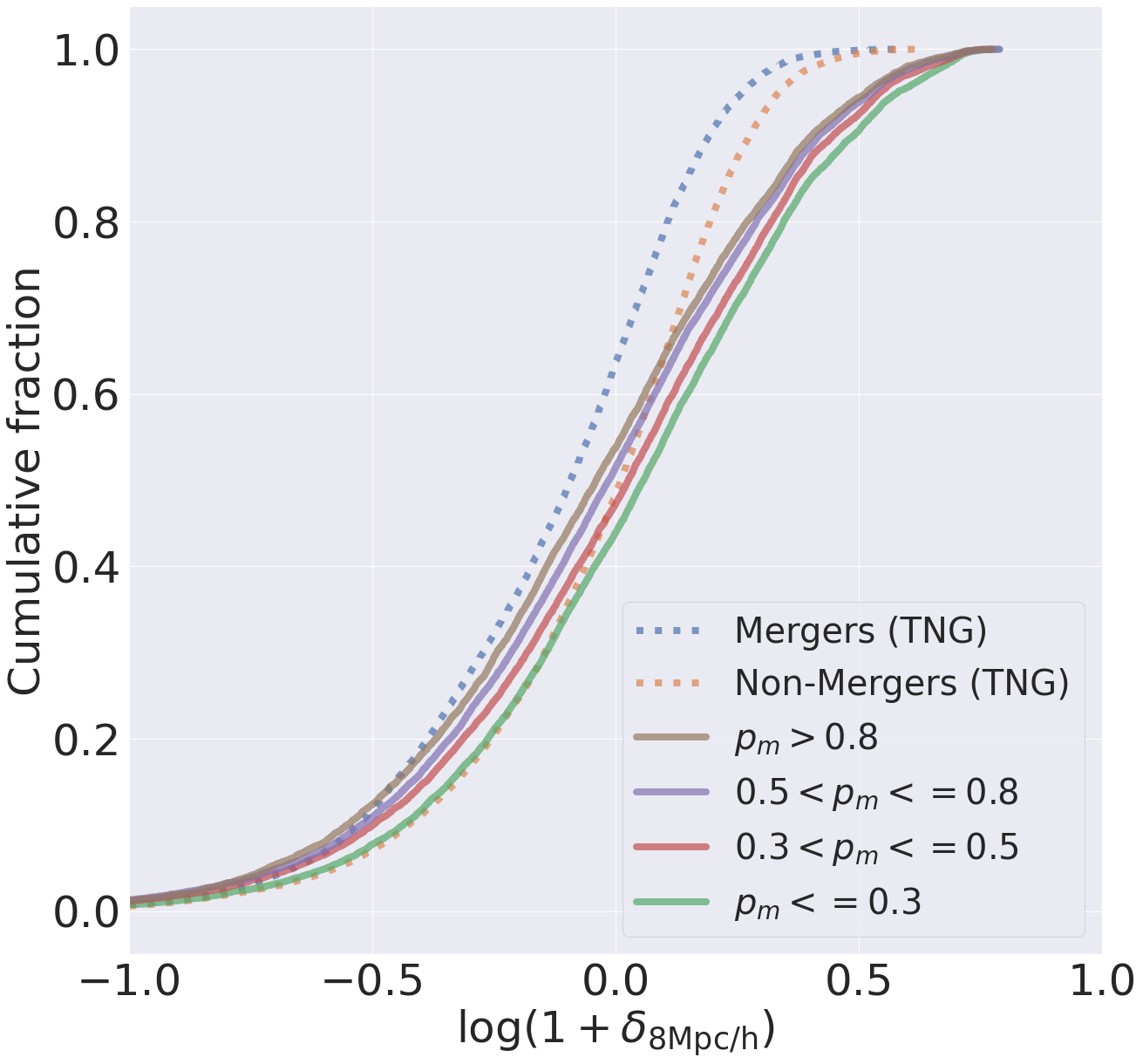}
            \caption[Network2]%
            {{\small Major companions\\ KS test (Simulation) statistic: 0.153% p-value: 0
            \\ KS test (Observation) statistic: 0.102}}% p-value: $3.591\times10^{-47}$}}    
            \label{fig:8Mmaj}
        \end{subfigure}
        \hfill
        \centering
        \begin{subfigure}[b]{0.475\textwidth}
            \centering
            \includegraphics[width=\textwidth]{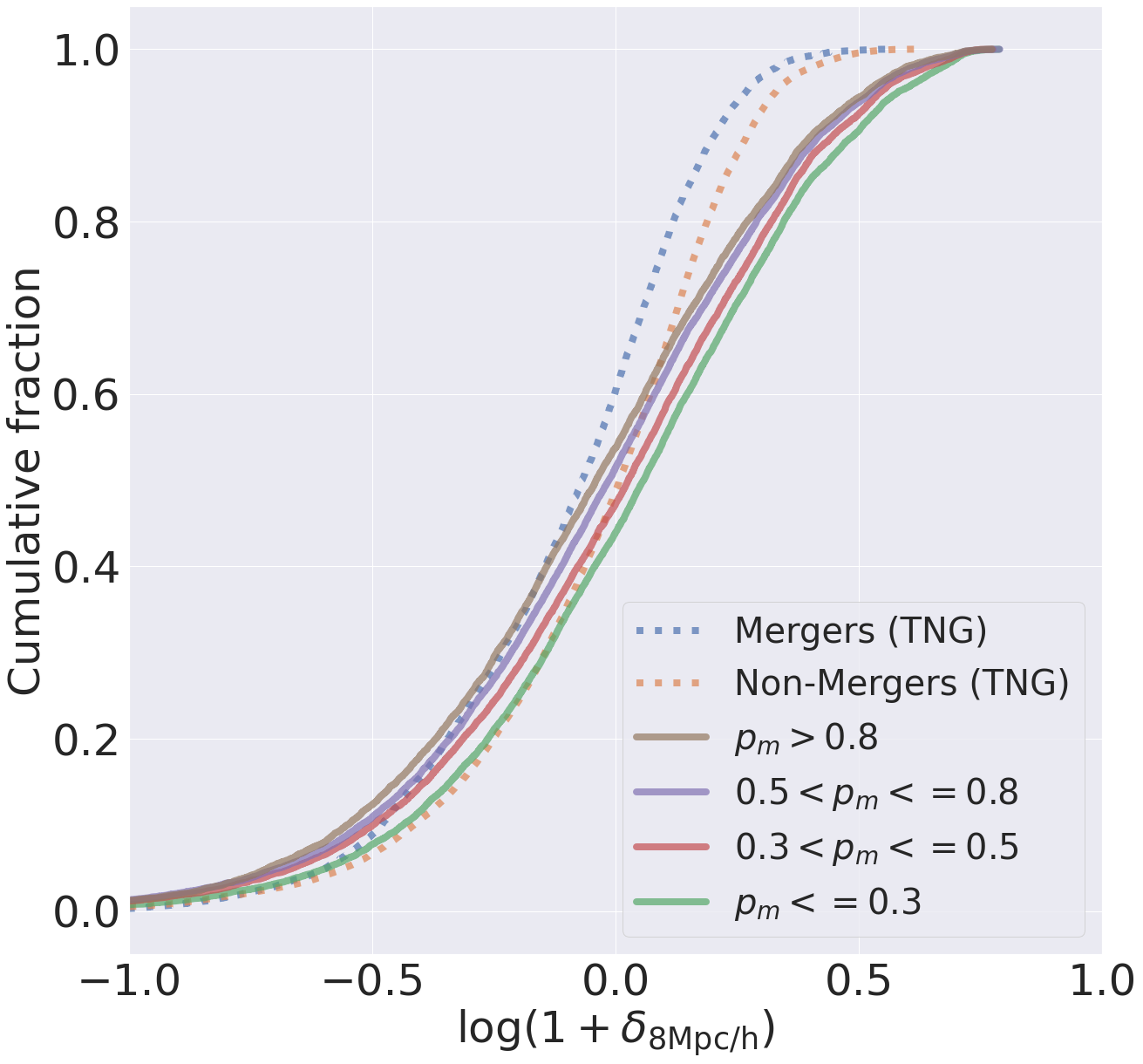}
            \caption[Network2]%
            {{\small Major and minor companions\\ KS test (Simulation) statistic: 0.123% p-value: $3.774\times10^{-286}$
            \\ KS test (Observation) statistic: 0.102}}% p-value: $3.591\times10^{-47}$}}    
            \label{fig:8Mmin}
        \end{subfigure}
        \hfill
        % \begin{subfigure}[b]{0.350\textwidth}  
        %     \centering 
        %     \includegraphics[width=\textwidth]{2Mpcmini.png}
        %     \caption[]%
        %     {{\small mini Neighbors}}      
        %     \label{fig:2Mmini}
        % \end{subfigure}
        \caption[ The average and standard deviation of critical parameters ]
         {\small Same as Fig. \ref{fig:5envhist}, but for stellar mass overdensities within a 8 Mpc radii spherical volume. Observational data is available for this parameter.} 
        \label{fig:8Mpcenvhist}
\end{figure*}
Figures \ref{fig:5envhist} through \ref{fig:8Mpcenvhist} show the cumulative histograms of environmental distributions of a total of 244,722 TNG50 and TNG100 mergers and non-mergers between snapshots 59 and 99 (redshifts $0.7>z>0.1$), grouped by redshift, further split into two subfigures, with the left figure accounting for companions with mass ratio > 1:4 (major companions) and the right figure accounting for companions with mass ratio > 1:10 (major and minor companions), for each parameter. The mergers and non-mergers were selected using the same timescale criteria used in creating the fine-tuning dataset for Zoobot, for a total of 17,877 mergers and  226,895 non-mergers.  The environmental parameters used are similar to the observations, being the stellar mass overdensities within a spherical volume with radii of 0.05, 0.1, 0.5, 1, 2, and 8 $h^{-1}$Mpc, as well as within the radii of the 5th nearest neighbor, and are calculated through similar computations as the environmental parameters in the observations in Section \ref{section:Data}. We also make the environmental distributions from the observations visible where the data is available, being the density within the 0.1 $h^{-1}$Mpc, 0.5 $h^{-1}$Mpc, 1 $h^{-1}$Mpc, 8 $h^{-1}$Mpc  radii spherical volume, and the radii of the 5th nearest neighbor. We split the observational data in 4 different merger probability bins, with $p_m$ <0.3, 0.3<$p_m$<0.5, 0.5<$p_m$<0.8, and 0.8<$p_m$.

We find 2 differing trends in the simulational data, depending on the scale of the environmental density parameter. For the environmental density parameters of the scale of 0.5 $h^{-1}$Mpc and larger, and for 5th nearest neighbor environments, a greater fraction of mergers lie in mass underdense environments, and a greater fraction of non-mergers lie in mass overdense environments, indicating a similar trend as that seen with the predictions on observational data, as the fraction of mergers in lower density environments increases with increasing merger probability bin. We also find that there are very little to no mergers found in the densest environments for each of these parameters. We also find that these trends are, in general, not sensitive to mass ratio, and hold true for all investigated redshift bins up to $z=0.7$. As such, we have a consistency in the relationship between galaxy mergers and environmental densities between observations and simulations at equivalent scales.

Conversely, we find that the above trend is reversed for the environmental density parameters within the spherical volume of 0.05 and 0.1 $h^{-1}$Mpc, found in Fig. \ref{fig:50kpcenvhist} and Fig. \ref{fig:100kpcenvhist}, respectively. That is, at these scales, we find that the majority of non-mergers are in the lowest density environments, as shown by the brown, red, and orange lines, and mergers are found in more dense environments, as shown by the blue, green, and purple lines. These trends are also not sensitive to mass ratio or redshift. However, this reversal does not occur in the observational data, as shown in Fig. \ref{fig:100kpcenvhist}. While the behavior of the low merger environment curves show a steeper gradient in underdense regions $(0.5<\log(1+\delta_{x})<0.0)$, the trend is qualitatively similar as the larger scales. Further investigation is required on the inconsistency between the observations and simulations at this scale, particularly as they are consistent at larger scales. We plan to investigate this in future works.

\section{Discussion}
\label{section:discussion}
Based on our merger probabilities and subsequent analysis in the previous section, our investigation finds that at the scales of 0.5 $h^{-1}$ Mpc and greater, i.e. close to the cluster scale and larger, merger galaxies are more prevalent in lower density environments, and higher density environments have a lower merger incidence, with this trend found both in observations and simulations. 
At the 0.1 $h^{-1}$Mpc and lower scales, i.e., close to the galaxy scale, there are more non-mergers in lower density environments in the simulations, however this reversal does not occur in our observational data.

Previous studies that have investigated the environmental dependence of merger activity have found varying results, with some works finding a greater merger fraction in higher density environments \citep{2012ApJ...754...26J}, and others finding merger fraction peaks in intermediate environments \citep{2009BAAA...52..225P}, and others finding that mergers are more likely to occur in lower density environments \citep{1998MNRAS.300..146G}. %cite
We have results showing that the merger prevalence can be increased at both high and low density environments, with the trend differing depending on the scale of the environmental parameter.

In the richest environments, such as in the center of clusters, velocity dispersions are high, at scales of $\sim1000$ km s$^{-1}$ \citep{1999ApJS..125...35S}. At these speeds, galaxy-galaxy interactions are likely to be elastic encounters, and mergers and infall are less likely to occur \citep{2002MNRAS.337..172K}. Such encounters can strip the gas required to fuel star formation events \citep{1972ApJ...176....1G}, and hence be a catalyst for dynamical evolution in cluster galaxies, but the final product of the interaction will likely not be a merger, resulting in lower merger incidence. As such, while higher density regions can have an increased pair fraction, but a lowered merger fraction \citep{2010ApJ...718.1158L}. Conversely, in lower density environments, accretion and merger events can occur more frequently. 

At a smaller scale, such as in the galaxy scale, over 90\% of non-merger galaxies are found in lower density environments, and higher density environments are mostly populated with mergers. At this scale, galaxies are likely to be at closer projected separation with its neighbors compared to at greater scales, and hence be part of a merger. As a result, higher densities at this scale means there is a greater likelihood of a neighbor being a physically connected merger, resulting in the reversal of the trend from the larger scales.

We suggest a possible reason for any disagreements with previous works, particularly in higher density environments, to be attributed to the difference in merger sample selection techniques.  For example, \citet{2010MNRAS.401.1552D}, a work based on visual identification of mergers, finds that while both mergers and non-mergers peak in intermediate density environments, mergers occupy slightly denser environments than non-mergers. However, \citet{2010MNRAS.407.1514E}, a work using spectroscopic pair selection, finds that a significant fraction of galaxy pairs are in higher density environments, but they also suggest that lower density environments are where mergers are likely to occur, a suggestion which is consistent with our results.

Different merger sample selection methods will lead to different merger samples, some with very little overlap \citep{2007ApJ...666..212D}, and as such, trends in physical properties such as environment may differ. Additionally, some methods may be susceptible to overestimation of merger galaxies and contamination.

Spectroscopic pair matching methods can contaminate merger samples with interlopers. In high velocity dispersion environments, using the line-of-sight velocities as a proxy for 3 dimensional velocities has limitations, and is strongly affected by interlopers \citep{2013ApJ...772...47S}, and galaxies may be considered merging pairs even if they are not physically connected. As such, merger catalogues created using these methods can overestimate the merger fraction.

Similarly, image-based classification techniques can also be contaminated with stellar and galactic chance projections, in both visual identification \citep{2010MNRAS.401.1552D, 2010MNRAS.401.1043D, 2019A&A...626A..49P} and quantitative morphologies \citep{2007ApJ...666..212D}. In visual identification, the lack of complete information for objects could lead to projected galaxy pairs and galaxy-star pairs to be incorrectly classified as mergers. Such projections and galaxy-star pairs can also pose difficulties when conducting studies with quantitative morphologies such as asymmetries. The source extraction algorithms used may incorrectly deblend galaxy images containing projections, leading to highly asymmetric non-mergers. Contaminations from projections are likely to be more frequent in higher density environments. 

Moreover, we note that our environment parameters discriminate by mass overdense and underdense regions, with no specification on group or cluster membership. While number overdense environments, such as groups and clusters, do have evidence of merger activity, particularly at \textit{z} > 1 \citep{2012ApJ...745..106L, 2017MNRAS.472.3512T, 2023MNRAS.523.2422L}, our parameters focus on the masses of these environments. In mass overdense environments, as our figures do show, there are still galaxies with high merger probabilities to be found. These galaxies are likely dry merger systems occurring in groups or cluster environments, particularly important for the formation of early-type brightest cluster galaxies (BCGs) in cluster centers \citep{2006ApJ...646..133M,2007ApJ...665L...9R,2008MNRAS.388.1537M,2008ApJ...683L..17T,2009MNRAS.396.2003L, 2010ApJ...718.1158L, 2013MNRAS.434.2856B, 2013MNRAS.433..825L, 2014MNRAS.442..589A}. However, at redshift \textit{z} < 1, many BCGs have finished their growth, and while there are BCGs that continue their growth through mergers at these redshifts, such cases are a minority \citep{2009Natur.458..603C, 2010ApJ...718...23S, 2011MNRAS.414..445S, 2015MNRAS.447.1491L,2018ApJ...853...47R}. Further, in the most mass overdense environments, at \textit{z} < 1, merger activity is suppressed due to increased velocity dispersions \citep{2014ApJ...797..127P}, leading to a lower merger fraction in these environments.
In mass underdense environments, such as in field environments, or the outskirts of clusters and groups, the likelihood of finding a merger increases \citep{1998MNRAS.300..146G,2019MNRAS.488.4169O}. This is also found at higher redshifts, both in the field \citep{2017ApJ...843..126D} and in outskirt \citep{2019A&A...623L..10K} regions. As such, the fraction of mergers in mass underdense regions should be greater than that in mass overdense regions, which is consistent with our results.

 Our environmental trends, particularly the decreased fraction of mergers in higher density environments, show that our morphology-based classification model is likely able to differentiate and give low merger probabilities to images of interlopers, i.e., non-merging pairs in higher density environments, an issue which needs corrections in both spectroscopy-based and image-based classification methods. We note that our environmental trends on observational data may harbor an implicit bias due to the morphology-density relation \citep{1980ApJ...236..351D}. The high velocity encounters in higher density environments can strip gas in galaxies, leading to visual morphologies exhibiting more smoother, spheroidal features. As such, galaxies in higher density environments are less likely to exhibit the tidal features that are present in mergers, resulting in such galaxies to be assigned low merger probabilities by our morphology-based merger identification model. However, we suggest that such a bias can be mitigated. First, the findings in \citet{2010MNRAS.401.1552D}, based on visual classification, suggest that mergers are more likely found in higher density environments, which contradicts what is found in comsmological n-body and hydrodynamical simulations. As the implicit biases due to the morphology density relation can be overlooked by visual classification studies, we suggest that a similar overlook is possible in our work. Second, the ground truth merger selection in our fine-tuning process is based on the time to the closest merger event, and is not be biased to any environments and morphlogies.

Investigations on the physical properties of the merger galaxies, such as projected separation and relative velocity differences with their neighbors, as well as colors, star formation rates and morphologies, are planned in future works, to determine if there are any environmental dependences on the properties of the mergers themselves in addition to merger incidence.

\section{Conclusion}
\label{section:conclusion}

In this work, we take a deep learning based approach for merger classification in Subaru HSC-SSP.  We fine-tune the pre-trained model Zoobot using synthetic HSC images of galaxies in the Illustris TNG simulations, then make predictions using the fine-tuned model on a sample of galaxies in HSC-SSP S21A wide cross-matched with SDSS and GAMA.
We find that the fine-tuning approach can achieve accuracies comparable to previous merger classification studies using simulational data, at a 76\% accuracy, as well as 80\% completion and precision. We achieved these results requiring a far smaller sample size than previous studies, sufficing with $\sim10^3$ training samples of each class. We also find that our morphology-based model is able to correctly predict both mergers and non-mergers of diverse appearances, mergers of differing mass ratios and stages, and can also distinguish to a degree between projections and true merging pairs. The merger rate found by our model is consistent with those of previous works if we adopt a `confident' threshold.

We will make the merger catalogue we produced in this work publicly available. We plan to classify all of HSC-SSP in the future and make the results publicly available as well.

Further, we studied the relationship between merger activity and multi-scale galaxy environments in both the simulation data of TNG and the observational data of HSC-SSP, and compared our results. We find two trends in the simulation data, and one trend in the observational data.
Both sets of data are in agreement that at scales of 0.5 $h^{-1}$Mpc and larger, merger galaxies favor lower density environments, and non-mergers favor higher density environments. 
However, below these scales, the simulation data finds that non-merger galaxies are most prevalent in lower density environments, and higher density environments favor mergers. 
In future works, we plan to investigate in both observations and simulations where the reversal in trend occurs, as well as the galaxy properties of the mergers themselves.

\begin{acknowledgements}
We would first and foremost like to thank the native Hawaiians for sharing the Maunakea mountain, a place of cultural, historical, and natural significance, allowing us access to a beautiful view of the Universe. Next, we would like to thank the anonymous referee for their comments and feedback in greatly improving the quality of this work.
The Hyper Suprime-Cam (HSC) collaboration includes the astronomical communities of Japan and Taiwan, and Princeton University.  The HSC instrumentation and software were developed by the National Astronomical Observatory of Japan (NAOJ), the Kavli Institute for the Physics and Mathematics of the Universe (Kavli IPMU), the University of Tokyo, the High Energy Accelerator Research Organization (KEK), the Academia Sinica Institute for Astronomy and Astrophysics in Taiwan (ASIAA), and Princeton University.  Funding was contributed by the FIRST program from the Japanese Cabinet Office, the Ministry of Education, Culture, Sports, Science and Technology (MEXT), the Japan Society for the Promotion of Science (JSPS), Japan Science and Technology Agency  (JST), the Toray Science  Foundation, NAOJ, Kavli IPMU, KEK, ASIAA, and Princeton University.

This paper is based [in part] on data collected at the Subaru Telescope and retrieved from the HSC data archive system, which is operated by Subaru Telescope and Astronomy Data Center (ADC) at NAOJ. Data analysis was in part carried out with the cooperation of Center for Computational Astrophysics (CfCA) at NAOJ.  We are honored and grateful for the opportunity of observing the Universe from Maunakea, which has the cultural, historical and natural significance in Hawaii.

This paper makes use of software developed for Vera C. Rubin Observatory. We thank the Rubin Observatory for making their code available as free software at http://pipelines.lsst.io/.

The Pan-STARRS1 Surveys (PS1) and the PS1 public science archive have been made possible through contributions by the Institute for Astronomy, the University of Hawaii, the Pan-STARRS Project Office, the Max Planck Society and its participating institutes, the Max Planck Institute for Astronomy, Heidelberg, and the Max Planck Institute for Extraterrestrial Physics, Garching, The Johns Hopkins University, Durham University, the University of Edinburgh, the Queen’s University Belfast, the Harvard-Smithsonian Center for Astrophysics, the Las Cumbres Observatory Global Telescope Network Incorporated, the National Central University of Taiwan, the Space Telescope Science Institute, the National Aeronautics and Space Administration under grant No. NNX08AR22G issued through the Planetary Science Division of the NASA Science Mission Directorate, the National Science Foundation grant No. AST-1238877, the University of Maryland, Eotvos Lorand University (ELTE), the Los Alamos National Laboratory, and the Gordon and Betty Moore Foundation.
            This work has been supported by the Japan Society for the Promotion of Science (JSPS) Grants-in-Aid for Scientific Research (21H01128). 
      This work has also been supported in part by the Collaboration Funding of the Institute of Statistical Mathematics ``New Perspective of the Cosmology Pioneered by the Fusion of Data Science and Physics''. 
      K.O. is supported by JSPS KAKENHI Grant Number JP23KJ1089.
      H.Y. was supported by JSPS KAKENHI Grant Number JP22K14072 and the Research Fund for International Young Scientists of NSFC (11950410492).
\end{acknowledgements}

% WARNING
%-------------------------------------------------------------------
% Please note that we have included the references to the file aa.dem in
% order to compile it, but we ask you to:
%
% - use BibTeX with the regular commands:
%   \bibliographystyle{aa} % style aa.bst
%   \bibliography{Yourfile} % your references Yourfile.bib
%
% - join the .bib files when you upload your source files
%-------------------------------------------------------------------
\bibliography{References}
\bibliographystyle{aa}
\appendix
\onecolumn
\section{Confusion matrices}
\label{appendix: A}
    \begin{figure*}[h]
    \centering
    \includegraphics[width=0.5\textwidth]{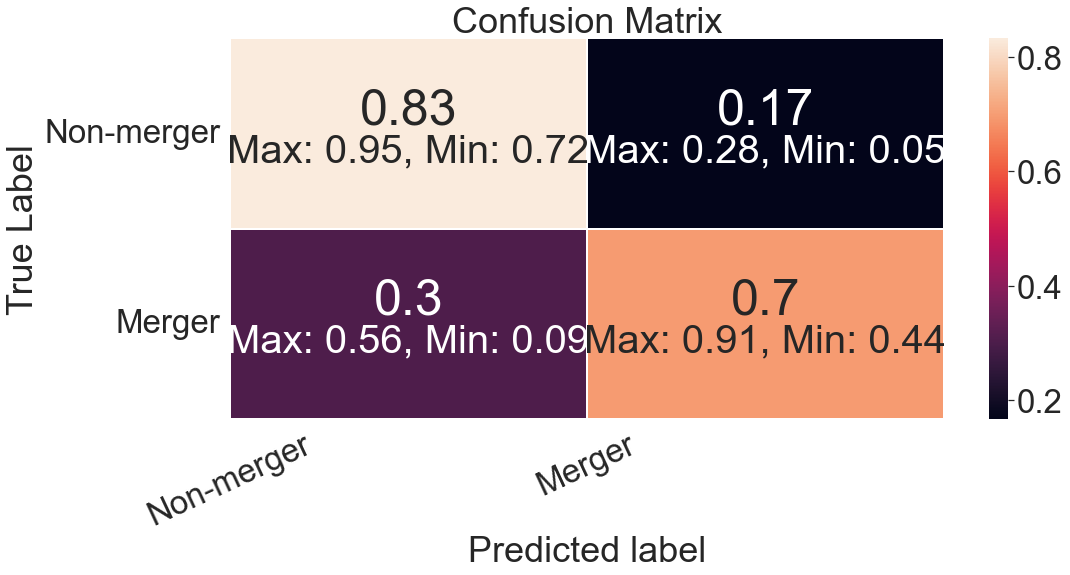}
    \caption{Total combined confusion matrix for the 10 runs of our Zoobot fine-tuning process. Each run fine-tuned a new model on a different set of training/validation data, and the confusion matrices are on 10 different sets of testing data.he stellar mass distributions for merger and non-merger galaxies used for fine-tuning Zoobot. The maximum and minimum values for each cell are also indicated.}
    \label{fig:Runs}
\end{figure*}
\newpage
\section{Predictions on simulation images}
\label{appendix: B}
\begin{figure*}[ht]
    \centering
    \hspace*{-3.5cm}\includegraphics[trim=0cm 3cm 0cm 0cm,scale=0.65]{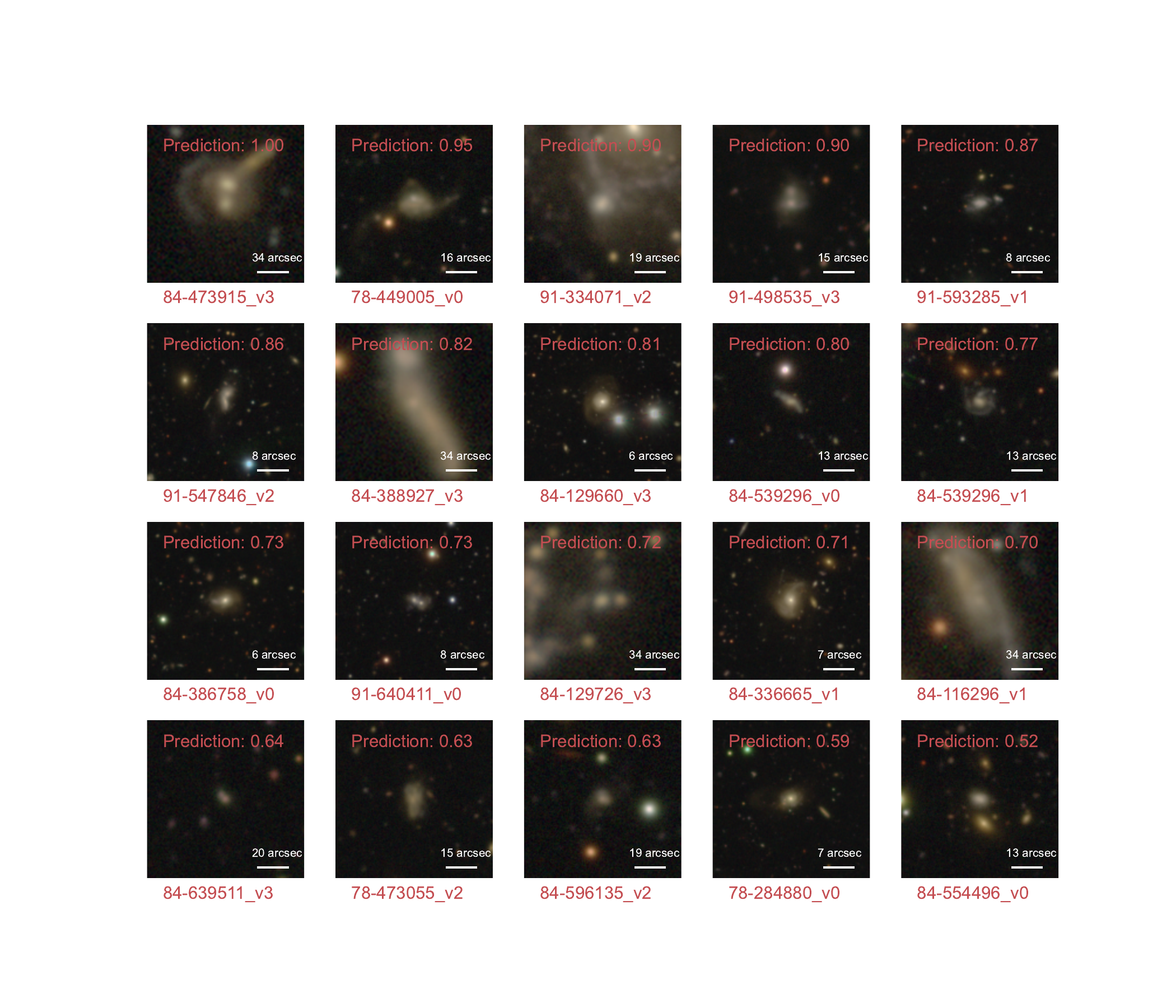}
    \caption{20 randomly drawn examples of true positive classifications (merger probability $>0.5$) from the 10 test sets, in descending order, with the probabilities indicated in the image, and the identification of the galaxy indicated below the image in the format [snapshot]-[ID]\textunderscore[viewing angle]. The model seems to be able to identify various merger features, including close companions and merger remnants such as rings.}
    \label{fig:TruePos}
\end{figure*}
\newpage
\begin{figure*}[ht]
    \centering
    \hspace*{-3.5cm}\includegraphics[trim=0cm 3cm 0cm 0cm,scale=0.65]{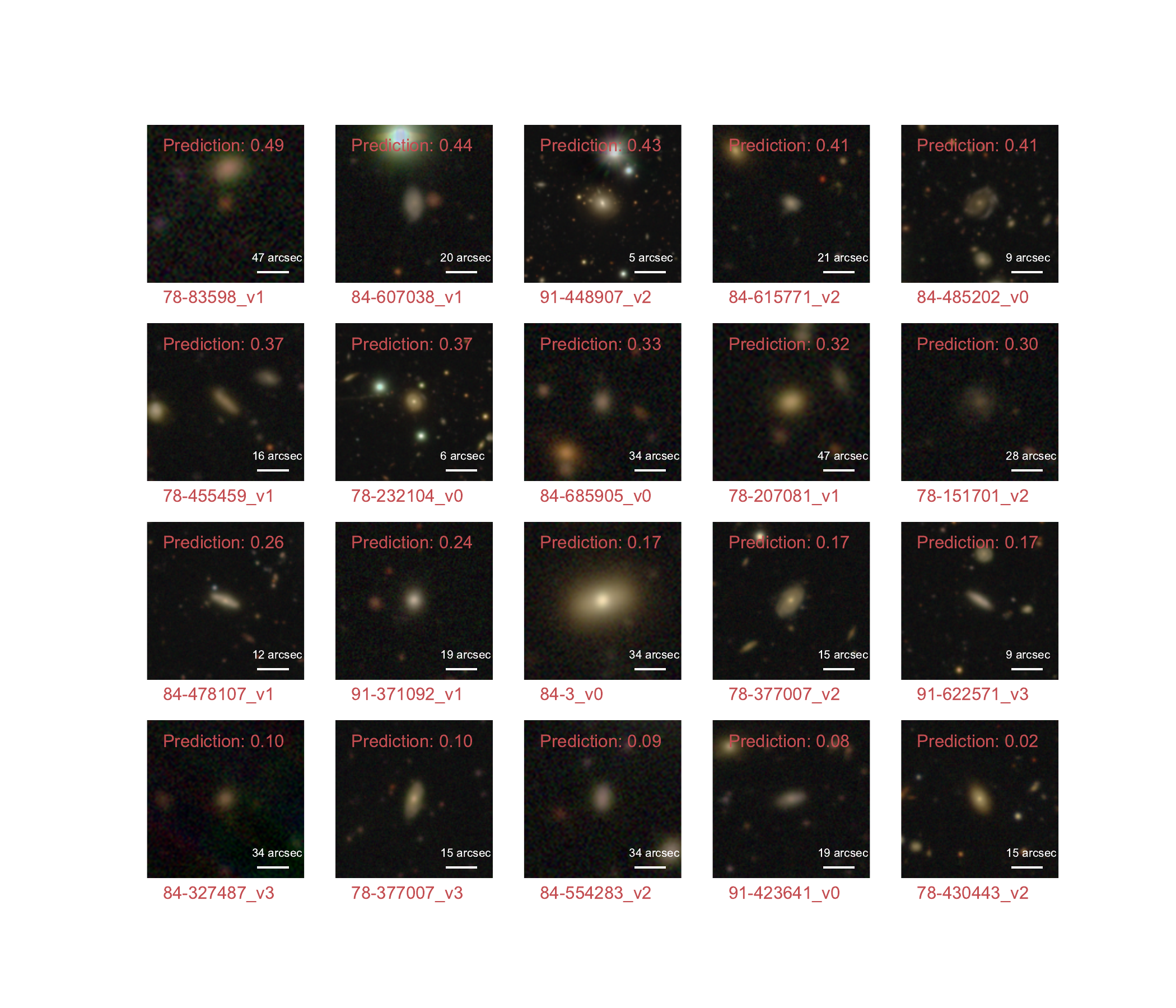}
    \caption{The same as Fig. \ref{fig:TruePos} but for true negatives. The model seems to be able to correctly identify some projections/overlaps as non-interacting. More importantly, a diverse appearance of non-mergers are correctly identified.}
    \label{fig:TrueNeg}
\end{figure*}
\newpage
\begin{figure*}[ht]
    \centering
    \hspace*{-3.5cm}\includegraphics[trim=0cm 3cm 0cm 0cm,scale=0.65]{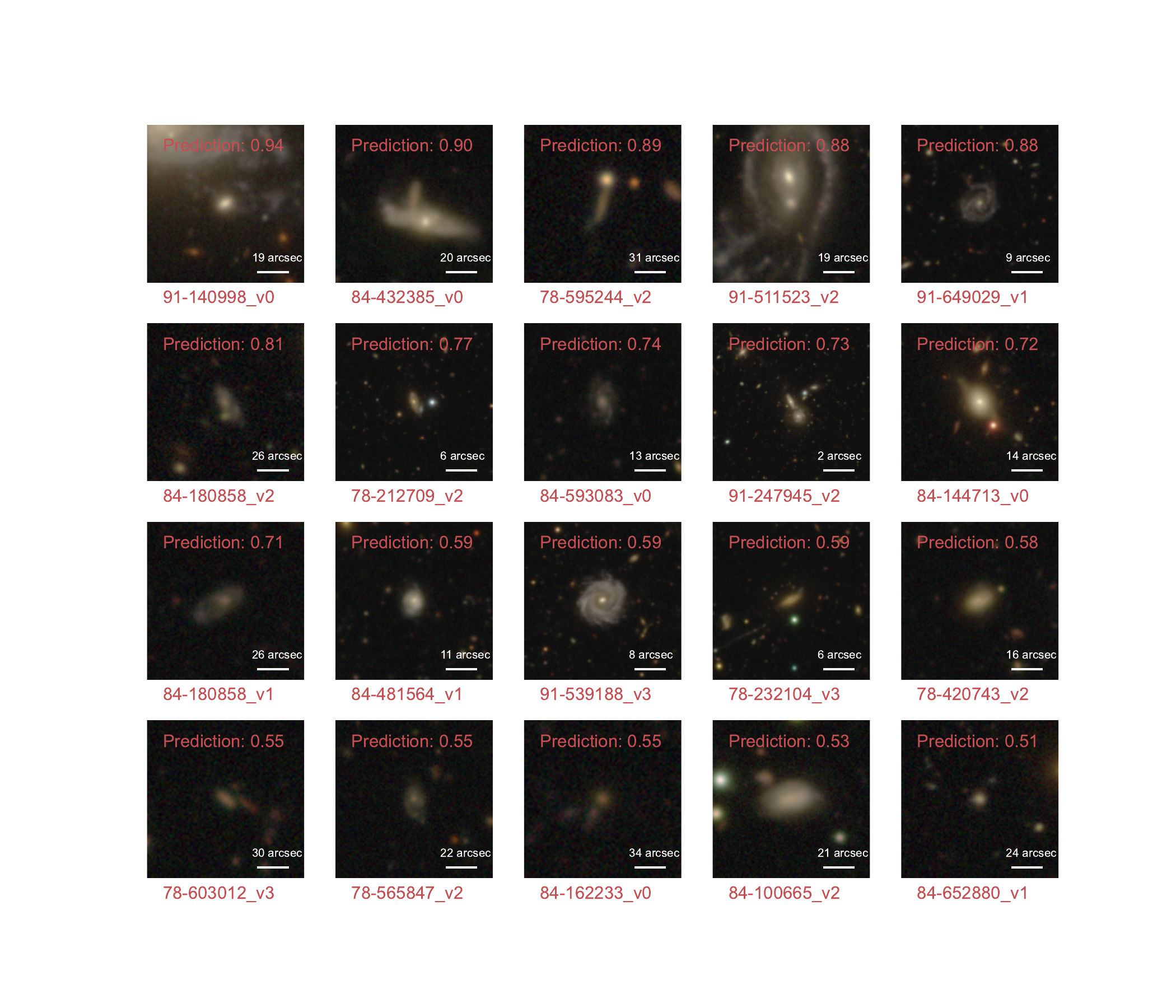}
    \caption{The same as Fig. \ref{fig:TruePos} but for false positives. Some close overlaps are incorrectly classified as a merging. Such galaxies, such as 84-432385\textunderscore v2 with a merger probability of 0.90, would also likely be classified as mergers by human visual identification.}
    \label{fig:FalsePos}
\end{figure*}
\newpage
\begin{figure*}[ht]
    \centering
    \hspace*{-3.5cm}\includegraphics[trim=0cm 3cm 0cm 0cm,scale=0.65]{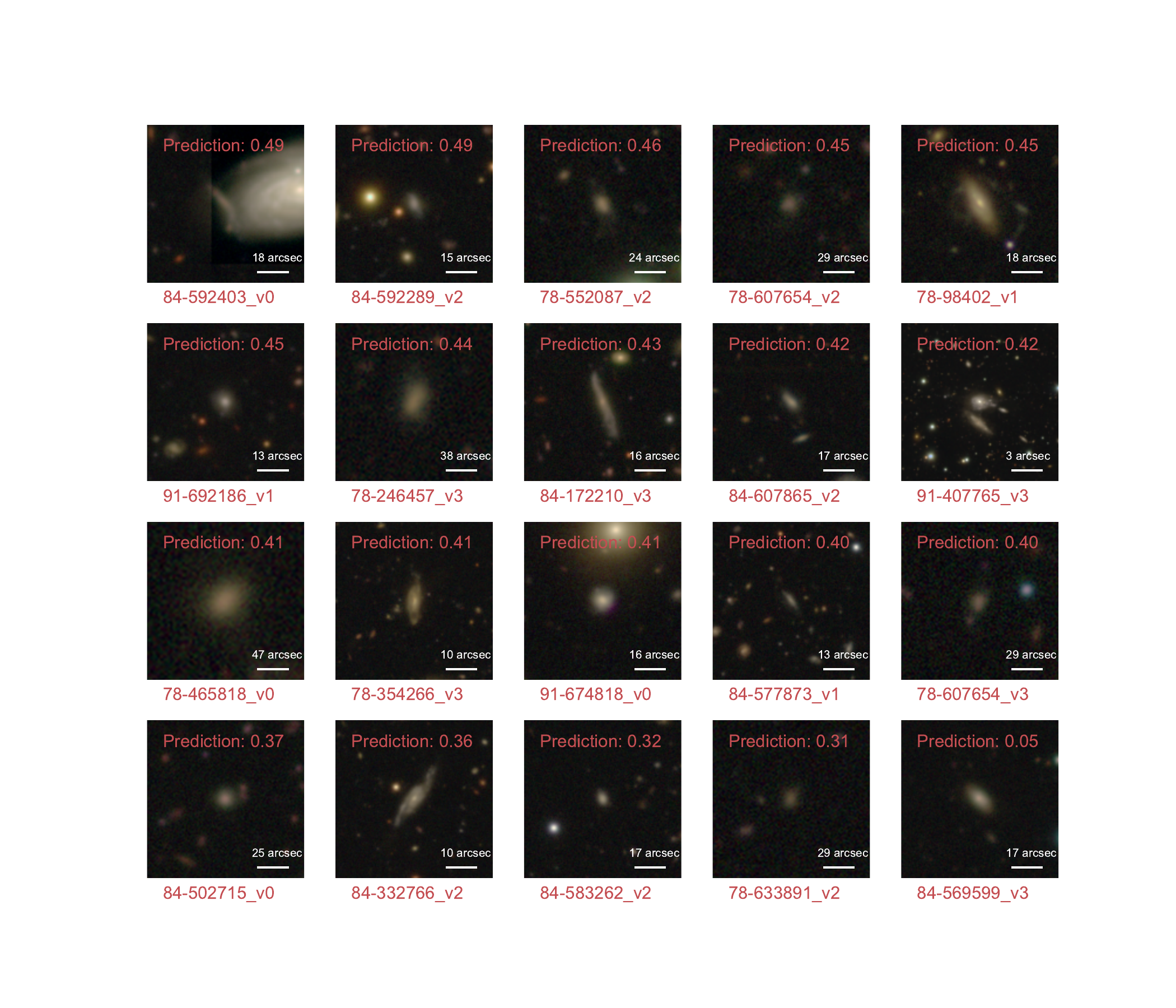}
    \caption{The same as Fig. \ref{fig:TruePos} but for false negatives. Many galaxies with lower probabilities, especially those in the bottom 2 rows are minor or mini mergers with the most recent/next merger close to the 0.5 Gyr threshold for merger selection.}
    \label{fig:FalseNeg}
\end{figure*}
\newpage

\section{Predictions on observation images}
\label{appendix: C}
\begin{figure*}[ht]
    \centering
    \hspace*{-3.5cm}\includegraphics[trim=0cm 3cm 0cm 0cm,scale=0.65]{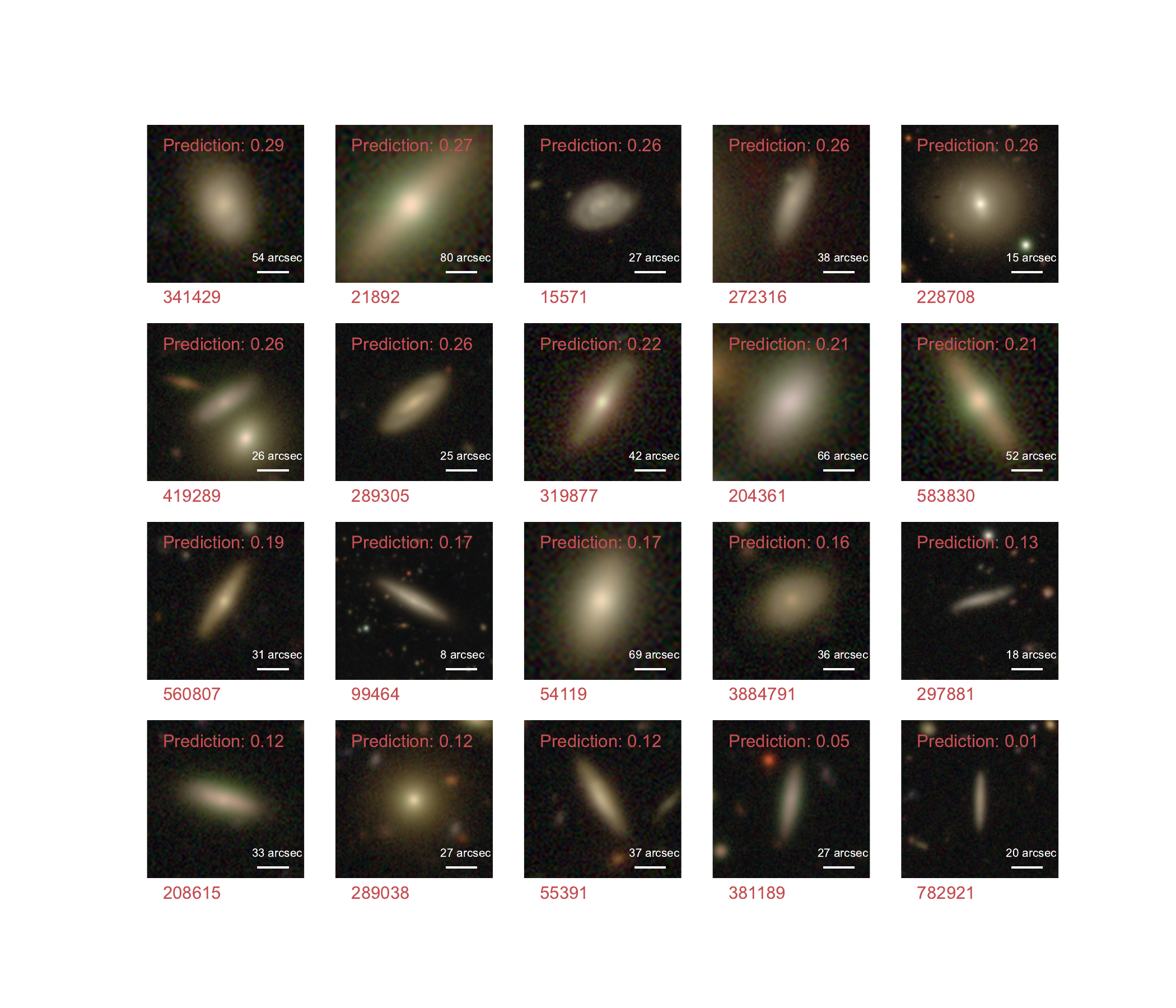}
    \caption{20 randomly drawn examples of GAMA galaxies with a merger probability $<0.3$, with merger probabilities in descending order. The merger probabilities are indicated in the image and the GAMA ID below the image.}
    \label{fig:GAMA03}
\end{figure*}

\begin{figure*}[ht]
    \centering
    \hspace*{-3.5cm}\includegraphics[trim=0cm 3cm 0cm 0cm,scale=0.65]{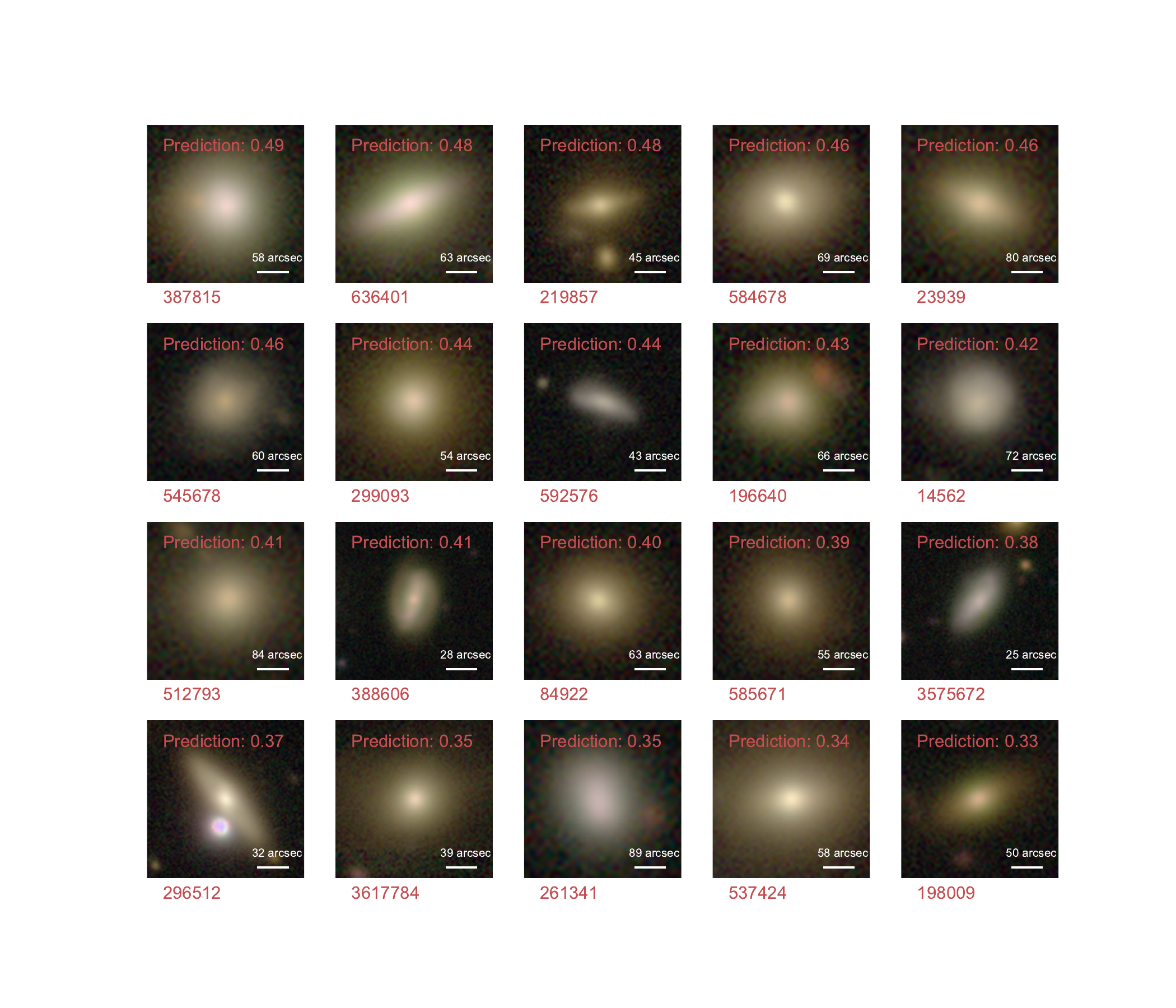}
    \caption{The same as Fig. \ref{fig:GAMA03} but for merger probability $>0.3$ and $<0.5$.}
    \label{fig:GAMA0305}
\end{figure*}

\begin{figure*}[ht]
    \centering
    \hspace*{-3.5cm}\includegraphics[trim=0cm 3cm 0cm 0cm,scale=0.65]{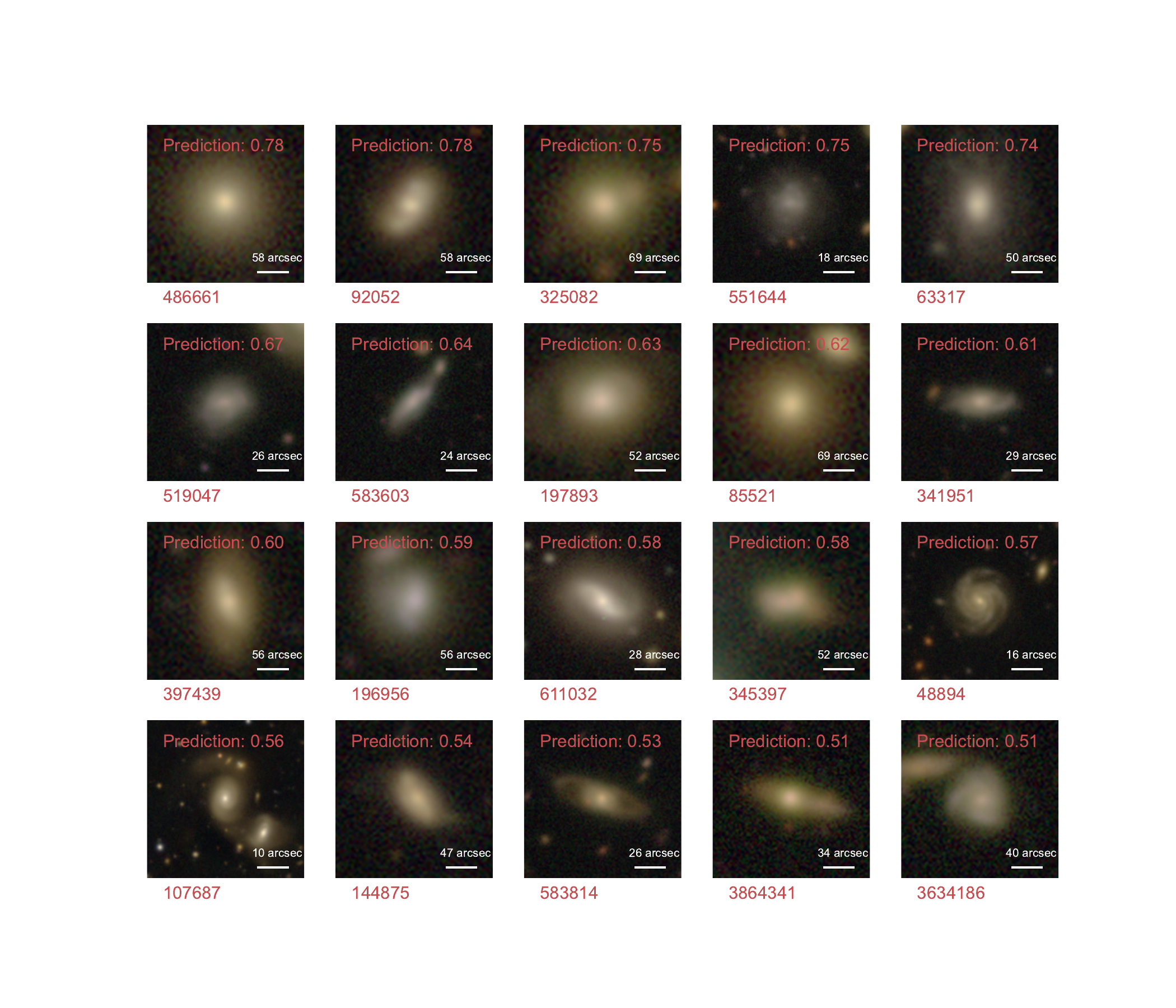}
    \caption{The same as Fig. \ref{fig:GAMA03} but for merger probability $>0.5$ and $<0.8$.}
    \label{fig:GAMA0508}
\end{figure*}

\begin{figure*}[ht]
    \centering
    \hspace*{-3.5cm}\includegraphics[trim=0cm 3cm 0cm 0cm,scale=0.65]{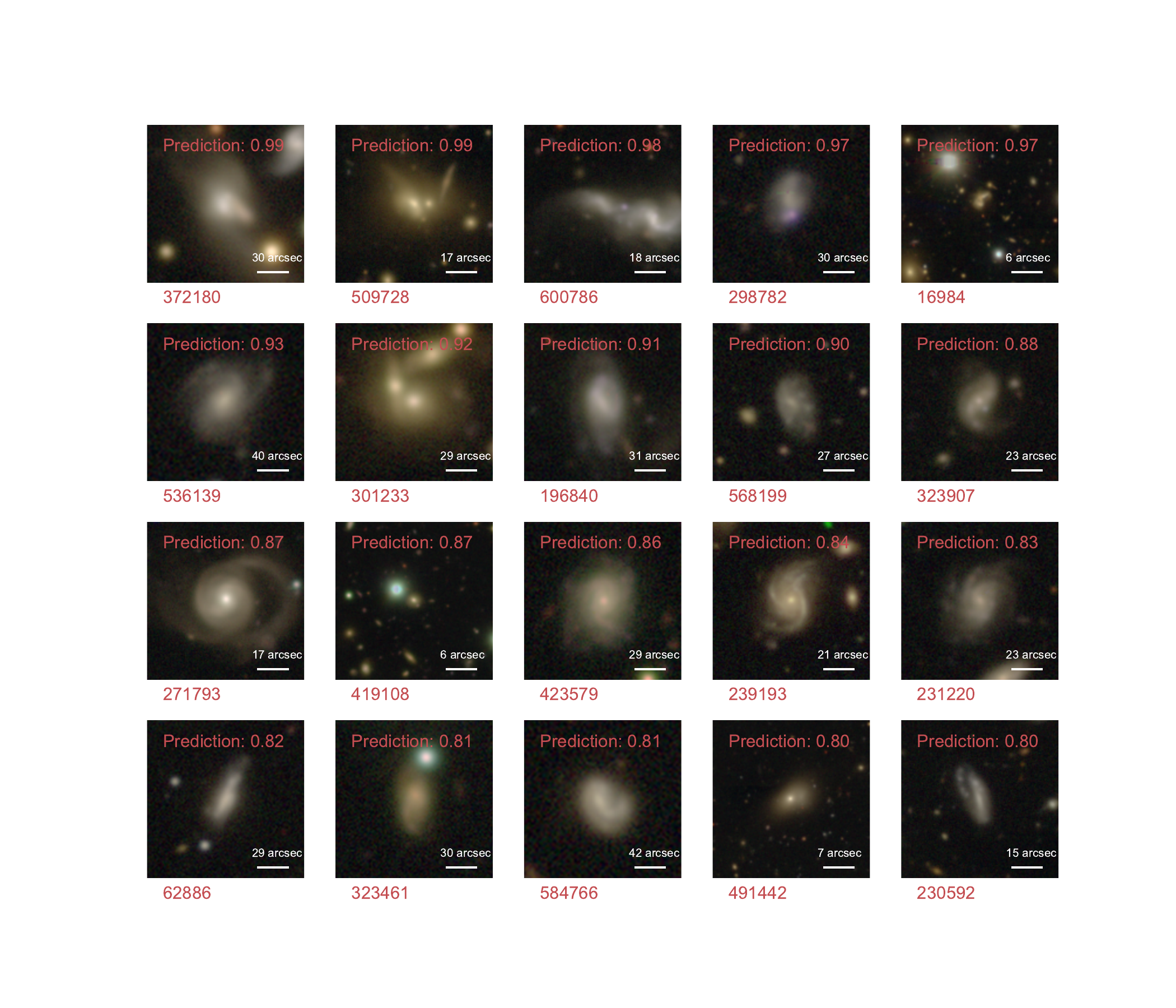}
    \caption{The same as Fig. \ref{fig:GAMA03} but for merger probability $>0.8$.}
    \label{fig:GAMA08}
\end{figure*}

\section{Mass-fixed environment trends}
\label{appendix: D}
\begin{figure*}
        \centering
        \begin{subfigure}[b]{0.475\textwidth}
            \centering
            \includegraphics[width=\textwidth]{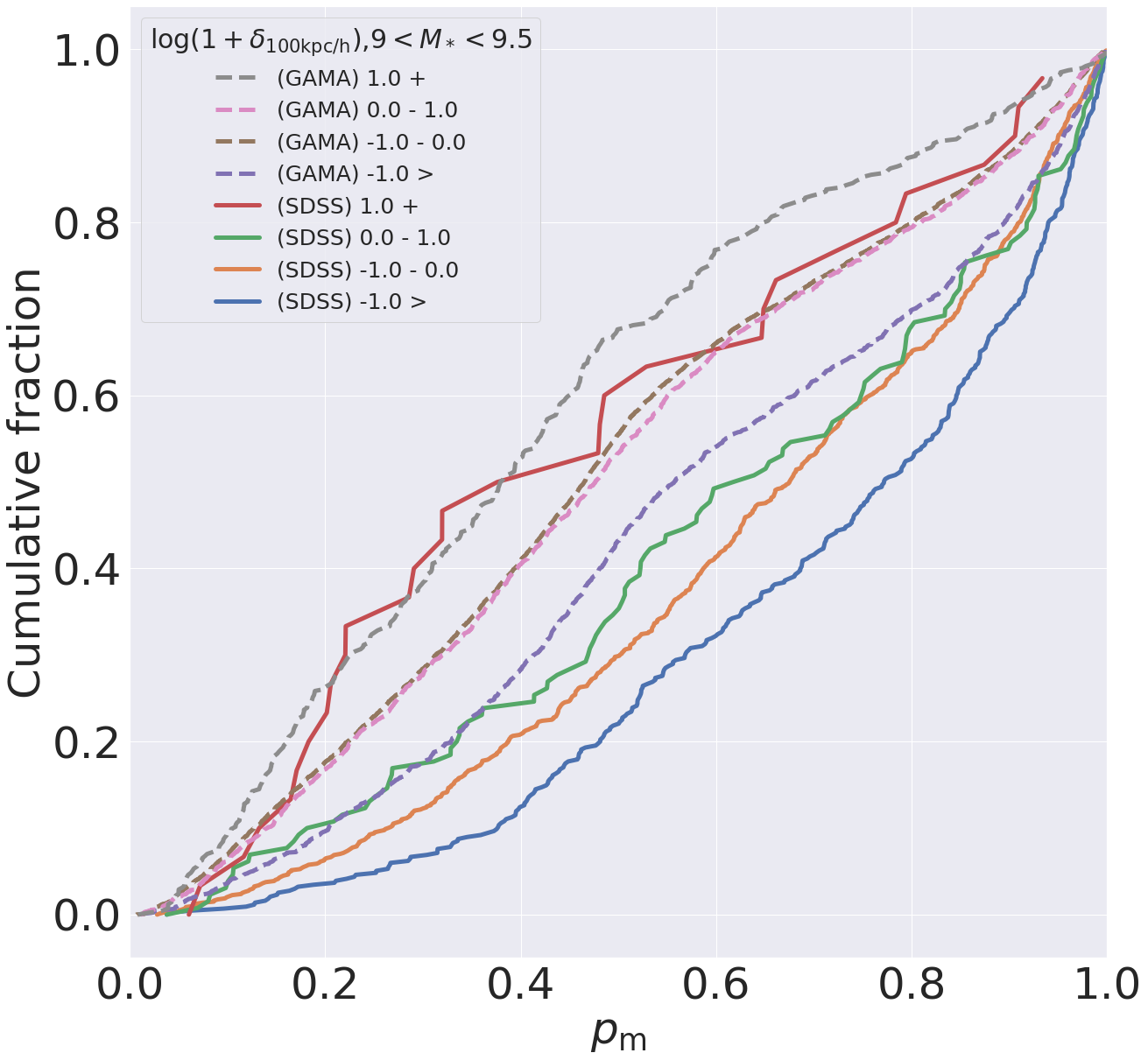}
            \caption[Network2]%
            {{\small $9 < \log(M_{*}) < 9.5$\\ KS test (SDSS) statistic: 0.432% p-value: $2.920\times10^{-5}$
            \\ KS test (GAMA) statistic: 0.263}}% p-value: $2.289\times10^{-20}$}}}    
            \label{fig:100kpc995}
        \end{subfigure}
        \hfill
        \centering
        \begin{subfigure}[b]{0.475\textwidth}
            \centering
            \includegraphics[width=\textwidth]{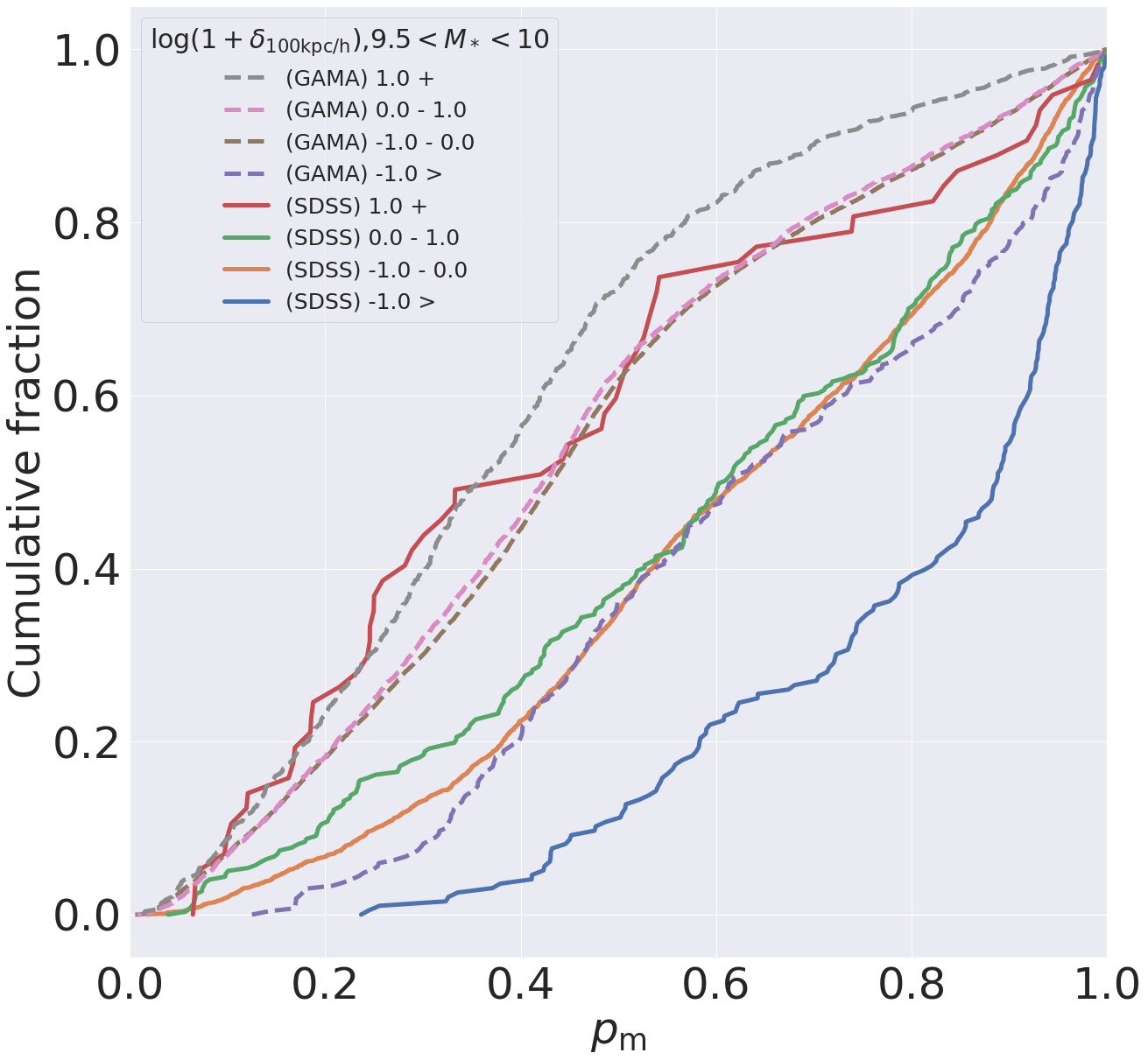}
            \caption[Network2]%
            {{\small $9.5 < \log(M_{*}) < 10$\\ KS test (SDSS) statistic: 0.601% p-value: $1.558\times10^{-15}$ 
            \\ KS test (GAMA) statistic: 0.378}}% p-value: $2.049\times10^{-26}$}}}    
            \label{fig:100kpc9510}
        \end{subfigure}
        \vskip\baselineskip
                \centering
        \begin{subfigure}[b]{0.475\textwidth}
            \centering
            \includegraphics[width=\textwidth]{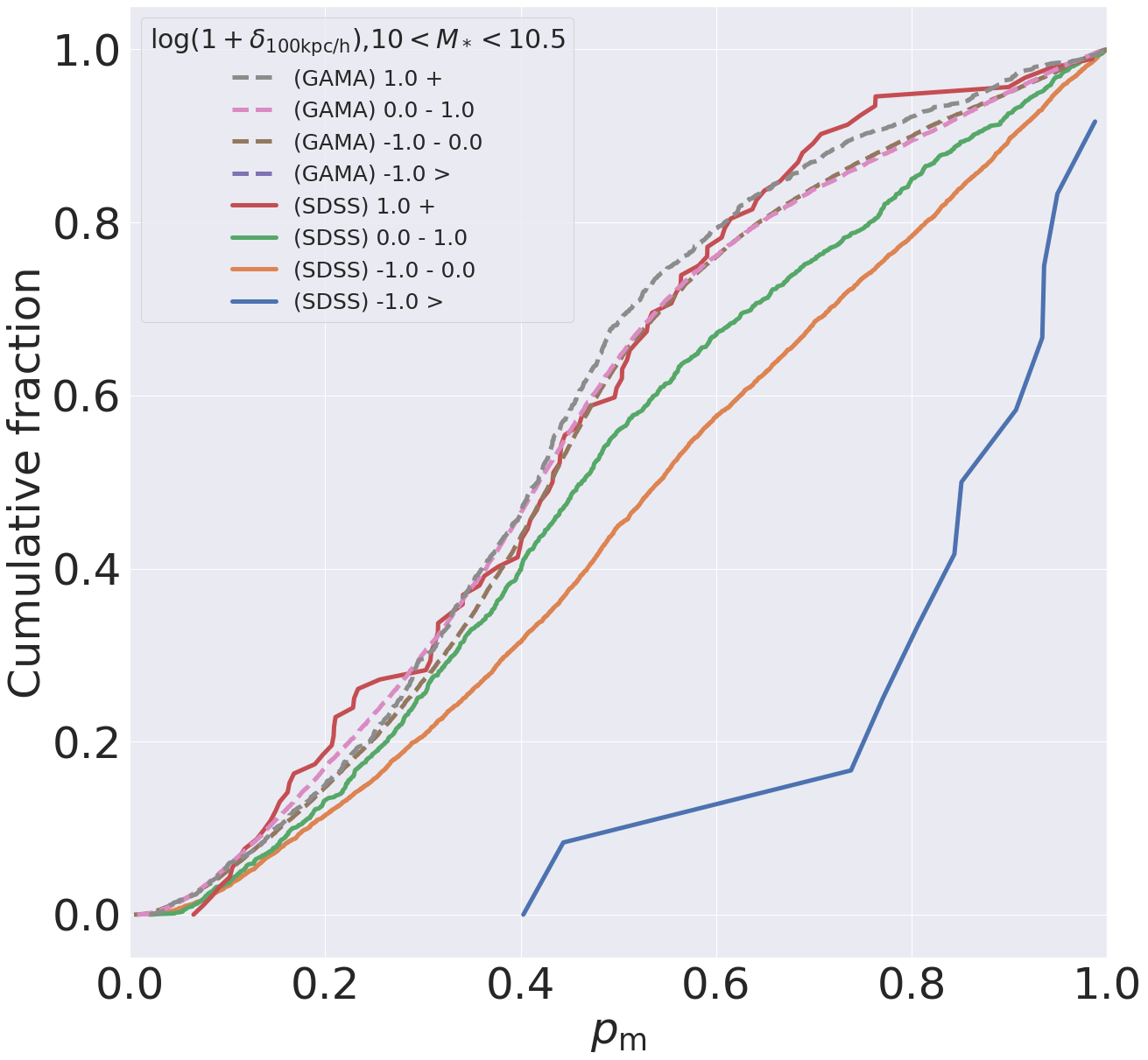}
            \caption[Network2]%
            {{\small $10 < \log(M_{*}) < 10.5$\\ KS test (SDSS) statistic: 0.757% p-value: $1.064\times10^{-6}$
            \\ KS test (GAMA) statistic: 0.053 }}%p-value: 0.001}}}    
            \label{fig:100kpc10105}
        \end{subfigure}
        \hfill
        \centering
        \begin{subfigure}[b]{0.475\textwidth}
            \centering
            \includegraphics[width=\textwidth]{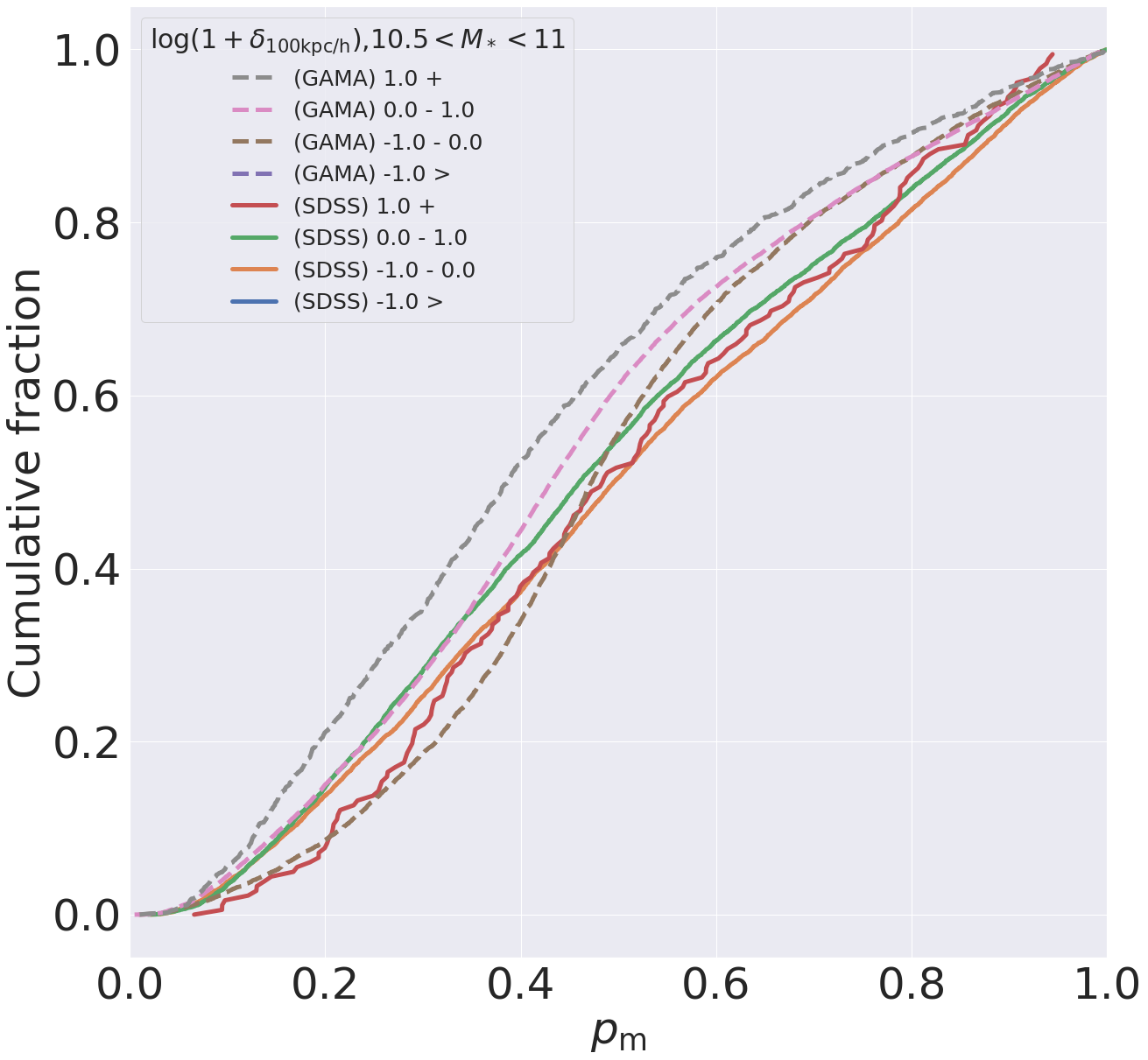}
            \caption[Network2]%
            {{\small $10.5 < \log(M_{*}) < 11$\\ KS test (SDSS) statistic: 0.064 p-value: 0.430\\ KS test (GAMA) statistic: 0.192}}% p-value: $2.372\times10^{-41}$}}}    
            \label{fig:100kpc10511}
        \end{subfigure}
        % \begin{subfigure}[b]{0.350\textwidth}  
        %     \centering 
        %     \includegraphics[width=\textwidth]{5thnearestmini.png}
        %     \caption[]%
        %     {{\small mini Neighbors}}    
        %     \label{fig:5mini}
        % \end{subfigure}
        \caption[ The average and standard deviation of critical parameters ]
        {\small Cumulative distribution curves of merger probabilities of HSC galaxies cross-matched with GAMA (dotted lines) and SDSS (solid lines) predicted by our fine-tuned model, similar to Fig. \ref{fig:envcumul}, but separated into different mass bins, as indicated in each figure. The KS test statistic between the most underdense and most overdense environments are also indicated for each subplot, with p-values available if > 0.05. Calculated for a 100 kpc radius aperture.}
        \label{fig:100kpcmassfixed}
\end{figure*}
\begin{figure*}
        \centering
        \begin{subfigure}[b]{0.475\textwidth}
            \centering
            \includegraphics[width=\textwidth]{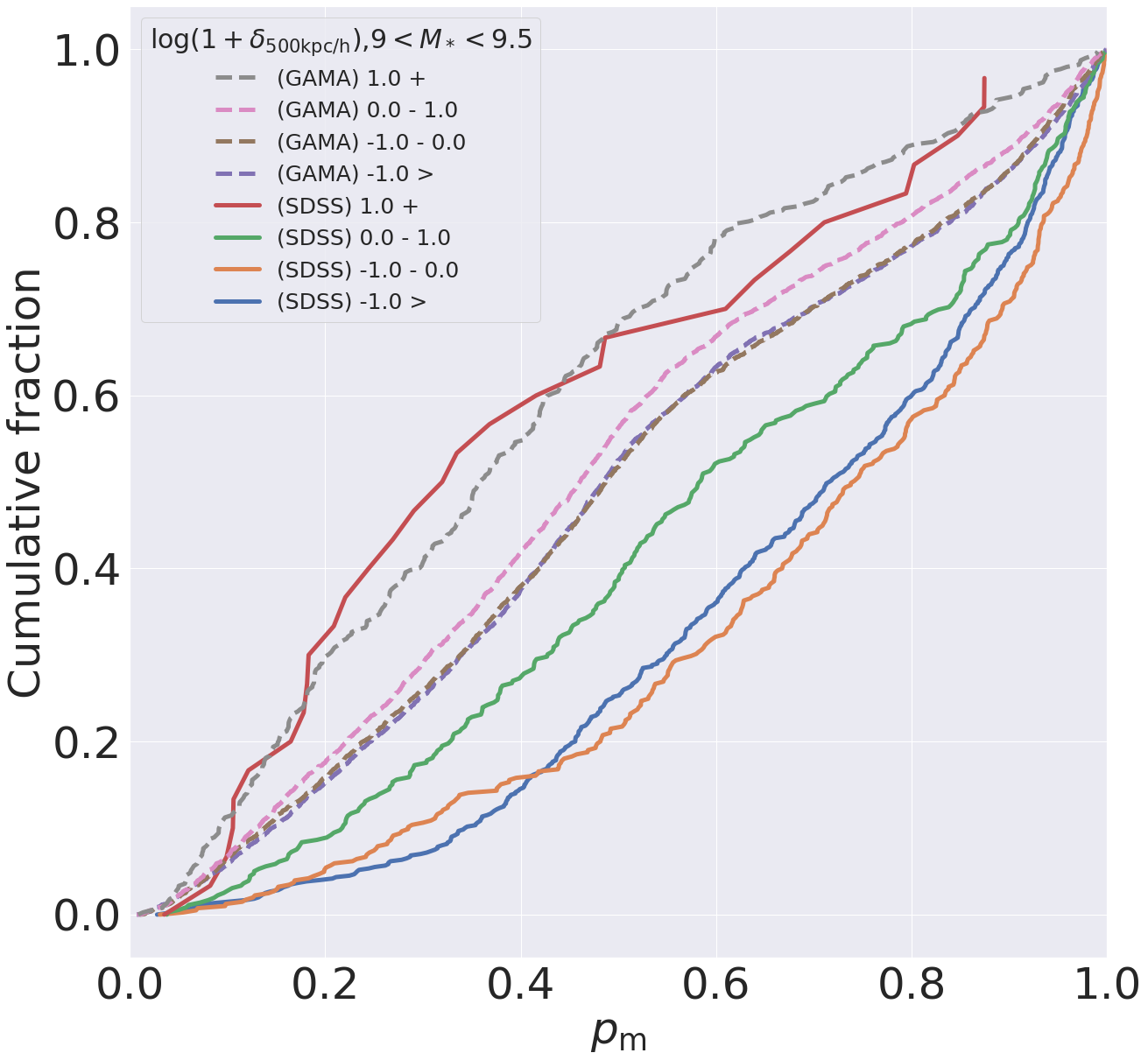}
            \caption[Network2]%
            {{\small $9 < \log(M_{*}) < 9.5$\\ KS test (SDSS) statistic: 0.483 %p-value: $1.241\times10^{-6}$
            \\ KS test (GAMA) statistic: 0.190}}% p-value: $1.095\times10^{-11}$}}}    
            \label{fig:500kpc995}
        \end{subfigure}
        \hfill
        \centering
        \begin{subfigure}[b]{0.475\textwidth}
            \centering
            \includegraphics[width=\textwidth]{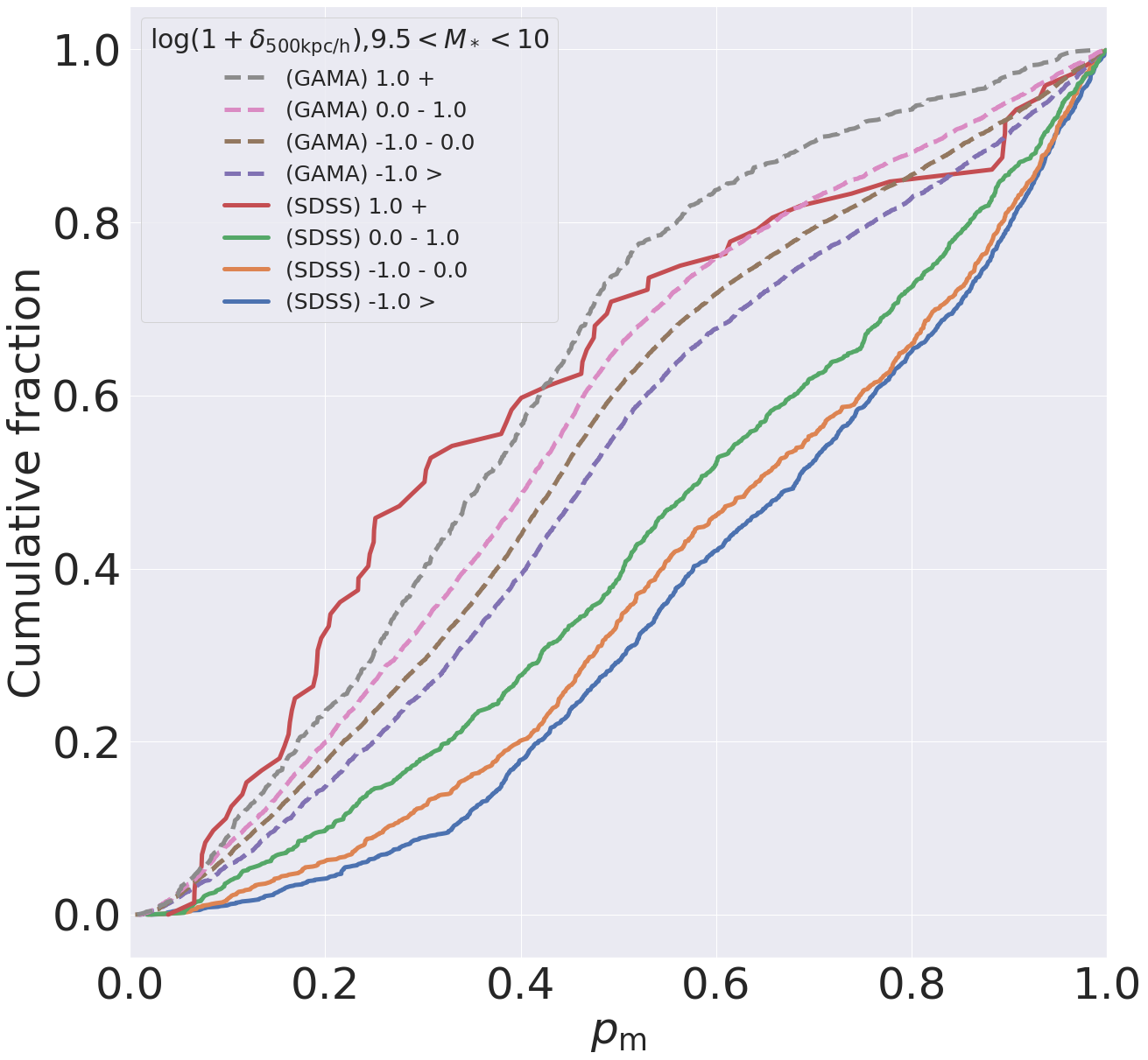}
            \caption[Network2]%
            {{\small $9.5 < \log(M_{*}) < 10$\\ KS test (SDSS) statistic: 0.455% p-value: $2.038\times10^{-13}$
            \\ KS test (GAMA) statistic: 0.193}}% p-value: $1.117\times10^{-24}$}}}    
            \label{fig:500kpc9510}
        \end{subfigure}
        \vskip\baselineskip
                \centering
        \begin{subfigure}[b]{0.475\textwidth}
            \centering
            \includegraphics[width=\textwidth]{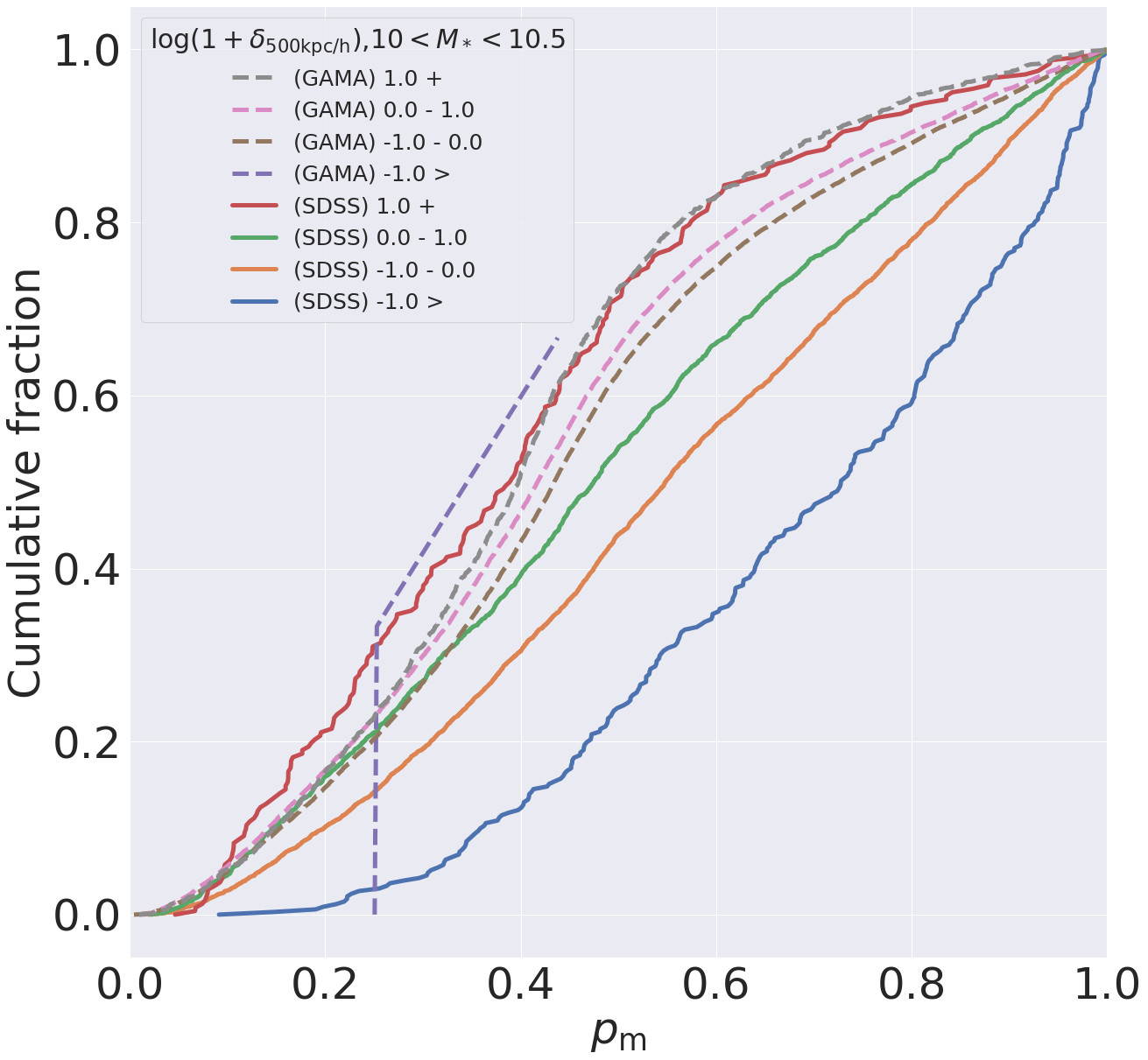}
            \caption[Network2]%
            {{\small $10 < \log(M_{*}) < 10.5$\\ KS test (SDSS) statistic: 0.492% p-value: $2.860\times10^{-31}$ 
            \\ KS test (GAMA) statistic: 0.434 p-value: 0.503}}
            \label{fig:500kpc10105}
        \end{subfigure}
        \hfill
        \centering
        \begin{subfigure}[b]{0.475\textwidth}
            \centering
            \includegraphics[width=\textwidth]{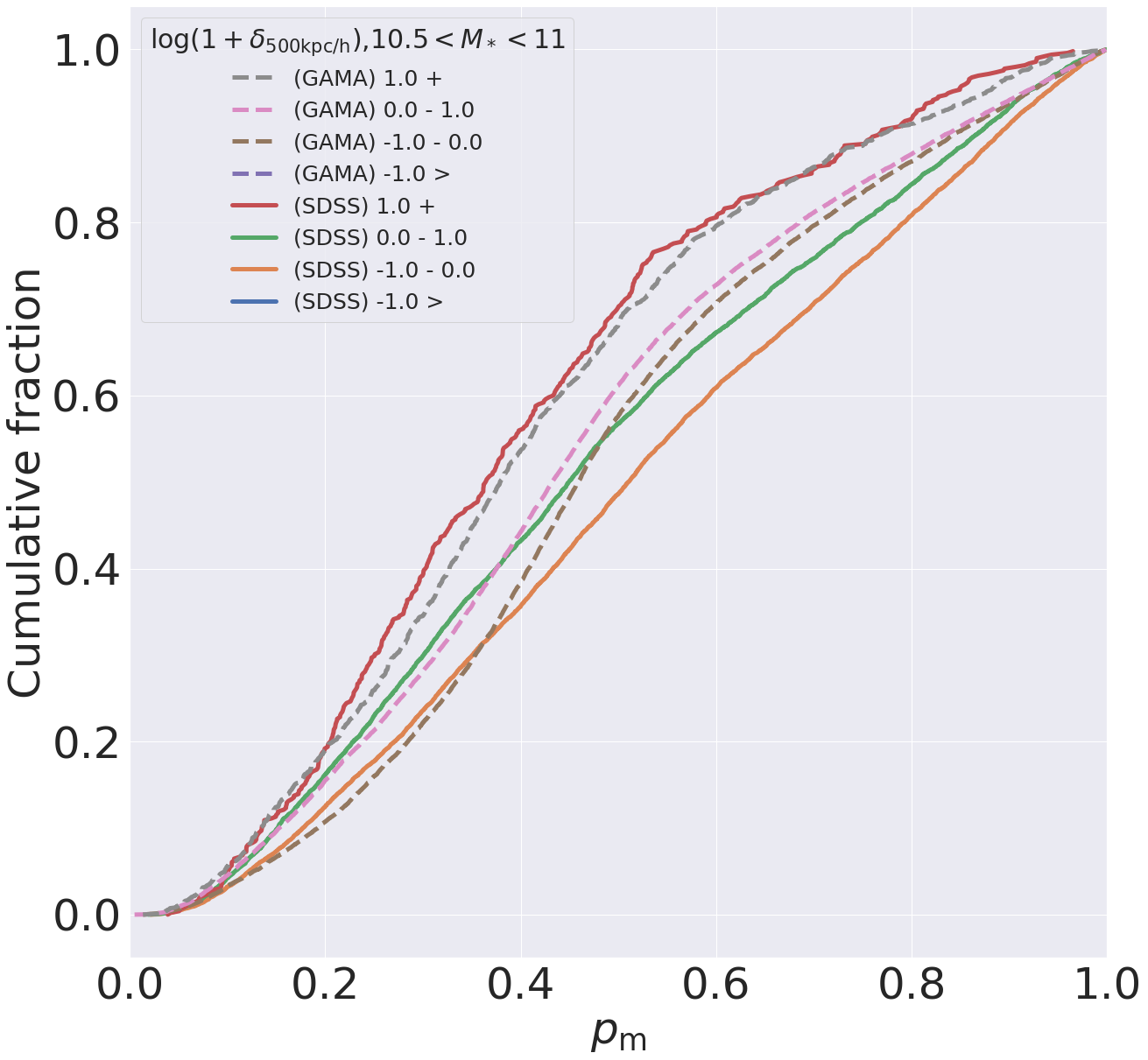}
            \caption[Network2]%
            {{\small $10.5 < \log(M_{*}) < 11$\\ KS test (SDSS) statistic: 0.233 %p-value: $4.058\times10^{-23}$ 
            \\ KS test (GAMA) statistic: 0.159}}% p-value: $5.867\times10^{-33}$}}}    
            \label{fig:500kpc10511}
        \end{subfigure}
        % \begin{subfigure}[b]{0.350\textwidth}  
        %     \centering 
        %     \includegraphics[width=\textwidth]{5thnearestmini.png}
        %     \caption[]%
        %     {{\small mini Neighbors}}    
        %     \label{fig:5mini}
        % \end{subfigure}
        \caption[ The average and standard deviation of critical parameters ]
        {\small The same as Fig. \ref{fig:100kpcmassfixed} but for a 500 kpc radius aperture.} 
        \label{fig:500kpcmassfixed}
\end{figure*}
\begin{figure*}
        \centering
        \begin{subfigure}[b]{0.475\textwidth}
            \centering
            \includegraphics[width=\textwidth]{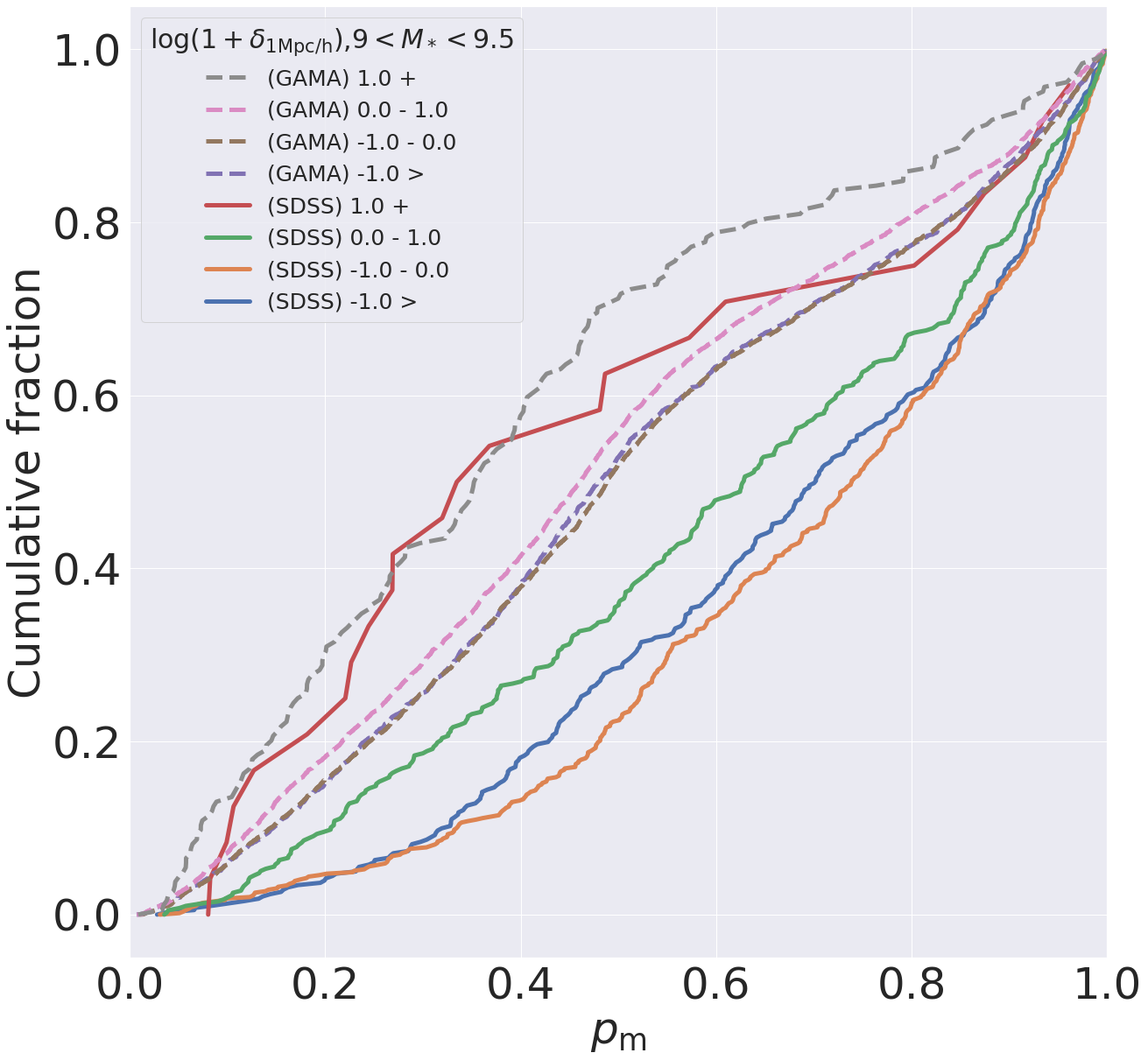}
            \caption[Network2]%
            {{\small $9 < \log(M_{*}) < 9.5$\\ KS test (SDSS) statistic: 0.436% p-value: $2.023\times10^{-4}$ 
            \\ KS test (GAMA) statistic: 0.211}}% p-value: $3.688\times10^{-7}$}}}    
            \label{fig:1Mpc995}
        \end{subfigure}
        \hfill
        \centering
        \begin{subfigure}[b]{0.475\textwidth}
            \centering
            \includegraphics[width=\textwidth]{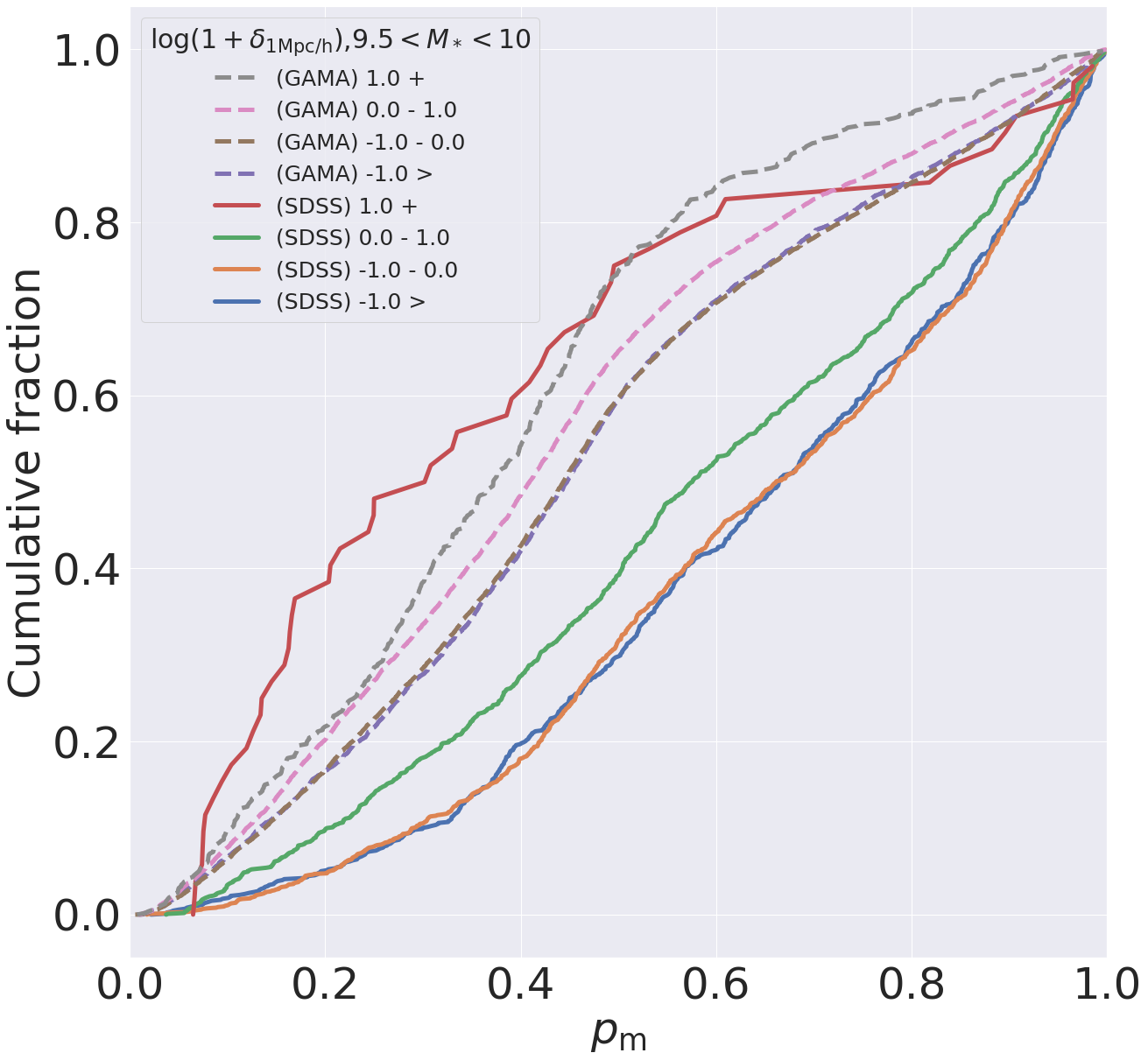}
            \caption[Network2]%
            {{\small $9.5 < \log(M_{*}) < 10$\\ KS test (SDSS) statistic: 0.473% p-value: $1.397\times10^{-10}$ 
            \\ KS test (GAMA) statistic: 0.156}}% p-value: $7.097\times10^{-11}$}}}    
            \label{fig:1Mpc9510}
        \end{subfigure}
        \vskip\baselineskip
                \centering
        \begin{subfigure}[b]{0.475\textwidth}
            \centering
            \includegraphics[width=\textwidth]{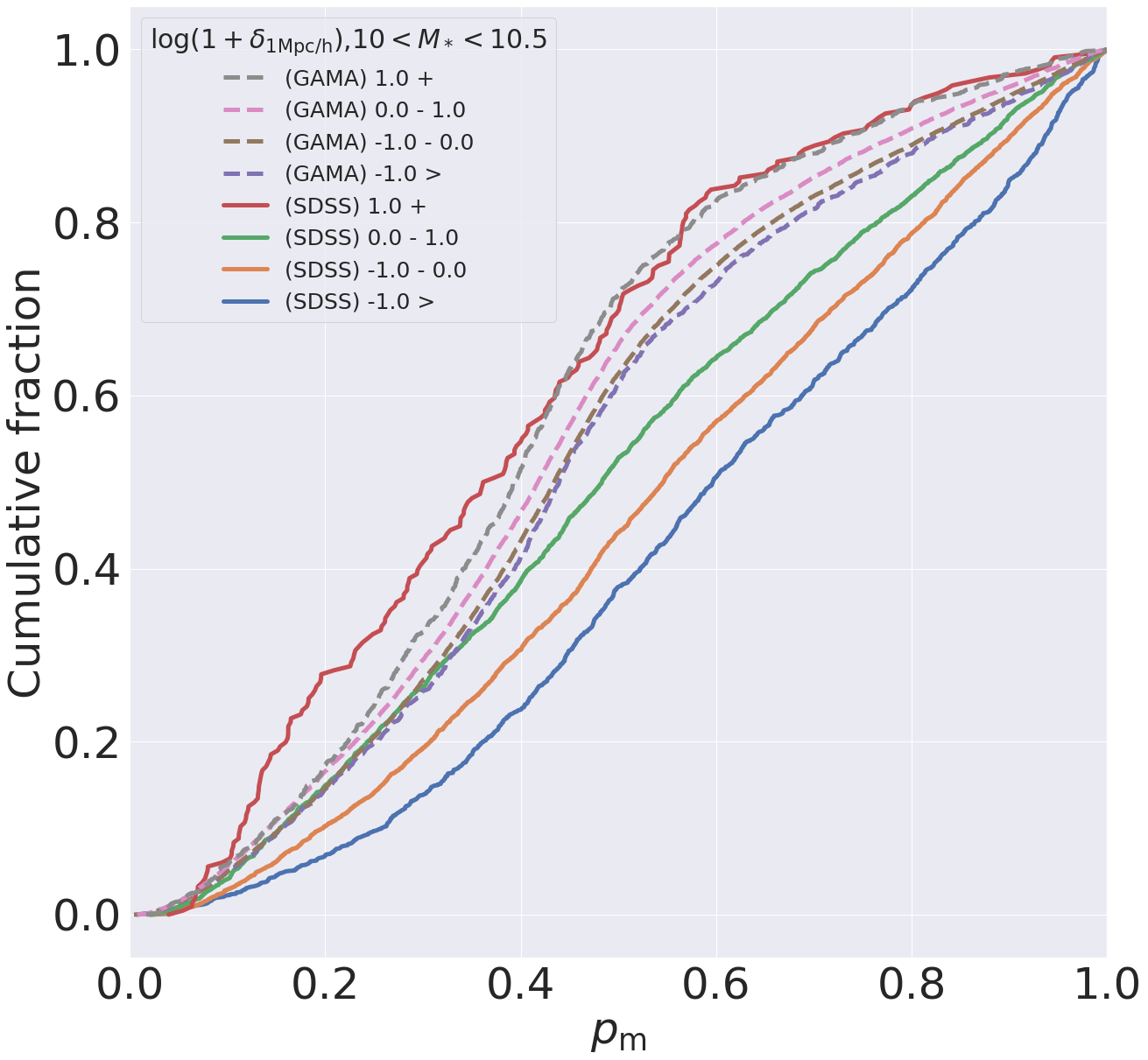}
            \caption[Network2]%
            {{\small $10 < \log(M_{*}) < 10.5$\\ KS test (SDSS) statistic: 0.352% p-value: $3.687\times10^{-21}$ 
            \\ KS test (GAMA) statistic: 0.113}}% p-value: $4.705\times10^{-8}$}}}    
            \label{fig:1Mpc10105}
        \end{subfigure}
        \hfill
        \centering
        \begin{subfigure}[b]{0.475\textwidth}
            \centering
            \includegraphics[width=\textwidth]{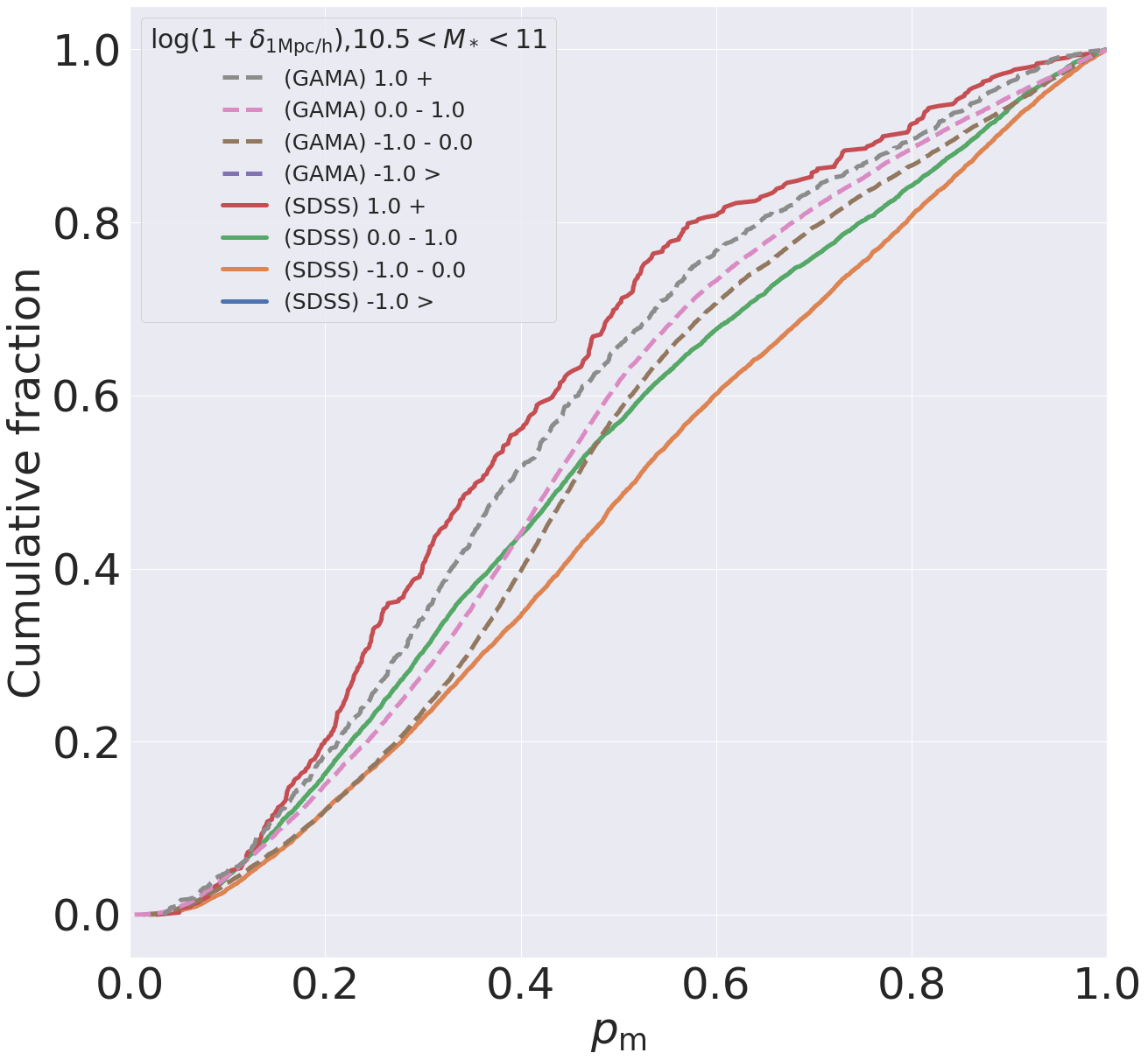}
            \caption[Network2]%
            {{\small $10.5 < \log(M_{*}) < 11$\\ KS test (SDSS) statistic: 0.241 %p-value: $2.377\times10^{-21}$ 
            \\ KS test (GAMA) statistic: 0.137}}% p-value: $4.489\times10^{-16}$}}}    
            \label{fig:1Mpc10511}
        \end{subfigure}
        % \begin{subfigure}[b]{0.350\textwidth}  
        %     \centering 
        %     \includegraphics[width=\textwidth]{5thnearestmini.png}
        %     \caption[]%
        %     {{\small mini Neighbors}}    
        %     \label{fig:5mini}
        % \end{subfigure}
        \caption[ The average and standard deviation of critical parameters ]
        {\small The same as Fig. \ref{fig:100kpcmassfixed} but for a 1 Mpc radius aperture} 
        \label{fig:1Mpcmassfixed}
\end{figure*}
\begin{figure*}
        \centering
        \begin{subfigure}[b]{0.475\textwidth}
            \centering
            \includegraphics[width=\textwidth]{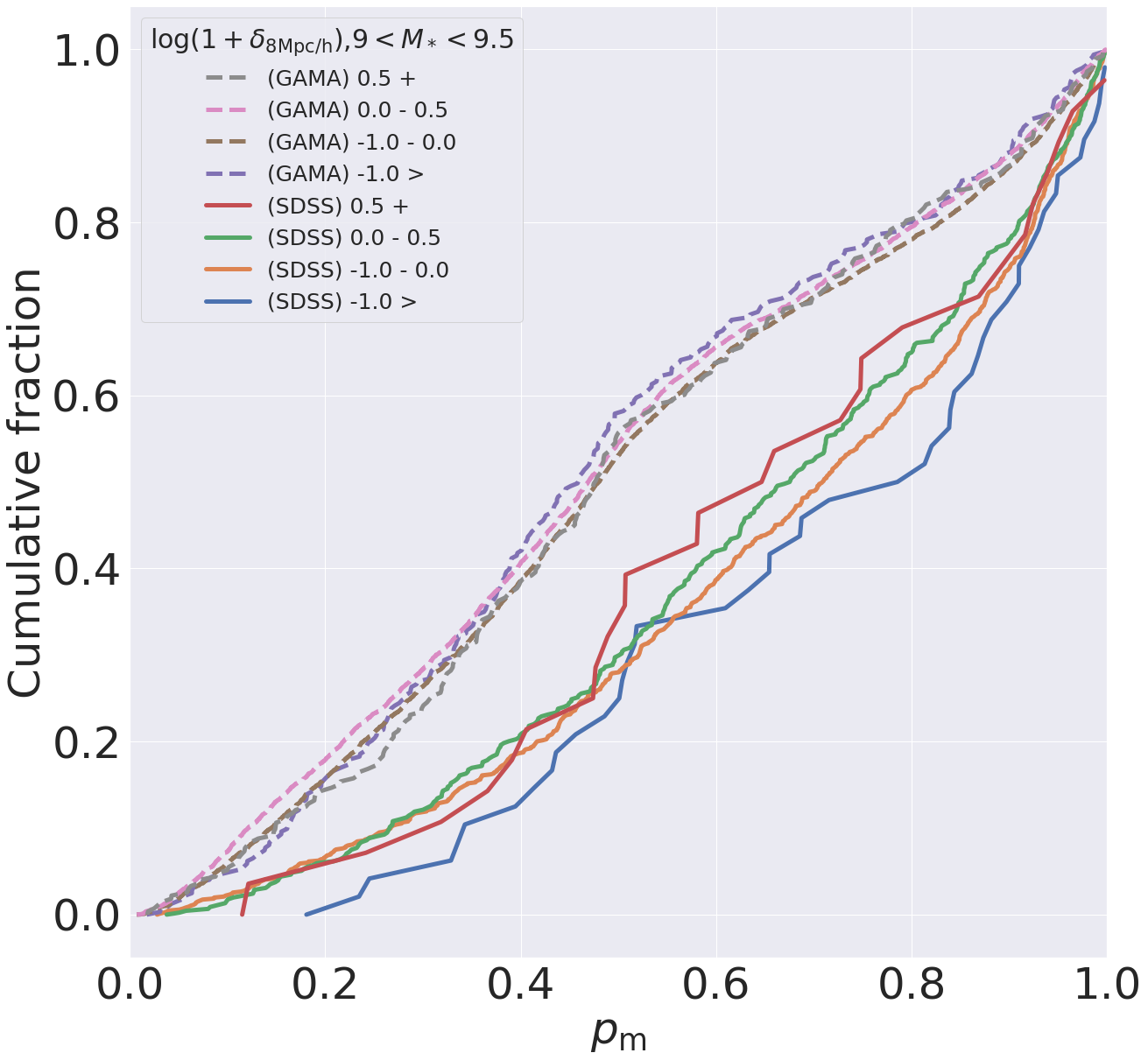}
            \caption[Network2]%
            {{\small $9 < \log(M_{*}) < 9.5$\\ KS test (SDSS) statistic: 0.193 p-value: 0.459 \\ KS test (GAMA) statistic: 0.053 p-value: 0.767}} 
            \label{fig:8Mpc995}
        \end{subfigure}
        \hfill
        \centering
        \begin{subfigure}[b]{0.475\textwidth}
            \centering
            \includegraphics[width=\textwidth]{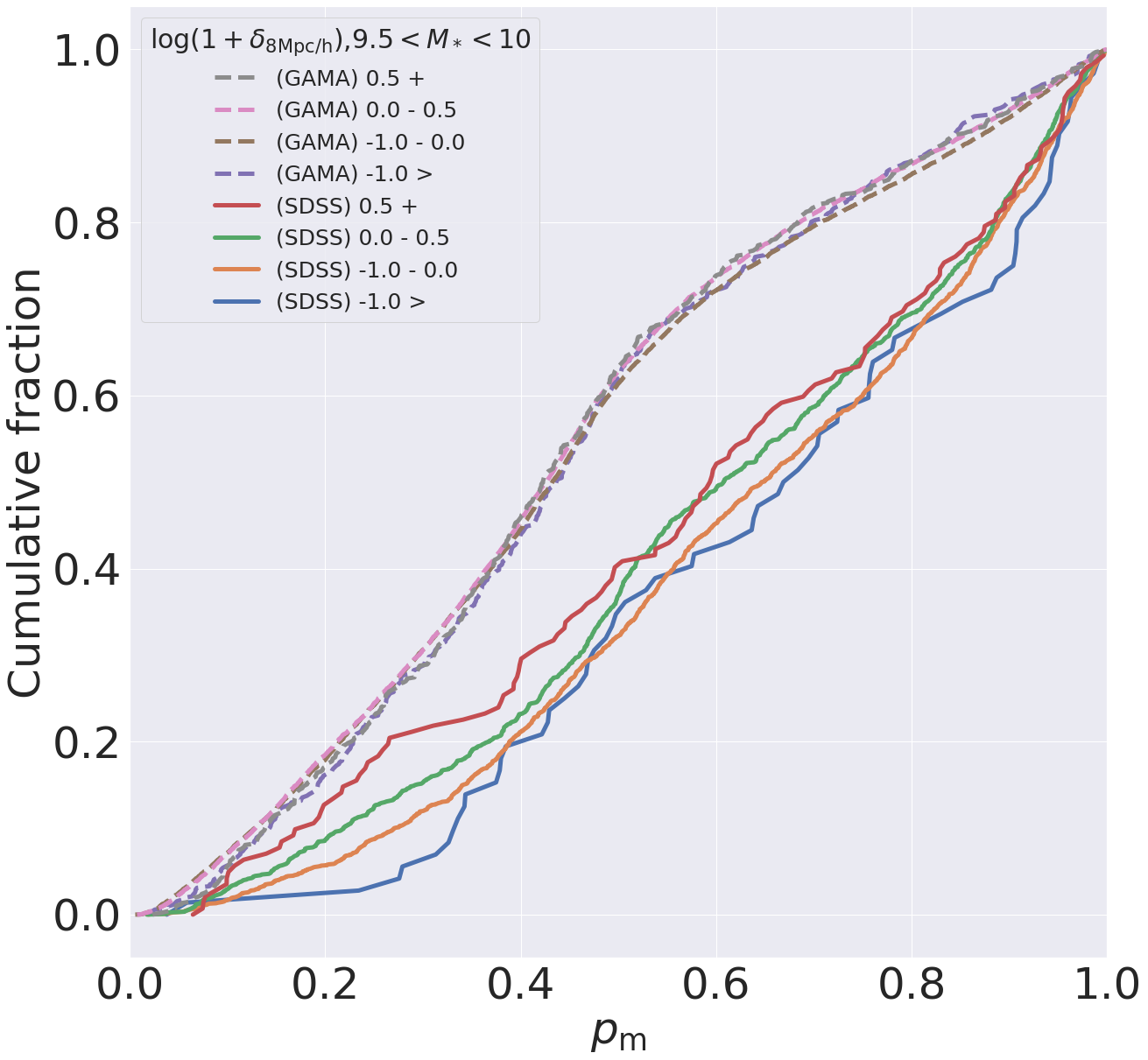}
            \caption[Network2]%
            {{\small $9.5 < \log(M_{*}) < 10$\\ KS test (SDSS) statistic: 0.170 p-value: 0.111 \\ KS test (GAMA) statistic: 0.036 p-value: 0.822}}    
            \label{fig:8Mpc9510}
        \end{subfigure}
        \vskip\baselineskip
                \centering
        \begin{subfigure}[b]{0.475\textwidth}
            \centering
            \includegraphics[width=\textwidth]{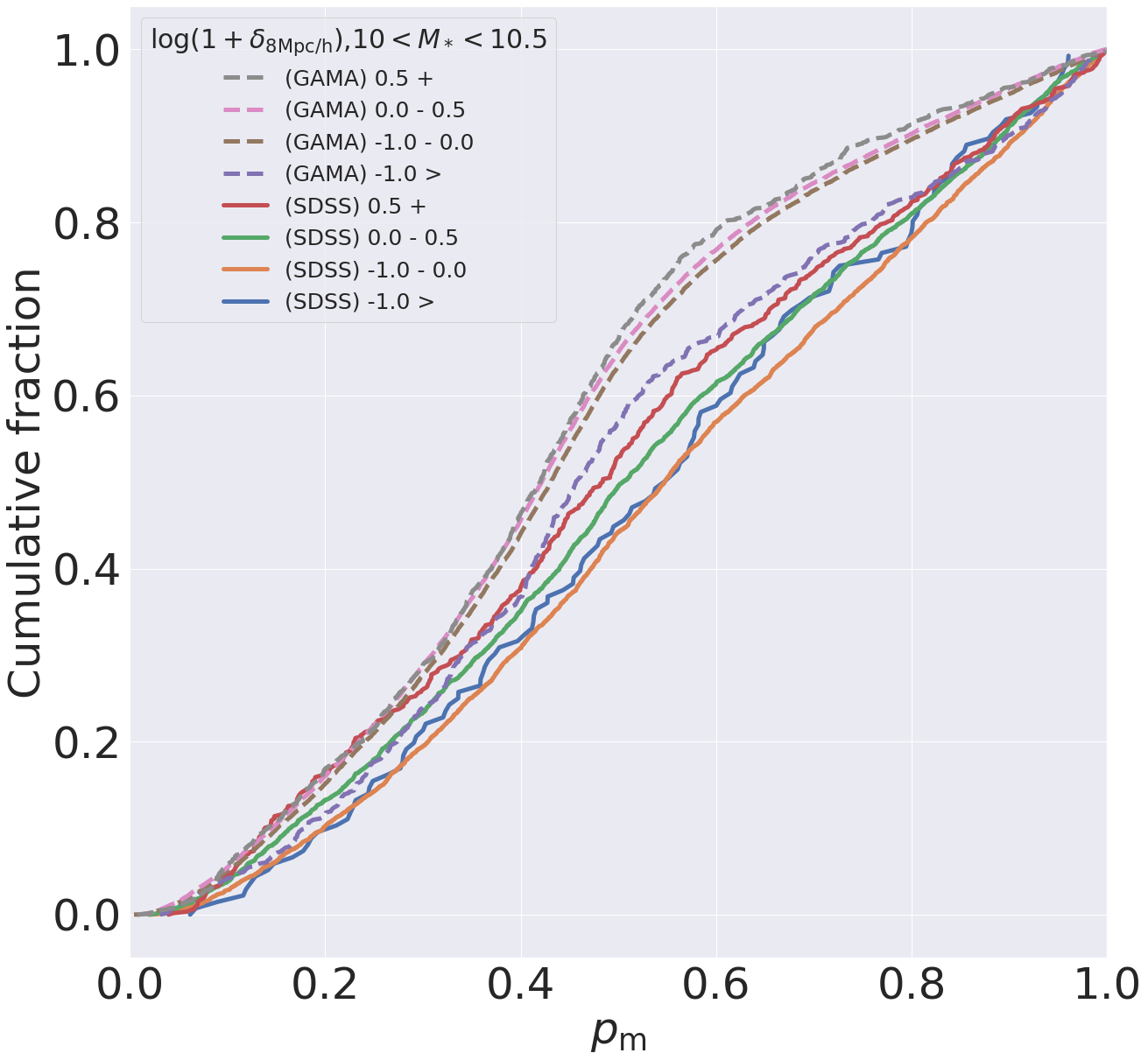}
            \caption[Network2]%
            {{\small $10 < \log(M_{*}) < 10.5$\\ KS test (SDSS) statistic: 0.107 p-value: 0.143 \\ KS test (GAMA) statistic: 0.125 p-value: $7.059\times10^{-7}$}}   
            \label{fig:8Mpc10105}
        \end{subfigure}
        \hfill
        \centering
        \begin{subfigure}[b]{0.475\textwidth}
            \centering
            \includegraphics[width=\textwidth]{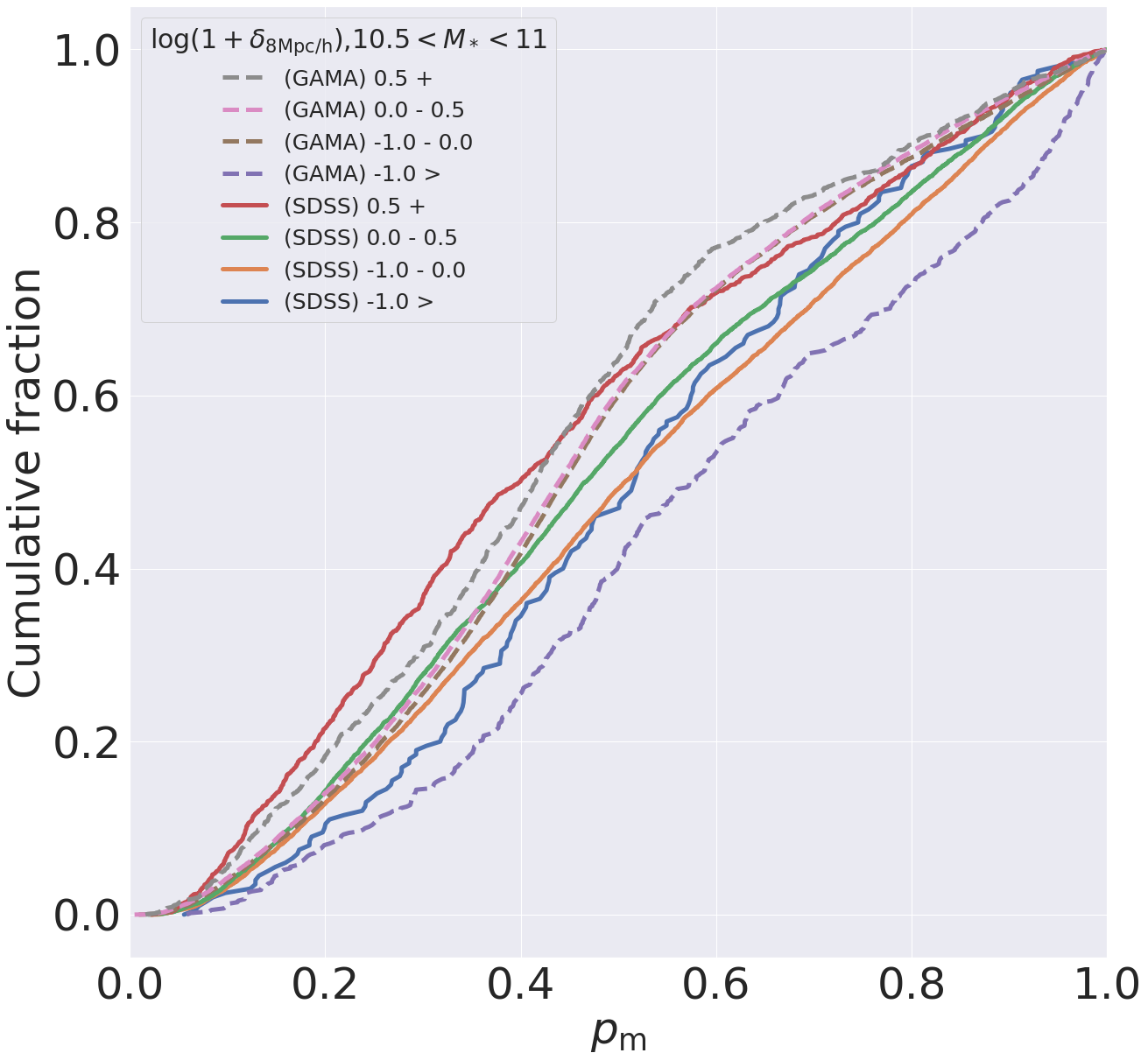}
            \caption[Network2]%
            {{\small $10.5 < \log(M_{*}) < 11$\\ KS test (SDSS) statistic: 0.200 %p-value: $1.461\times10^{-6}$ 
            \\ KS test (GAMA) statistic: 0.255}}% p-value: $1.179\times10^{-23}$}}}    
            \label{fig:8Mpc10511}
        \end{subfigure}
        % \begin{subfigure}[b]{0.350\textwidth}  
        %     \centering 
        %     \includegraphics[width=\textwidth]{5thnearestmini.png}
        %     \caption[]%
        %     {{\small mini Neighbors}}    
        %     \label{fig:5mini}
        % \end{subfigure}
        \caption[ The average and standard deviation of critical parameters ]
        {\small The same as Fig. \ref{fig:100kpcmassfixed} but for a 8 Mpc radius aperture} 
        \label{fig:8Mpcmassfixed}
\end{figure*}
\begin{figure*}
        \centering
        \begin{subfigure}[b]{0.475\textwidth}
            \centering
            \includegraphics[width=\textwidth]{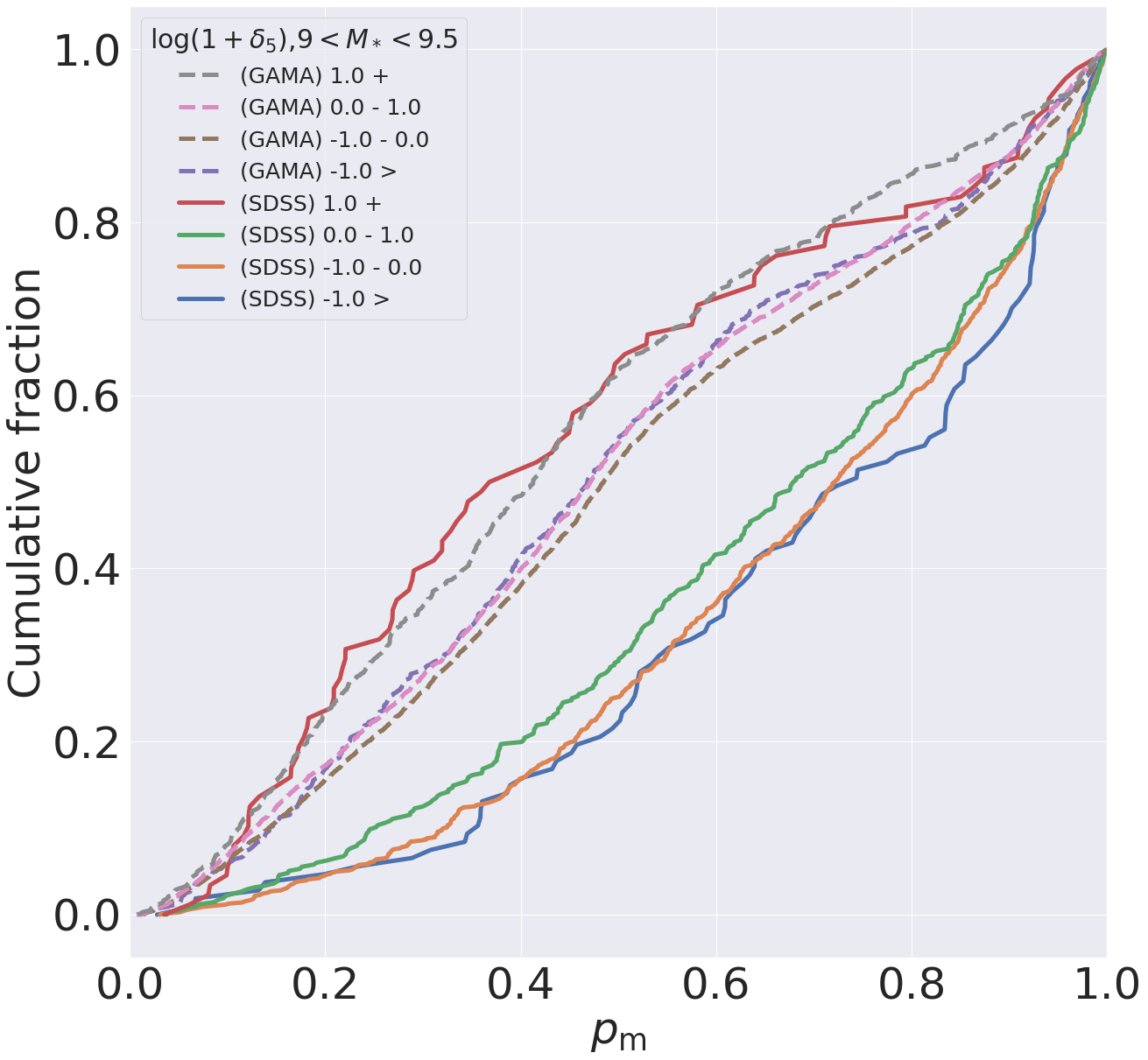}
            \caption[Network2]%
            {{\small $9 < \log(M_{*}) < 9.5$\\ KS test (SDSS) statistic: 0.423 %p-value: $2.632\times10^{-8}$ 
            \\ KS test (GAMA) statistic: 0.094}}% p-value: 0.003}}}    
            \label{fig:5thnearest995}
        \end{subfigure}
        \hfill
        \centering
        \begin{subfigure}[b]{0.475\textwidth}
            \centering
            \includegraphics[width=\textwidth]{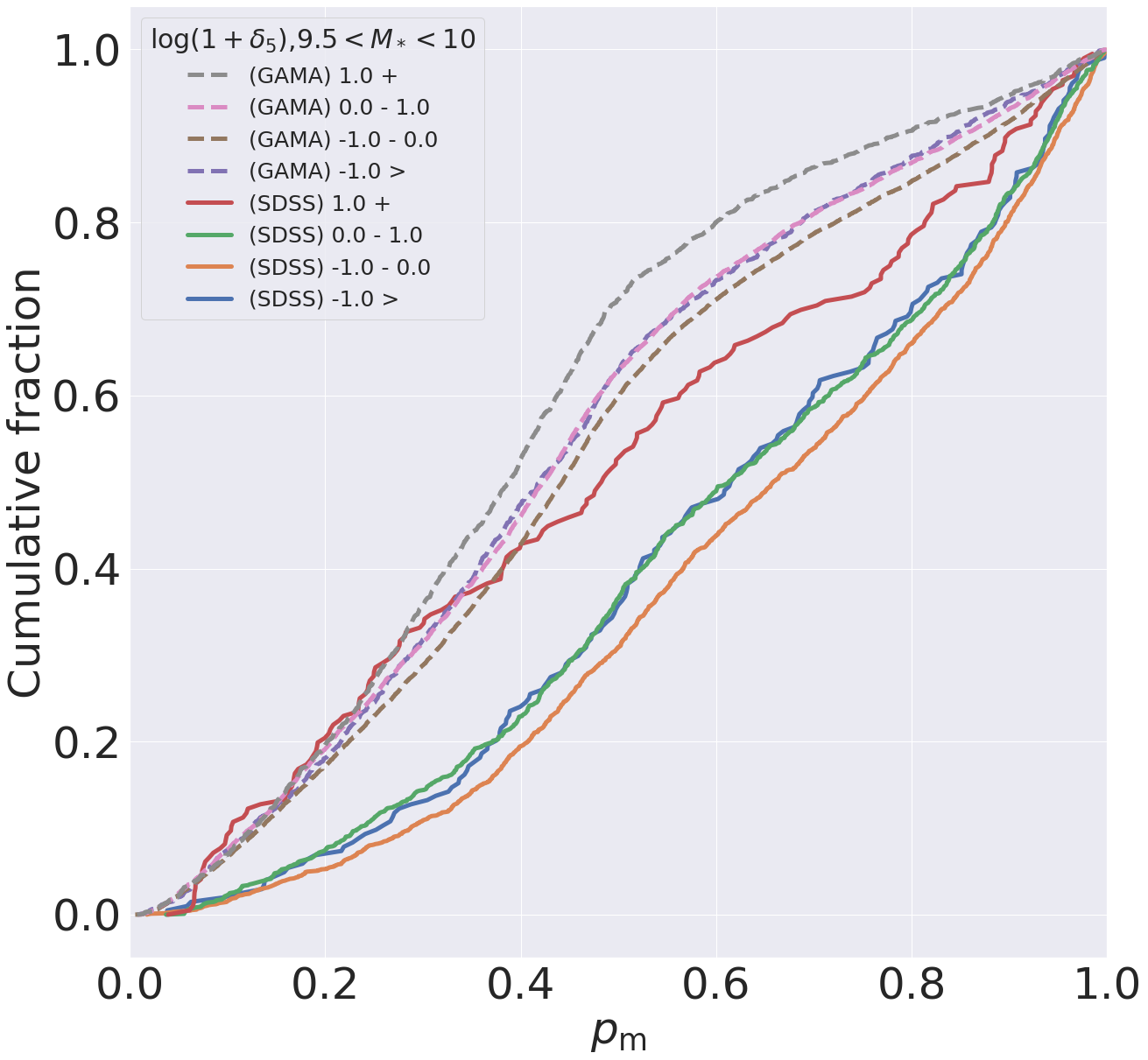}
            \caption[Network2]%
            {{\small $9.5 < \log(M_{*}) < 10$\\ KS test (SDSS) statistic: 0.220 %p-value: $9.292\times10^{-5}$ 
            \\ KS test (GAMA) statistic: 0.091}}% p-value: $2.383\times10^{-6}$}}}    
            \label{fig:5thnearest9510}
        \end{subfigure}
        \vskip\baselineskip
                \centering
        \begin{subfigure}[b]{0.475\textwidth}
            \centering
            \includegraphics[width=\textwidth]{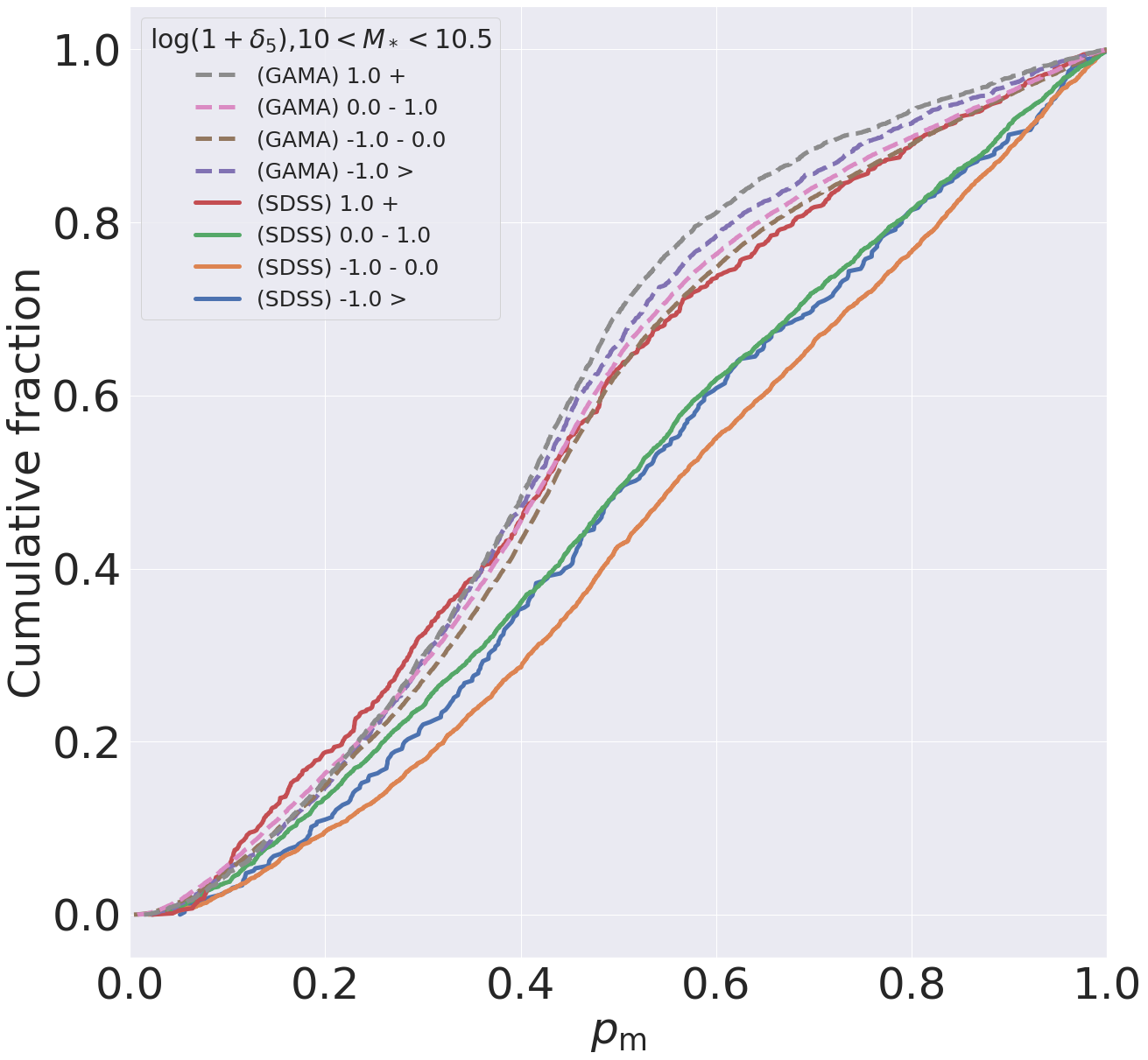}
            \caption[Network2]%
            {{\small $10 < \log(M_{*}) < 10.5$\\ KS test (SDSS) statistic: 0.152 %p-value: $3.325\times10^{-5}$ 
            \\ KS test (GAMA) statistic: 0.037 p-value: 0.045}}    
            \label{fig:5thnearest10105}
        \end{subfigure}
        \hfill
        \centering
        \begin{subfigure}[b]{0.475\textwidth}
            \centering
            \includegraphics[width=\textwidth]{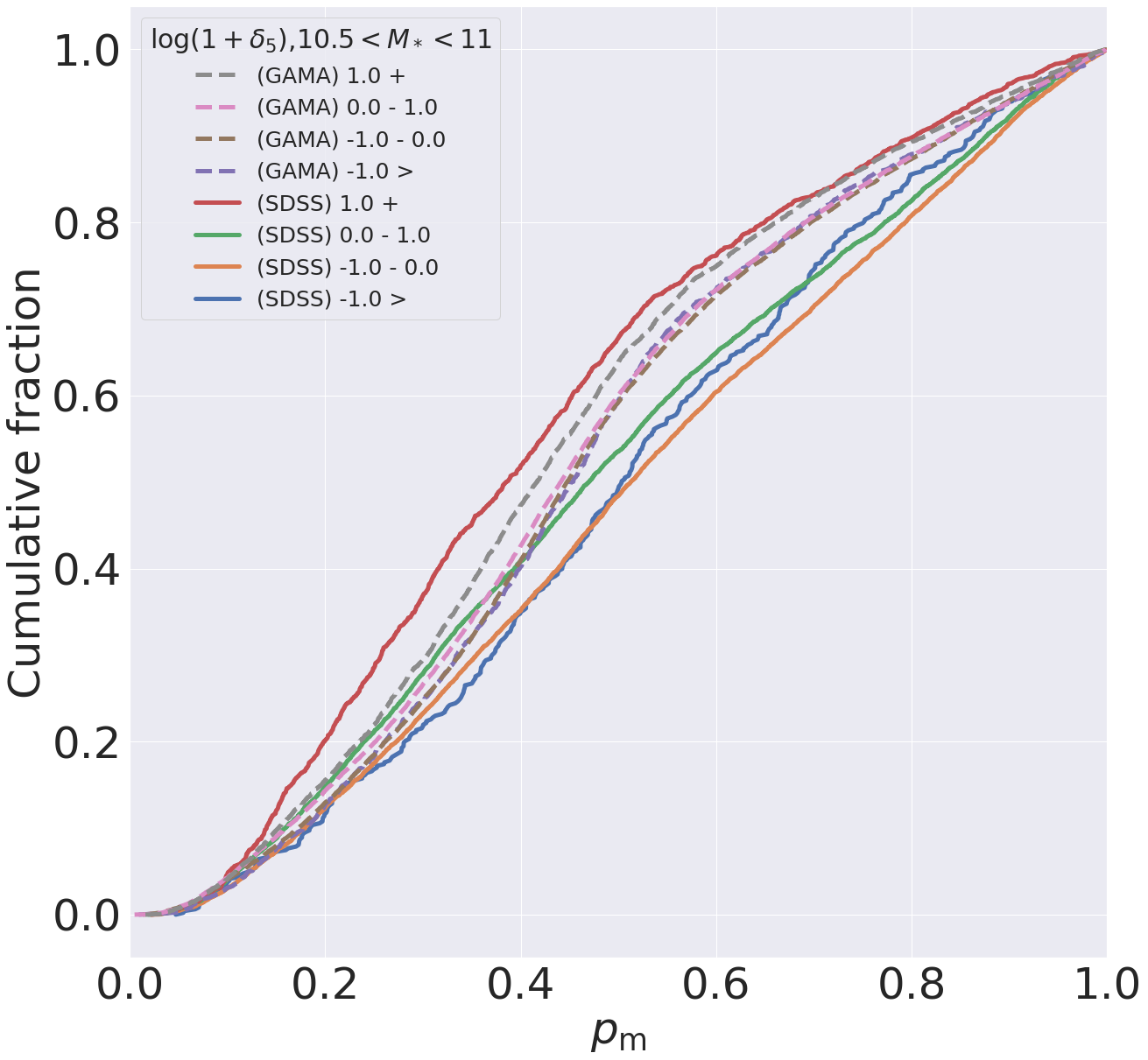}
            \caption[Network2]%
            {{\small $10.5 < \log(M_{*}) < 11$\\ KS test (SDSS) statistic: 0.190 %p-value: $1.653\times10^{-12}$ 
            \\ KS test (GAMA) statistic: 0.075}}% p-value: $2.087\times10^{-6}$}}}    
            \label{fig:5thnearest10511}
        \end{subfigure}
        % \begin{subfigure}[b]{0.350\textwidth}  
        %     \centering 
        %     \includegraphics[width=\textwidth]{5thnearestmini.png}
        %     \caption[]%
        %     {{\small mini Neighbors}}    
        %     \label{fig:5mini}
        % \end{subfigure}
        \caption[ The average and standard deviation of critical parameters ]
        {\small The same as Fig. \ref{fig:100kpcmassfixed} but for an aperture of radius up to the 5th nearest neighbor.}
        \label{fig:5envhistmassfixed}
\end{figure*}

% \begin{figure}[h!]

%     \centering
%     \begin{subfigure}[b]{0.5\textwidth}
%         \includegraphics[width=\textwidth]{observationacc.png}
%         \subcaption{Accuracy}
%         \label{fig:oacc}
%     \end{subfigure}
% %
%     \begin{subfigure}[b]{0.5\textwidth}
%         \includegraphics[width=\textwidth]{observationloss.png}
%         \subcaption{Loss}
%         \label{fig:oloss}
%     \end{subfigure}
% %
%     \begin{subfigure}[b]{0.5\textwidth}
%         \includegraphics[width=\textwidth]{observationvalacc.png}
%         \subcaption{Validation accuracy}
%         \label{fig:ovalacc}
%     \end{subfigure}
%     \begin{subfigure}[b]{0.5\textwidth}
%         \includegraphics[width=\textwidth]{observationvalloss.png}
%         \subcaption{Validation loss}
%         \label{fig:ovalloss}
%     \end{subfigure}
    
%     \caption{Learning curves for accuracy, loss, validation accuracy, validation loss from fine-tuning Zoobot with TNG images for purpose of making predictions on observations. Training and validation datasets were split at 70\% and 30\%, respectively.}
%     \label{fig:ometrics}
% \end{figure}%

\end{document}